\documentclass{article}
\usepackage{tabulary,tabularx,graphicx,times,caption,fancyhdr,amsfonts,amssymb,amsbsy,latexsym,amsmath}
\usepackage[utf8]{inputenc}
\usepackage{url,multirow,morefloats,floatflt,cancel,tfrupee,textcomp,colortbl,xcolor,pifont}
\usepackage[nointegrals]{wasysym}
\usepackage{longtable}
\usepackage{float}
\usepackage{caption}
\usepackage{subcaption}
\usepackage{hyperref}       
\usepackage[noabbrev]{cleveref}
\usepackage{array}
\usepackage{float,xcolor}
\usepackage{arxiv}
\usepackage[section]{placeins}
\usepackage[utf8]{inputenc} 
\usepackage[T1]{fontenc}    

\usepackage{url}            
\usepackage{booktabs}       
\usepackage{amsfonts}       
\usepackage{nicefrac}       
\usepackage{microtype}      
\usepackage{lipsum}		
\usepackage{graphicx}
\usepackage{natbib}
\usepackage{doi}

\title{Exploring the Dynamics of Fungal Cellular Automata}

\date{September 30, 2022}	

\author{ \href{https://orcid.org/0000-0003-2426-7634}{\includegraphics[scale=0.06]{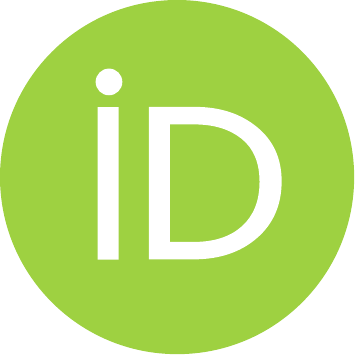}\hspace{1mm}Carlos S.~ Sepúlveda}\thanks{Directorate of Programs, Research and Development, Chilean Navy, email: csepulvedam@armada.cl} \\
	Facultad de Ingeniería y Ciencias\\
	Universidad Adolfo Ibáñez\\
	Santiago, Chile\\
	\texttt{carlos.sepulveda@alumnos.uai.cl} \\
	\And
	Eric Goles \\
	Facultad de Ingeniería y Ciencias\\
	Universidad Adolfo Ibáñez\\
	Santiago, Chile\\
	\texttt{eric.chacc@uai.cl} \\
	\AND
	Martín Ríos-Wilson\\
	Facultad de Ingeniería y Ciencias\\
	Universidad Adolfo Ibáñez\\
	Santiago, Chile\\
	\texttt{martin.rios@uai.cl} \\
	\And
	Andrew Adamatzky \\
	Unconventional Computing Laboratory\\
	UWE Bristol \\
	Bristol, UK \\
	\texttt{andrew.adamatzky@uwe.ac.uk} \\
}

\hypersetup{
	pdftitle={Exploring the Dynamics of Fungal Cellular Automata},
	pdfsubject={nlim.CG, cs.ET},
	pdfauthor={Carlos S.~Sep\'ulveda, Eric Goles, Mart\'in R\'ios-Wilson, Andrew Adamatzky},
	pdfkeywords={cellular automata, unconventional computing, discrete dynamical systems, complex systems},
}

\begin{document}
	\maketitle
	
	\begin{abstract}
		Cells in a fungal hyphae are separated by internal walls (septa). The septa have tiny pores that allow cytoplasm flowing between cells. Cells can close their septa blocking the flow if they are injured, preventing fluid loss from the rest of filament. This action is achieved by special organelles called Woronin bodies. Using the controllable pores as an inspiration we advance one and two-dimensional cellular automata into Elementary fungal cellular automata (EFCA) and Majority fungal automata (MFA) by adding a concept of Woronin bodies to the cell state transition rules. EFCA is a cellular automaton where the communications between neighboring cells can be blocked by the activation of the Woronin bodies (Wb), allowing or blocking the flow of information (represented by a cytoplasm and chemical elements it carries) between them. We explore a novel version of the fungal automata where the evolution of the system is only affected by the activation of the Wb.  We explore two case studies:  the Elementary Fungal Cellular Automata (EFCA), which is a direct application of this variant for elementary cellular automata rules, and the Majority Fungal Automata (MFA), which correspond to an application of the Wb to two dimensional automaton with majority rule with Von Neumann neighborhood. By studying the EFCA model, we analyze how the 256 elementary cellular automata rules are affected by the activation of Wb in different modes, increasing the complexity on applied rule in some cases. Also we explore how a consensus over MFA is affected when the continuous flow of information is interrupted due to the activation of Woronin bodies.
	\end{abstract}


	\section{Introduction}\label{doc:intro}
	
	The fungi kingdom is one of the widest spread form of life on earth. Without them our planet landscape would be totally different. Life on land has evolved with the participation of fungi and would collapse without their continued activities~\cite{watkinson2015fungi}. Fungal morphology is based on hyphae, which are long and branching filaments. Collectively called mycelium, they form the vegetative body on fungi. Hyphae are formed by one or more cells enclosed by a tubular cell wall. In Ascomycota, one of the several divisions of the fungi, hyphae are divided into compartments by internal cross-walls named septum, formed by centripetal growth of the cell wall and crossed by a perforation through which cytoplasmic organelles can pass. An electron-dense protein body called Woronin body (Wb), is present on either side of the septa, regulating the opening and closing of the septal pores which is used to reduce or cut the flow of cytoplasm and organelles between cells compartments when the hypha is ruptured~\cite{maheshwari2016fungi}.
	
	Fungal physiology and behaviour gave rise to a novel field of fungal computing and fungal electronics~\cite{adamatzky2018towards,adamatzky2020boolean, beasley2021mem}. Whilst experimental laboratory prototyping of fungi-based computing devices is underway it is imperative to establish a wider theoretical background for fungal computing. This is why we drawn our attention to developing formal models of fungal automata. First steps in these theoretical designs have been done in \cite{goles2020computational,goles2021generating, adamatzky2020fungal}. In present paper we advance the ideas into the dynamical properties of a model based on the elementary cellular automata (ECA). This is a classic model that describes a vast variety of natural dynamical phenomena over the years. One of the aspect that is most interesting of ECA model is that, even when it is based in a simple set of local rules that can be quite straightforward, its global dynamical behavior can be extremely complex. In fact, since the end of the 1970's, the behavioural complexity emerging from normative simplicity motivated a wide range of researcher to understand and classify ECA rules according to their dynamic behavior and computational capabilities \cite{wolfram1984cellular, wolfram2018cellular}. A first well known example is Wolfram's classification which considers different criteria to cluster rules in four groups according to their dynamical behavior starting from random initial conditions \cite{wolfram1983statistical}.  A second way of grouping the elementary rules is according to their equivalences up to simple transformations of their local transitions, such as \emph{reflection}, \emph{conjugation} and the combination of both. There are 88 elementary cellular automata that are non-equivalent up to these transformations.  A third important example of rule classification is in the study of the computational capabilities of a given rule \cite{moore2002computational,moore1997majority,griffeath1996life}.  For instance, rule 110 is capable of representing universal Turing computation in its dynamics \cite{neary2006p}.
	Within this framework, different cellular automata models, inspired from fungi hyphae behavior have been recently proposed. For instance, in ~\cite{adamatzky2020fungal} the one-dimensional fungal automaton has been introduced. This model is based on the composition of two elementary cellular automaton functions, one controlling the activation state of the Wb and the other controlling the automaton evolution. In addition, in~\cite{goles2020computational}, the authors have extended the concept to two dimensions implementing a fungal sandpile automata, and shown the computational universality of this FCA.
	
	We present novel variant of the one-dimensional fungal automaton model, called elementary fungal cellular automata (EFCA). By numerical simulations of this model, we exhaustively explore the impact of Wb activation on the dynamics of 88 non-equivalent ECA rules. In addition, based on the fungal sandpile automata, we develop a two dimensional cellular automaton ruled by the well-known majority rule cellular automaton with the Von-Neumann neighborhood.  We focus on observing how different choices for the activation of Wb can produce different dynamical behavior on both cases. We accomplish this task by proposing different metrics such as the magnetization and Hamming distance in order to compare the original rules with the ones in which the Wb are activated.  Finally, for majority rule, we study how the besides  Wb activation, which temporally block information flow the consensus of the network is not affected for strict majority rule and the consensus skew is accelerated in the skew majority rule.
	
	The paper is organized as follows.  First we present the set-up for our simulations, including the metrics we used for our analysis. There we discuss  the simulation set-up for our analysis of one-dimensional elementary cellular automata rules and the simulation set-up for the majority fungal automata rules. Then we analyze the results of  the numerical experiments respecting the same organization of the latter section.	 
	
	\section{Preliminaries}
	
	\subsection{Elementary Fungal Cellular Automata}\label{doc:1FCA}
\begin{figure}[!htb]
		\centering
		\includegraphics[width=0.7\linewidth]{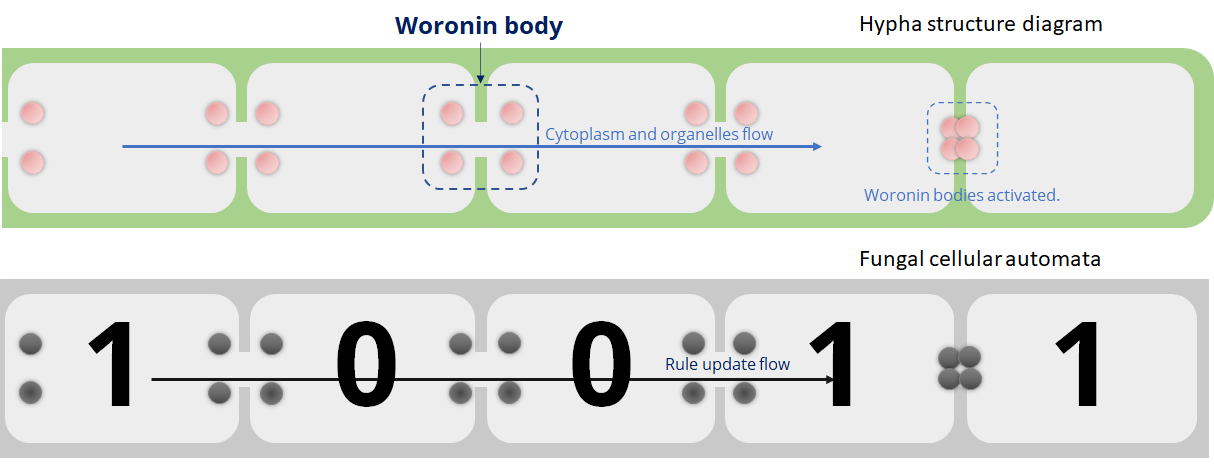}
		\caption{Fungal cellular automata structure based on fungal hypha. At the top is the biological scheme of the hypha. At the bottom  is a one-dimensional FCA, where each cell has a Woronin body (Wb) that can be activate/deactivated blocking or allowing to see the current state value of his neighbors. }
		\label{fig:hyphafca}
	\end{figure}

	An elementary fungal cellular automaton (EFCA) is an elementary cellular automata (ECA) where adjacent cells can cut the flow of information~\footnote{The rule according which the automaton changes his current state} between them by the activation of Wb (see Figure~\ref{fig:hyphafca}). This means that cells with activated Wb can't see the current state of their neighbors, which produces a miss information and ambiguity in applying the ECA rules and updating the cells' states. 
	
	FCA were first proposed on~\cite{adamatzky2020fungal}, implemented as one-dimensional cellular automata~\footnote{A ring of size $N$ which update his current state according to a rule} with cell binary state and governed by two ECA rules, one for the cell state transition rule $f(\cdot)$ and other the activation of Wb $g(\cdot)$. In this manner, they implemented two species of FCA on which the activation of Wb is given by $w^{t+1} = { g\left({u\left({x}\right)}^{t}\right)}$, where $u{\left({x}\right)}^t$ represents the neighborhood of cell $x$ at instant $t$ and the update of cell $x$ for the first specie, is given by equation~\ref{eq:Adm1} and for the second specie by equation~\ref{eq:Adm2}.
	\begin{align}
		\label{eq:Adm1} x^{t+1}_{i} = \begin{cases}
			0	& \text{if } w^t = 1 \\
			f\left({u\left({x_i}\right)}^t\right) & \text{otherwise}
		\end{cases}	
		\\
		\label{eq:Adm2} x^{t+1}_{i} = \begin{cases}		
			x^t	& \text{if } w^t = 1 \\		
			f\left({u\left({x_i}\right)}^t\right) & \text{otherwise}
		\end{cases} 
	\end{align}	
	Our approach is  different. First the Wb state is not governed by current cell state, instead is externally controlled, having states $Q=\lbrace 0,1\rbrace $, representing deactivation and activation respectively. 
	Our formulation for the update of cell state is the following:
	Each cell $x_i$ has a unique index $i \in \lbrace 0,1,\ldots, n \rbrace$ in a ring of size $n$, with $n \in \mathbb{N}$, and a Wb ${Wb}_i$. If the Wb is activated, $Q:{Wb}_{i} = 1$, there is no communication between cell $x_i$ and cell $x_{i+1}$.
	Special case is when $i = n$, due the implementation is in a ring there will be no communication between $x_n$ and $x_0$.
	Every cell $x_i$ has two neighbors at distance 1, $x_{i-1}$ to the left and $x_{i+1}$ to the right. The function $f(\cdot)$ is then applied to the triplet which leads to a traditional ECA. Considering four cells $\left({x_{i-1}, x_{i}, x_{i+1}, x_{i+2}}\right)$ and the activation of Woronin body at position $i$, when the rule  $f(x_{i-1}, x_{i}, x_{i+1})$ is applied we will not able to see the content of cell at  $i+1$ position due to the Wb activation and therefore  we will face an  ambiguity of information at the left of the Woronin body activation. Then, applying the rule to the next triplet $f(x_{i}, x_{i+1}, x_{i+2})$ will produce an ambiguity information to the right of Wb activation. We can manage the information ambiguity in both sides of Wb activation when $f(x_{i-1}, x_{i}, 0) = f(x_{i-1}, x_{i}, 1)$ for rules at left of activated Wb and $f(0, x_{i+1}, x_{i+2}) = f(1, x_{i+1}, x_{i+2})$ if not, we remain in actual state. The explained behavior is equivalent to application of the rule at left side of Wb activation and other, different, rule at the right side of Wb. The formalization of this behavior, that describes the evolution of EFCA from instant $t$ to instant $t+1$ is given by equation~\ref{eq:csm}, as follow:
	\begin{equation}
		\label{eq:csm}
		x^{t+1}_{i} = \begin{cases}
			& \text{if $w^{t}_{i} =1$ and $f \left(x^{t}_{i-1}, x^{t}_{i}, 1\right) \neq f\left(x^{t}_{i-1}, x^{t}_{i}, 0\right)$}\\
			x^{t}_{i}	& \qquad\qquad\qquad\text{ or}\\
			& \text{if $w^{t}_{i} =1$ and  $f\left(1, x^{t}_{i}, x^{t}_{i+1}\right) \neq f\left(0, x^{t}_{i}, x^{t}_{i+1}\right)$}\\
			\\
			f(x^{t}_{i-1}, x^{t}_{i}, x^{t}_{i+1}) & \text{otherwise}
		\end{cases}
	\end{equation}
	
	Here $f(\cdot)$ is the ECA rule. In cell whose Wb is not activated, $Q:{Wb}_i = 0$, the state update will be given by $x^{t+1}_{i} = f(x^{t}_{i-1}, x^{t}_{i}, x^{t}_{i+1})$ (his neighbors at distance 1). 
	
	\subsection{Majority fungal automata}\label{doc:2FCA}
	
	A two dimensional cellular automaton is a torus in which the value of a cell $x_{\left( i,j\right)}$ will be $1$ or $0$ depending on the value of a function $F(\cdot)$ whose apply over neighbors of $x_{\left( i,j\right)}$. We use von Neumann neighborhood of range 1 which is defined as follows:
	\begin{equation}
		\label{eq:vn}
		V\left(i_0,j_0\right) = \left\{{\left({i,j}\right): \mid{i-i_0}\mid + \mid{j-j_0}\mid \leq 1}\right\}
	\end{equation}
	The function $F(\cdot)$ determines the value of a cell based on his neighbors is the majority, meaning that the value of cell $x_{\left( i,j\right)}$ will be (zero or one) the most frequent value among neighborhood. In other words, the opinion of a subject will be the opinion of majority. Two types of majority are possible depending on how we deal with ties (equal number of zeros and ones among neighbors). We will refer to skew majority when in presence of a tie we force the value of cell $x_{\left( i,j\right)}$ to one (or zero depending on how the skew is defined). We will use majority when in presence of the tie the cell value will remain in his current state (no change of opinion). The study of this type of totalistic function is useful to understand consensus achievable over a network. 
	
\begin{figure}[!htb]
		\centering
		\begin{subfigure}[b]{.45\textwidth}
			\centering
			\includegraphics[width= \textwidth]{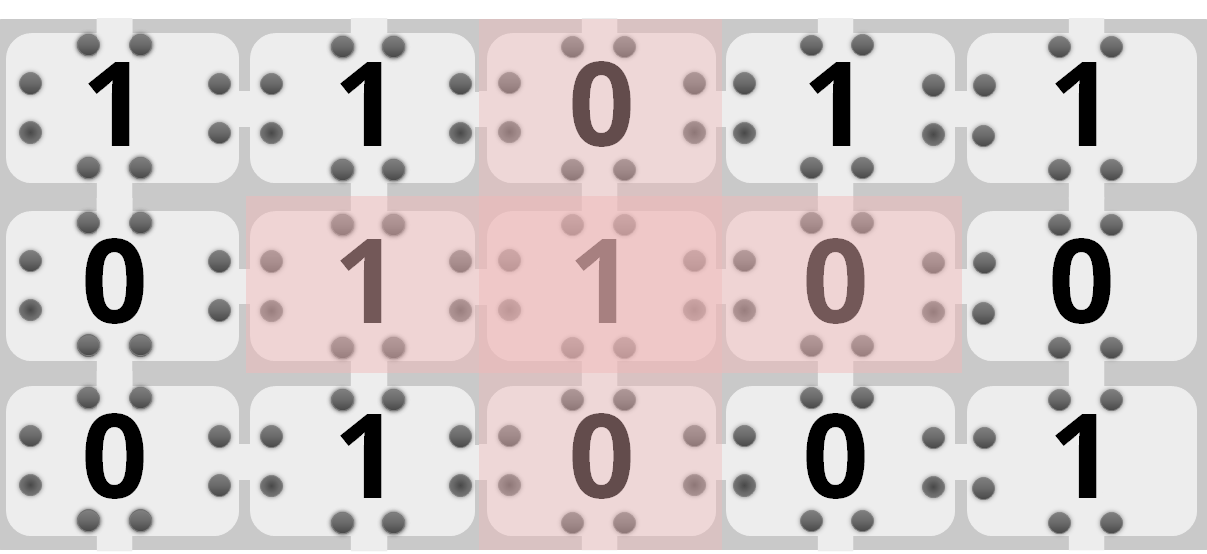}
			\caption{Full Von neumann neighborhood where none Woronin body is activated}
			\label{fig: _all}
		\end{subfigure}
		\hfill
		\begin{subfigure}[b]{.45\textwidth}
			\centering
			\includegraphics[width=\textwidth]{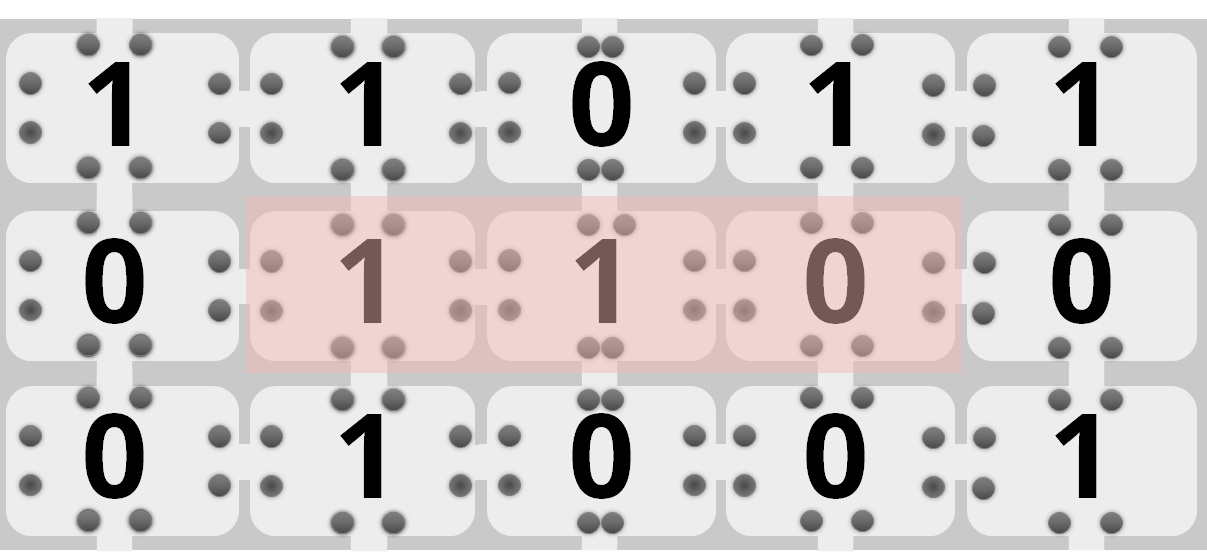}
			\caption{Von neumann neighborhood where vertical Woronin body are activated}
			\label{fig: _V}
		\end{subfigure}
		\newline
		\begin{subfigure}[b]{.45\linewidth}
			\centering
			\includegraphics[width=1\linewidth]{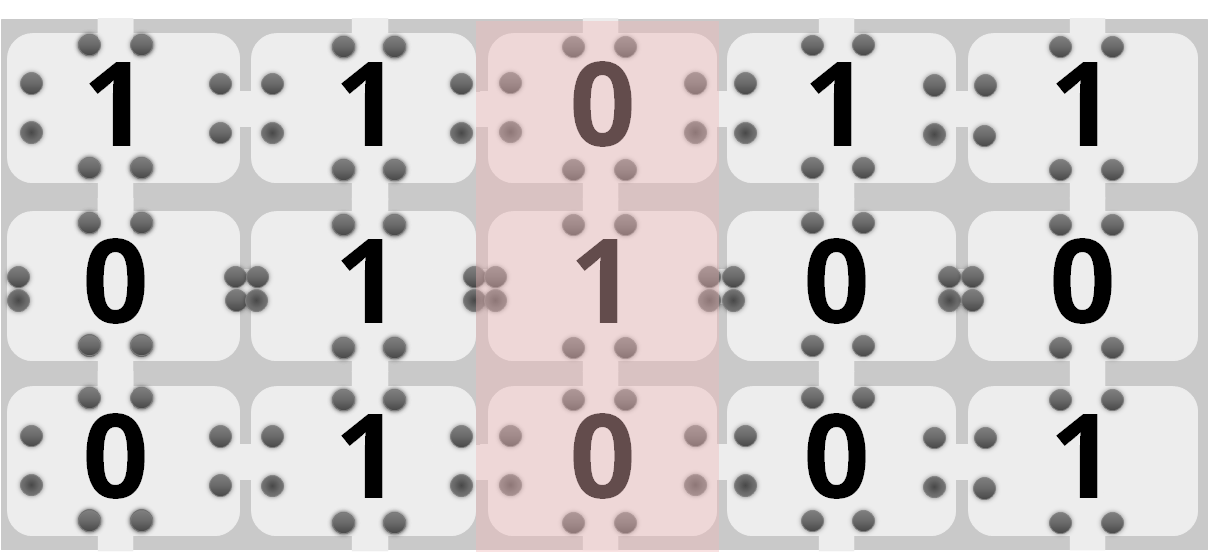}
			\caption{Von neumann neighborhood where horizontal Woronin body are activated}
			\label{fig: _H}
		\end{subfigure}	
		\caption{Majority fungal automata. Depending of which Wb are activate at time $t$ the neighbors able to see his current state change. The majority function in invariable, which means that takes the majority in this reduce neighbor.}
		\label{fig:2d-fca}
	\end{figure}

	In Majority fungal automaton (MFA) each cell $x_{\left( i,j\right)}$ has horizontal and vertical Wbs ${Wb}_{\left( i,j\right)}$ which can be activated by analogy with Wb activated in EFCA. The activation modifies neighbors that cell $x_{\left( i,j\right)}$ can see (Figure~\ref{fig:2d-fca}). Following the implementation used in~\cite{goles2020computational} we activated Wb groups by rows or columns. In this way, each cell  has a state $Q:\lbrace 0,1\rbrace $ which is updated according to a global function $F(\cdot)$. At each time step $t$ we could open or close all horizontal or vertical Wb, thus the MFA becomes a tuple $MFA = \left\langle {\mathbb{Z}^ 2, Q, V, F,w} \right\rangle 
	$, where $V$ is the von Neumann neighborhood, $F$ a global function (majority in this case), $w$ is a finite word on the alphabet H,V (horizontal, vertical) denoting which Wb are activated at each iteration of automata evolution.	

	\sloppy To implement majority function over a von Neumann neighborhood we denoted $V_{\left( i,j\right)} = \left[x_{\left( i-1,j\right)}, x_{\left( i+1,j\right)}, x_{\left( i,j-1\right)}, x_{\left( i,j+1\right)} \right]$, over this set  we apply the operator $S$ (sum of all elements due we work with zero and ones values) getting the following functions when none Wb is activated:
	\begin{center}
		\begin{itemize}
			\item \textbf{Strict Majority}: $\begin{cases}
				1 & \text{ if } S\left( {V_{\left( i,j\right)}}\right) > 2, \\
				x_{(i,j)}  & \text{ if }  S\left( {V_{\left( i,j\right)}}\right) = 2, \\
				0 &  \text{otherwise}
			\end{cases}$
			
			\item \textbf{Skew Majority}: $\begin{cases}
				1 & \text{ if } S\left( {V_{\left( i,j\right)}}\right) \geq 2, \\
				0 & \text{otherwise}
			\end{cases}$
		\end{itemize}
	\end{center}
	When the Wb is activated (vertical or horizontal) the neighbors of  $x_{\left( i,j\right)}$ are reduced at half, so we have two options that is  a conservative approach --- we take the majority of this reduced set or we sustain the original threshold which is equivalent to skew to the opposite value. When Wbs are activated the following functions are employed:
	\begin{itemize}
		\item \textbf{Strict Majority}: $\begin{cases}
			1 & \text{ if } S\left( {V_{\left( i,j\right)}}\right) > 1, \\
			x_{(i,j)}  & \text{ if }  S\left( {V_{\left( i,j\right)}}\right) = 1, \\
			0 &  \text{otherwise}
		\end{cases}$
		
		\item \textbf{Skew Majority}: $\begin{cases}
			1 & \text{ if } S\left( {V_{\left( i,j\right)}}\right) = 2, \\
			0 & \text{otherwise}
		\end{cases}$
	\end{itemize}

	
	
	\section{Simulation description and metrics}
	Several types of numerical experiments were carried on FCA and MFA to evaluate the impact of Wba on the behavior of the dynamics related to each cellular automata rule.
	We describe how the experiments were implemented and also the metrics we used to study them.
	
	\subsection{EFCA rules}
	\paragraph{First simulation}\label{doc:MFA} EFCA were implemented on a ring of size 100, with a random initial condition. Then the automata were evolving according to the specified rules for 99 generations given a final matrix of $100 \times 100$. Based on the complete review of ECA rules classification in~\cite{martinez2013note}, we only used the 88 non-equivalent rules of ECA rule-space. Every cell is binary, so we call magnetization the number of cells whose state is one. Initial EFCA state is randomly selected employing $\left[1, 10, 30, 60 \right]$ magnetizations. For every selected initial state five different Wb activation were explored, plus a base case where none Wb is activated (this corresponds to a traditional ECA). The modes in which Wbs were activated are following:
	\begin{itemize}
		\item 1Wb-on: Only the Wb between the last and first element of the ring was activated during 99 generations. The rule behaves in a constraint array instead of a ring.
		\item 4Wb-on: Four Wbs  regularly placed every N/4 of ring length (N) and activated during all generations.		
		\item 4Wb-mod: The same four Wbs previously mentioned, but with cycles of activation/deactivation every 10 steps.	
		\item 4Wb-c\&r (cut and release):
		The same four Wbs activated by 20 first steps and then deactivated until the final step.		
		\item allWb-on:
		every cell with its Wbs is activated to see how rule acts in a single cell in isolation.
	\end{itemize}	
	
	In order to get insights into how the activation of Wb affects EFCA, we measure the magnetization index $m$, defined at $t$ instant as:
	\begin{equation}
		\label{eq:magIdx}
		m_{t} = \sum_{i=0}^{n}{x_{i}^t}
	\end{equation}
	
	Also relative hamming distance $\delta_t$ between two consecutive generations is measured,  defined as:
	\begin{equation}
		\label{eq:relHam}
		\begin{split}
			&\delta_t = \frac{\Delta(X_{t-1}, X_{t})}{n} \quad \text{with}\quad t \geq 1, n \in \mathbb{N} \quad \text{and} \\ 
			& \Delta(X_{t-1}, X_{t}) = X_{t-1} \oplus X_{t}
		\end{split}
	\end{equation}
	.
	
	\paragraph{Exhaustive simulation with one Wb activated}
	In order to get a more complete dynamic comprehension of Wb effect over the rule dynamics, we run the following numerical experiment. We implement EFCA rules on a ring of size 12 and for every possible initial input state (this is $2^{12} = 4096$ configurations) we measure the hamming distance and magnetization index between the original ECA rule (none Wb activated) and the same rule with one Wb activated. This is done for ECA rules $\left[30, 32, 90, 110, 150 \right] $. Also for each configuration we search for cycles and when they are found the period and transient are  measured. For every configuration we also build the evolution graph, which is a directed graph where nodes represent the state configuration and  edges indicate the next generation state in the evolution. This kind of graph let us visualize graphically the occurrence of cycles and fixed points when they exists. 
	
	\paragraph{Simulating more cells with one Wb activated}
	Mixing the two previous experiment we use same four rules of the second trial, with Wb activation of the previous experiment but in an increased ring size of 32. The initial state of the EFCA is randomly selected from the $2^{32}$ possible initial condition making a sample of 6500 different initial configuration. The same metrics than in previous experiment were calculated adding the relative Hamming distance from initial to final state and magnetization variation, also from initial to final state. These two new measures are defined as shown in equations~\ref{eq:relHdb2e} and~\ref{eq:varMb2e}.
	\begin{equation}
		\begin{split}
			\label{eq:relHdb2e}
			& \delta_{\text{init state}} = \frac{\Delta(X_{t_0}, X_{t})}{n} \quad \text{with} \quad t \geq 1, n \in \mathbb{N} \quad \text{and} \\ 
			& \Delta(X_{t_0}, X_{t}) = X_{t_0} \oplus X_{t}
		\end{split}
	\end{equation}
	
	\begin{equation}
		\label{eq:varMb2e}
		\Delta m_{\text{init}-t} = \sum_{i=0}^{n}{x_{i}^{t_0}} - \sum_{i=0}^{n}{x_{i}^{t}}
	\end{equation}
	
	where $x^{t_0}$ and $x^{t}$ are the initial and final state of the EFCA.
	
	\subsection{Majority Fungal Automata}	
	MFA were implemented in a torus of size $100 \times 100$ and we let evolve during 99 generations as we did with EFCA. Evolution of the MFA was done using the global function previously described over Von Neumann neighborhood. Our global function was implemented to be invariant respect Woronin bodies activation, this mean that when Wb were activated the function (e.g. majority) was applied	 to smaller Von Neumann neighborhood, due to the activation restricting the number of neighbors that is possible to see. Initial condition of the MFA is randomly choosen using different magnetizations (number of cells whose state is one) up to a maximum of 9000, which is 90\% of the torus.
	In every step only one kind of Wbs (horizontal or vertical), all of them, are activated. To describe the activation sequence of Wbs we use words on the alphabet ${H,V}$. For example, if we close horizontal Wb at current step, and in the next step we switch them by verticals this will be expressed as: $HV$, if we repeat this sequence until the final step an additional star (*) will be added, having a notation in the form of $\left({HV}\right)^{*}$. We do not study the behavior when only vertical or horizontal Wbs are activated, due to this case being similar to a typical ECA with the following rules:
	\begin{itemize}
		\item Strict Majority transform into ECA rule 232.
		\item Majority (skew) transform into ECA rule 250.
	\end{itemize}
	To investigate how the MFA dynamics is affected by behavuour of Wbs, we set up five different modes of Wb activation defined as follow:
	\begin{itemize}
		\item \textbf{Modulation},  switching  between open and close of Horizontal and  Vertical Woronin bodies at every step $\left({HV}\right)^{*}$
		
		\item \textbf{H2V2}, an activation of Wb but every two steps $\left({HHVV}\right)^{*}$
		
		\item \textbf{H4V4}, an activation of Wb but every four steps $\left({HHHHVVVV}\right)^{*}$
		
		\item \textbf{cut \& release}, an  activation of Wb at every step $\left({HV}\right)^{20}$, but only for the first 20 generations, then all the Wbs are open until the next generation.
		
		\item \textbf{Rand}, a random activation of Wb at every step $\left({HV}\right)^{20}$. This means that in each step a vertical or horizontal Wb is activated with probability 0.5.
	\end{itemize}
	
	To quantify the possible change, we defined the following inter and intra-mode metrics:
	\begin{itemize}
		
		\item \textbf{(intra) Relative Hamming distance}, is the hamming distance between MFA self initial and final state divided by $N^2$, where $N$ is the size of MFA (square matrix $N \times N$), this can be expressed as:
		\begin{equation}
			\label{eq:intraHd}
			\delta_{\text{intra}}= \frac{X_{t_0}^{F,w} \oplus X_{t_n}^{F,w}}{N^2} 			
		\end{equation}
		
		where $X$ is a square matrix of size $m \times m$, $F$ is the global activation function in use, $w$ represents the word that describe the activation sequence of Wb and the step of evolution of the MFA goes from $t_0, t_1, \ldots, t_n$.
		
		\item \textbf{(inter) Relative Hamming distance}, is the hamming distance between final state of control MFA (without activation of Wb at any time) and the others MFA with different Wb activation mode, both with the same initial state and global function. This can be described as:
		\begin{equation}
			\label{eq:interHd}
			\delta_{\text{inter}} = \frac{{X_{t_n}^{F,w_0}} \oplus {X_{t_n}^{F,w_k}}}{N^2}
		\end{equation}
		
		where $X$ is a square matrix of size $m \times m$, $F$ is one of the two global activation functions, $w_0 = \emptyset$ which means none Wb activated, $w_k$ represent a mode in the set $w = \left\lbrace {H1V1}, {H2V2}, {H4V4}, \text{cut\&rel} \right\rbrace $
		
		\item \textbf{(intra) Magnetization index}, quantifies the variation in cells with one as value, between MFA self initial and final state. This can be write as:
		\begin{equation}
			\label{eq:intraMI}
			m_{\text{intra}} = \sum_{i=0}^{m}{\sum_{j=0}^{m}{x_{t_0}^{F,w}(i,j)}} - \sum_{i=0}^{m}{\sum_{j=0}^{m}}{x_{t_n}^{F,w}(i,j)}
		\end{equation}
		
		where $x(i,j)$ is the element in position $(i,j)$ of the square matrix $X$. 
		
		\item \textbf{(inter) Magnetization index}, quantifies the variation of cells with one as value between final state of control MFA (without activation of Wb at any time) and the others MFA with different Wb activation mode, both with the same initial state and global function, according to the following:
		\begin{equation}
			\label{eq:interMI}
			m_{\text{inter}} = \sum_{i=0}^{m}{\sum_{j=0}^{m}{x_{t_n}^{F,w_0}(i,j)}} - \sum_{i=0}^{m}{\sum_{j=0}^{m}{x_{t_n}^{F,w_k}(i,j)}}
		\end{equation}
	\end{itemize}
	Our global functions are both majority. A natural question that arises from this fact is how the consensus, meaning the agreement over the network, is affected by the activation of Wb. With the aim of investigate this phenomena we define the consensus index, which is the mean MFA state at a time $t$, but changing all zero values by $-1$. The following equation describes this index:
	
	\begin{equation}
		\label{eq:ci}
		\begin{aligned}
			\begin{split}
				{\text{Cid}(x_t)} = \frac{\sum_{i=0}^{m}{\sum_{j=0}^{m}{x_{t_n}^{F,w}(i,j)[0 \mapsto -1]}}}{N^2}
			\end{split}
		\end{aligned}
	\end{equation} 	
	Where $$x_{t_n}^{F,w}(i,j)[0 \mapsto -1] =  \begin{cases} 
		x_{t_n}^{F,w}(i,j) & \text{ if }   x_{t_n}^{F,w}(i,j) = 1 \\
		-1 &  \text{ otherwise. } 
	\end{cases}$$ 
	
	By using this index, we seek to capture two different type of phenomena:  a) we measure the possible fluctuation in consensus between two different timestamps $t_i, t_j$, and b) we study the average consensus reached or see how different Wb activations could change the consensus of the network. This is specially interesting in knowledge areas such as: social dynamics, networking and even blockchain, to determine if consensus is affected by the temporal interruption on the communication flow for Wb activation or if we can accelerate.

	\section{Results}
	\subsection{EFCA rules}
\begin{figure}[!htb]
		\centering
		\begin{subfigure}{.75\linewidth}
			\centering
			\includegraphics[width=.85\linewidth]{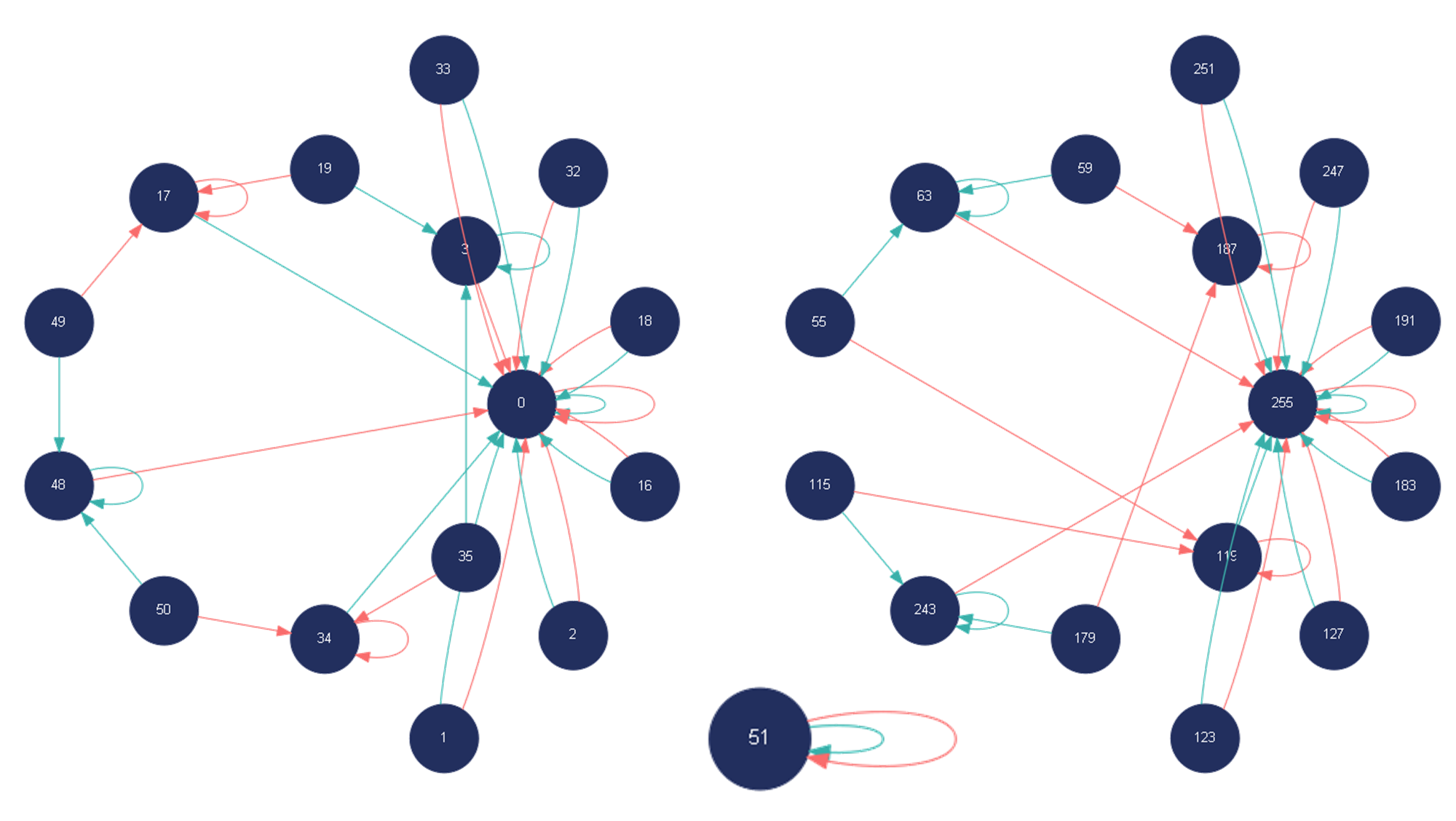}  
			\caption{Graph representing left and right rules for ECA rules. Rule 0, 51, and 255 are the strongly connected components of the graph representing left and right rules.}
			\label{fig:3-attractors}
		\end{subfigure}
		\newline
		\begin{subfigure}{.75\linewidth}
			\centering
			\includegraphics[width=.9\linewidth]{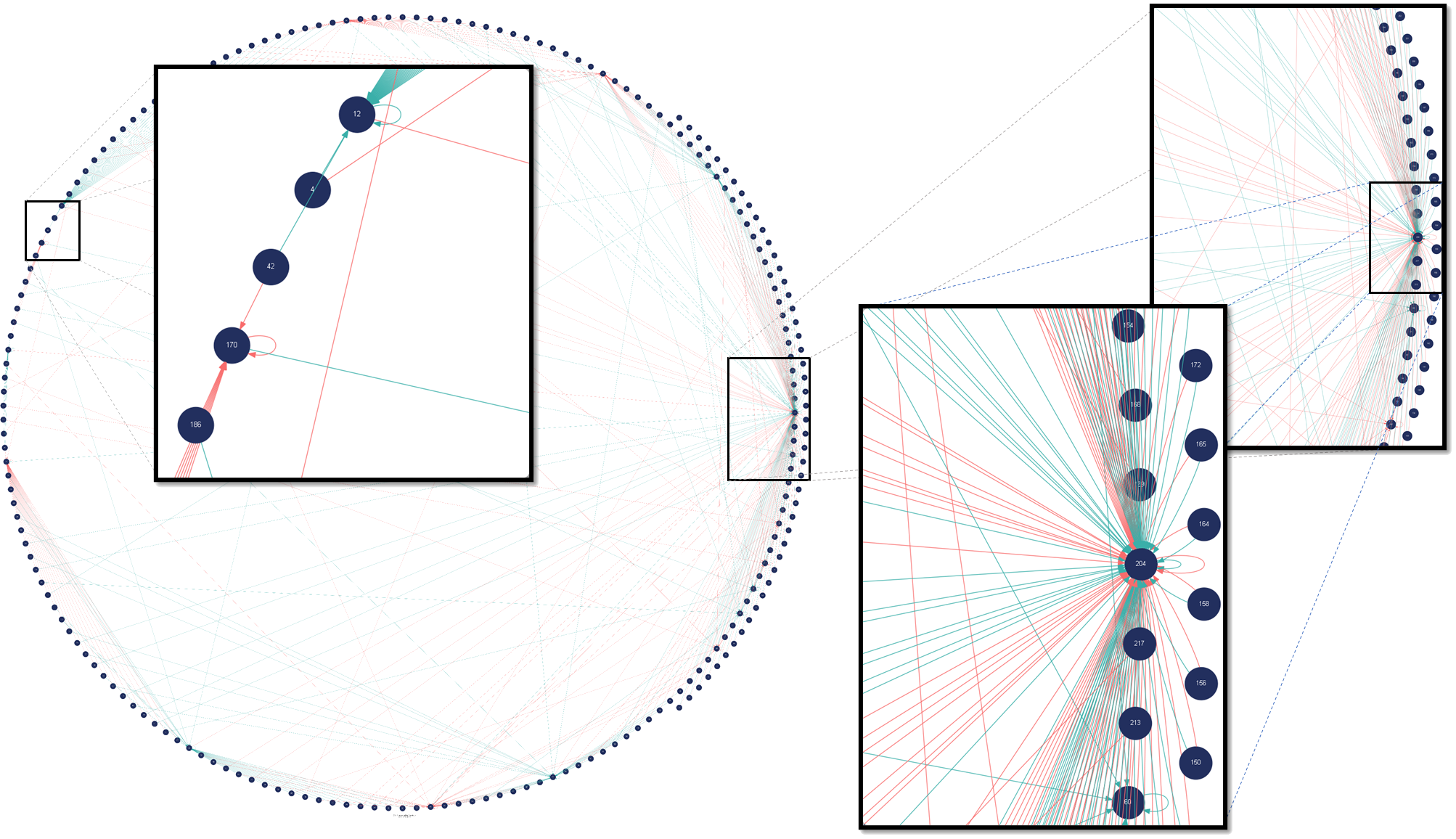}  
			\caption{Rule 204 (the identity) as the node with biggest in-degree (sink)}
			\label{fig:204-attractor}
		\end{subfigure}
		\caption{Four strongly connected components on the graph representing the fungal automata rule space. Every node (in blue) represents a rule, when a Wb is activated, the rule will change its behavior at the left (in red) or at the right (in green) of Wb, leading to other rule dynamic.}
		\label{fig:attractors}
	\end{figure}

	Elementary Cellular Automata (ECA) rules were classified by Wolfram in~\cite{wolfram1983statistical} into four classes dependending on their behavior starting from uniform (class $1$) and follow by periodic, chaotic and complex (class $4$). This classification naturally leads to  the question in  whether our approach induces changes in the class classification. 
	To answer that question and to get a better representation of new system dynamics induced by the Wb activation, we create a graph in which the nodes represent every of the  $255$ ECA rules. In this graph, each rule represented by a node has an edge towards its his left (in red color) and right (in green color) equivalent EFCA rule. This graph is shown in Figure~\ref{fig:attractors} where rules $\left[0, 51, 204, 255 \right]$ are the only ones with loops representing the strongly connected components of the graph. The rule $51$ has two incoming self-edges representing that no matter how many Wb are activated or in which mode its left and right rules are not only the same but actually the same rule $51$. On the other hand rule $204$ (identity rule) has the biggest incoming degree (acts as a sink), meaning that the majority of rules will have a path in the graph connecting them to rule $204$. In order to illustrate some of the observations that can be inferred from the graph, consider the following example.  An EFCA rule is defined over a ring of length $N$ with only one Wb activated at position $i$ and rule $110$ is the  update function for the next state. In all the cells not affected by the Wb activation, the rule $110$ will be applied normally, as if  the original ECA rule. For the triplets of cells  that are affected by Wb in his rightmost bit, instead of applying rule $110$, rule $204$ will be applied and, on the other hand, for the triplet  affected in his leftmost bit by Wb, rule $238$ will be applied. A complete list in a table format of rule's equivalence (left and right) is given in Table~\ref{tab:eqRule}. 
	
	We note from the previous results that rules $57$ and $99$ (which are equivalent between them and belong to Wolfram's class 2) are the only ones in which Wb activation increases class complexity at the right and left rules, transforming it into class $3$ rules. The rest of rules are summarized in Table~\ref{tab:1}
	
	\begin{table}
		\centering
		\caption{Summary of Wolfram's class classification changes for Wb activation}
		\begin{tabularx}{\textwidth}{m{0.31\textwidth} m{0.07\textwidth} m{0.55\textwidth}}
			\hline
			\textbf{Case}  & \centering{\textbf{Qty}} & \textbf{Rules} \\
			\hline
			Right and left rule increase in class complexity & \centering{2} & 57, 99 \\
			\hline
			Right and left rule don't change class complexity & \centering{128} & 0, 4, 5, 6, 7, 12, 13, 14, 15, 19, 20, 21, 23, 28, 29, 31, 32, 35, 36, 37, 42, 43, 44, 49, 50, 51, 55, 59, 68, 69, 70, 71, 72, 73, 74, 76, 77, 78, 79, 84, 85, 87, 88, 91, 92, 93, 94, 95, 100, 104, 108, 109, 112, 113, 115, 128, 132, 133, 134, 140, 141, 142, 143, 148, 156, 157, 158, 159, 160, 164, 170, 171, 172, 173, 178, 179, 196, 197, 198, 199, 200, 201, 202, 203, 204, 205, 206, 207, 212, 213, 214, 215, 216, 217, 218, 219, 220, 221, 222, 223, 228, 229, 232, 233, 236, 237, 240, 241, 250, 251, 254, 255 \\
			\hline
			Right rule increases, left remains class complexity & \centering{16} & 8, 40, 52, 53, 58, 61, 67, 83, 136, 163, 168, 211, 234, 235, 238, 239 \\
			\hline
			Right rule remains, left rule increase class complexity & \centering{16} & 25, 27, 38, 39, 64, 96, 103, 114, 155, 177, 192, 224, 248, 249, 252, 253 \\
			\hline
			Right rule decreases, left remains class complexity & \centering{26} & 17, 34, 65, 66, 80, 81, 82, 102, 116, 117, 119, 125, 153, 162, 180, 181, 186, 187, 188, 189, 194, 208, 209, 210, 244, 245 \\
			\hline
			Right rule remains, left rule decrease class complexity & \centering{26} & 3, 9, 10, 11, 24, 26, 46, 47, 48, 60, 63, 111, 138, 139, 152, 154, 166, 167, 174, 175, 176, 195, 230, 231, 242, 243 \\
			\hline
			Right rule increases, left rule decrease class complexity & \centering{4} & 56, 62, 131, 227 \\
			\hline
			Right rule decreases, left rule increase class complexity & \centering{4} & 98, 118, 145, 185 \\
			\hline
			Right and left rule decrease class complexity & \centering{50} & 1, 2, 16, 18, 22, 30, 33, 41, 45, 54, 75, 86, 89, 90, 97, 101, 105, 106, 107, 110, 120, 121, 122, 123, 124, 126, 127, 129, 130, 135, 137, 144, 146, 147, 149, 150, 151, 161, 165, 169, 182, 183, 184, 190, 191, 193, 225, 226, 246, 247 \\
			\hline
		\end{tabularx}
		\label{tab:1}
	\end{table}
	\subsubsection{First simulation}
\begin{figure}[!htb]
		\centering
		\begin{subfigure}[b]{.27\textwidth}
			\centering
			\includegraphics[width= \textwidth]{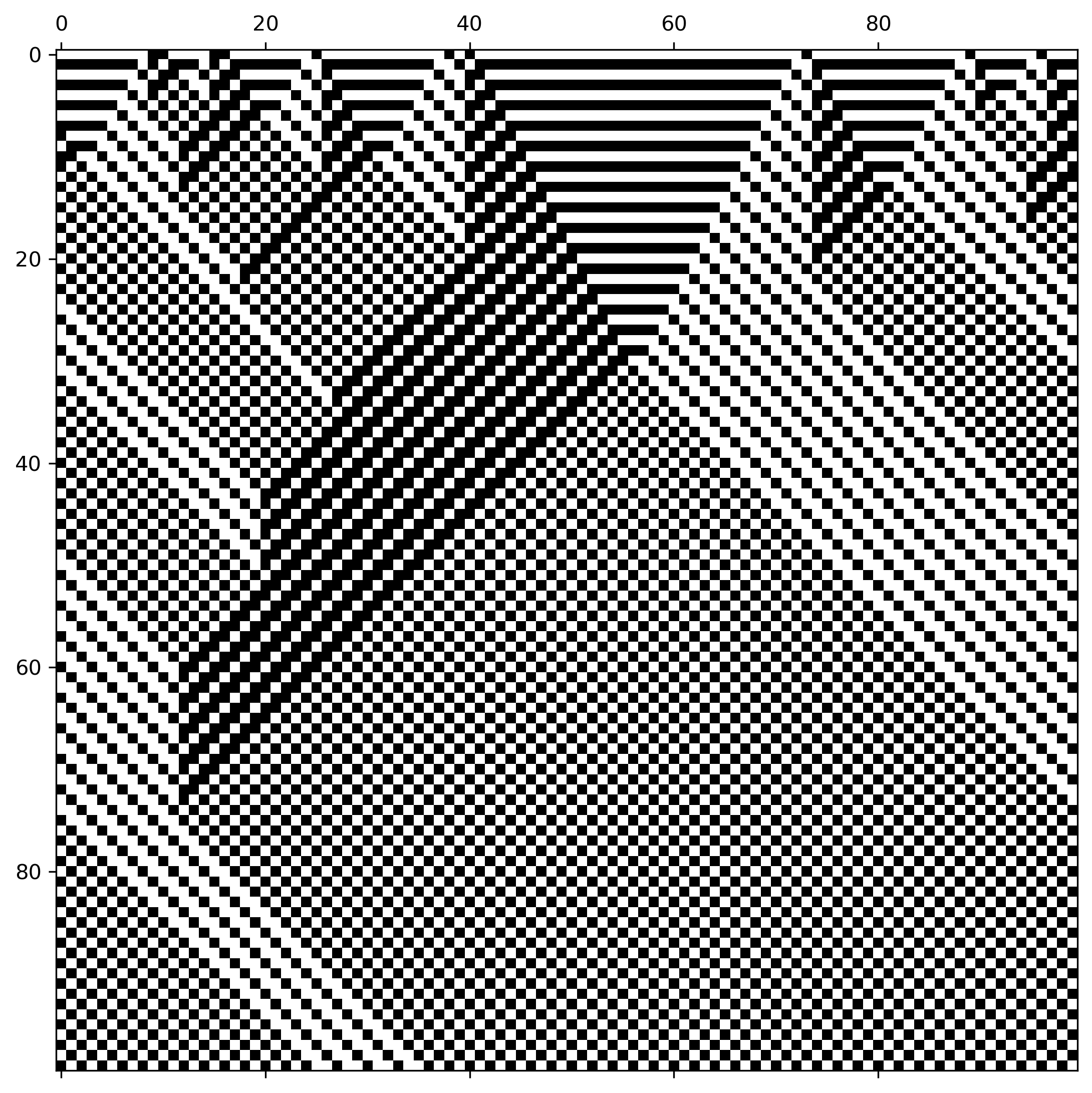}
			\caption{Control rule (none Wb)}
			\label{fig:57ctrl}
		\end{subfigure}
		\hfill
		\begin{subfigure}[b]{.27\textwidth}
			\centering
			\includegraphics[width=\textwidth]{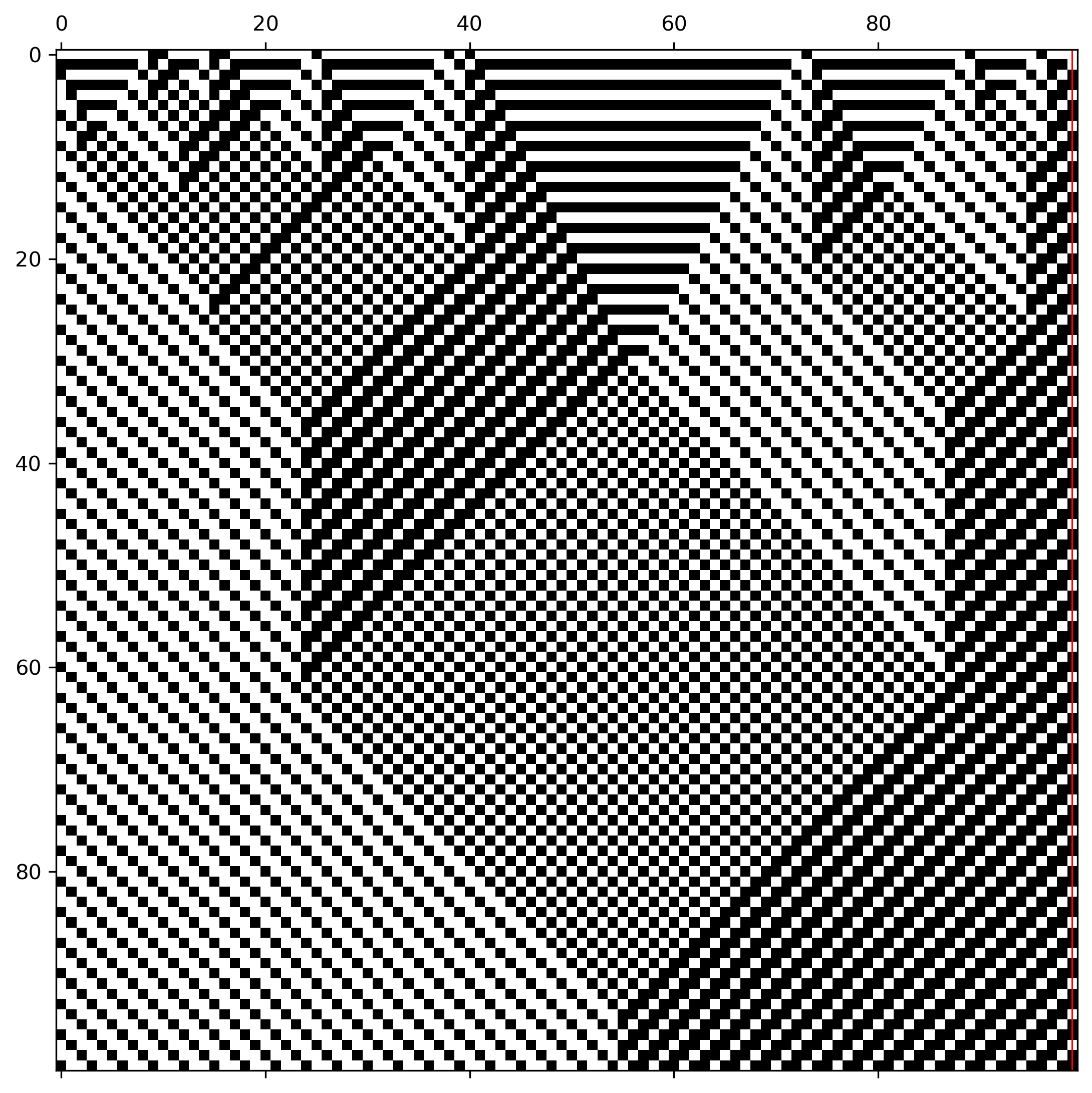}
			\caption{1Wb}
			\label{fig:571wb}
		\end{subfigure}
		\hfill
		\begin{subfigure}[b]{.27\textwidth}
			\centering
			\includegraphics[width=\textwidth]{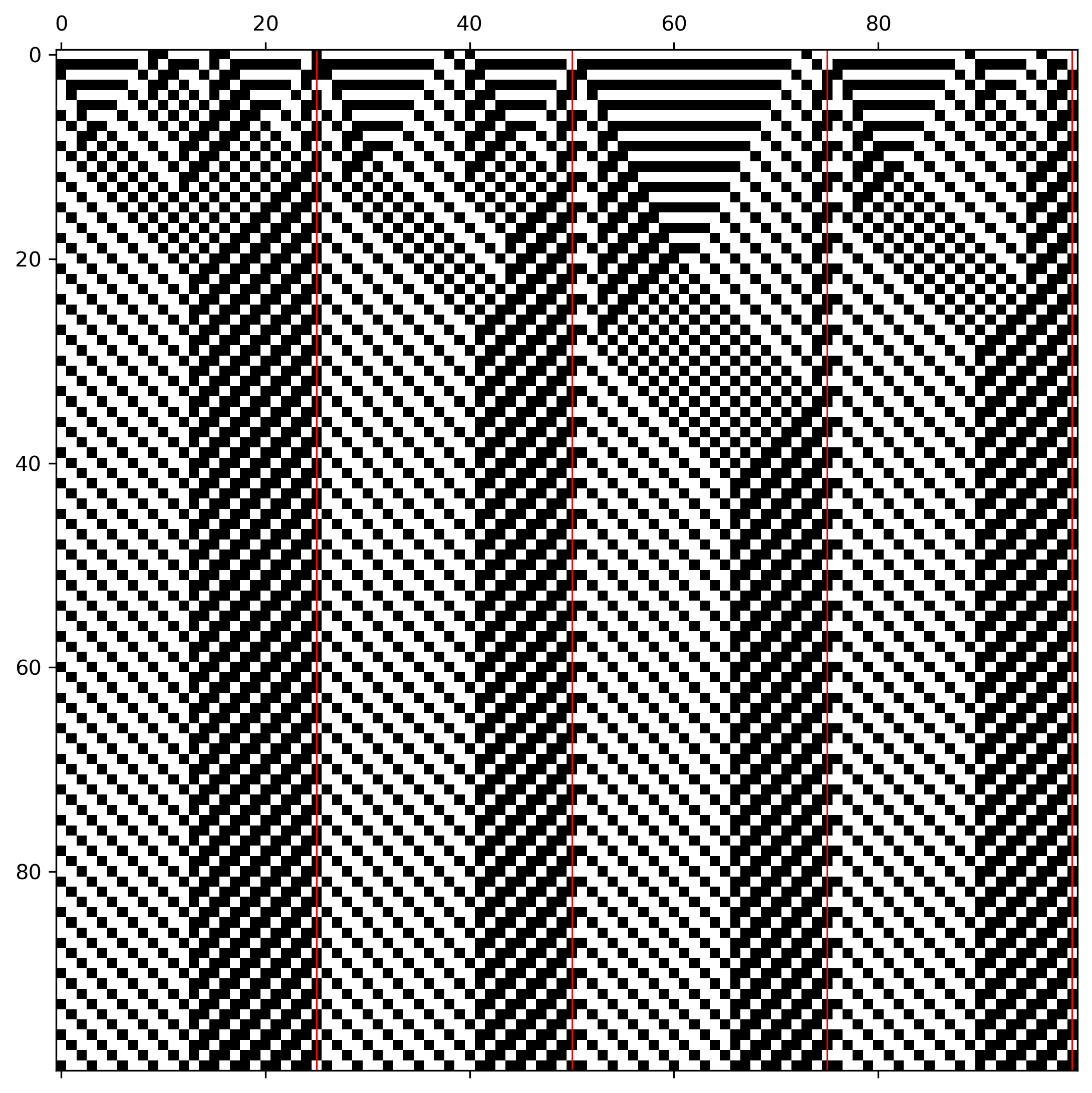}
			\caption{4Wb}
			\label{fig:574wb}
		\end{subfigure}
		\newline
		\begin{subfigure}[b]{.27\textwidth}
			\centering
			\includegraphics[width= \textwidth]{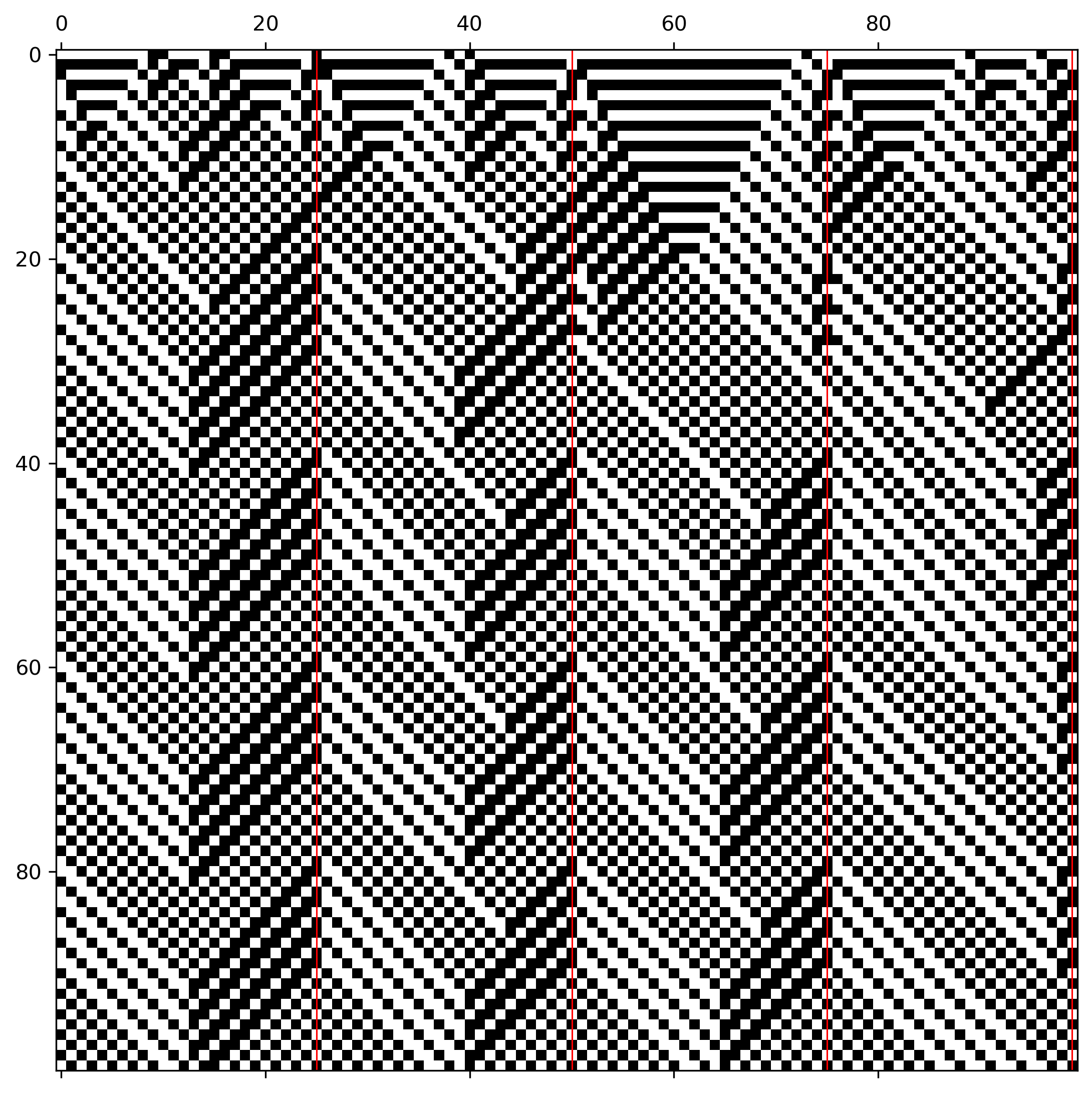}
			\caption{4Wb-mod}
			\label{fig:57mod}
		\end{subfigure}
		\hfill
		\begin{subfigure}[b]{.27\textwidth}
			\centering
			\includegraphics[width=\textwidth]{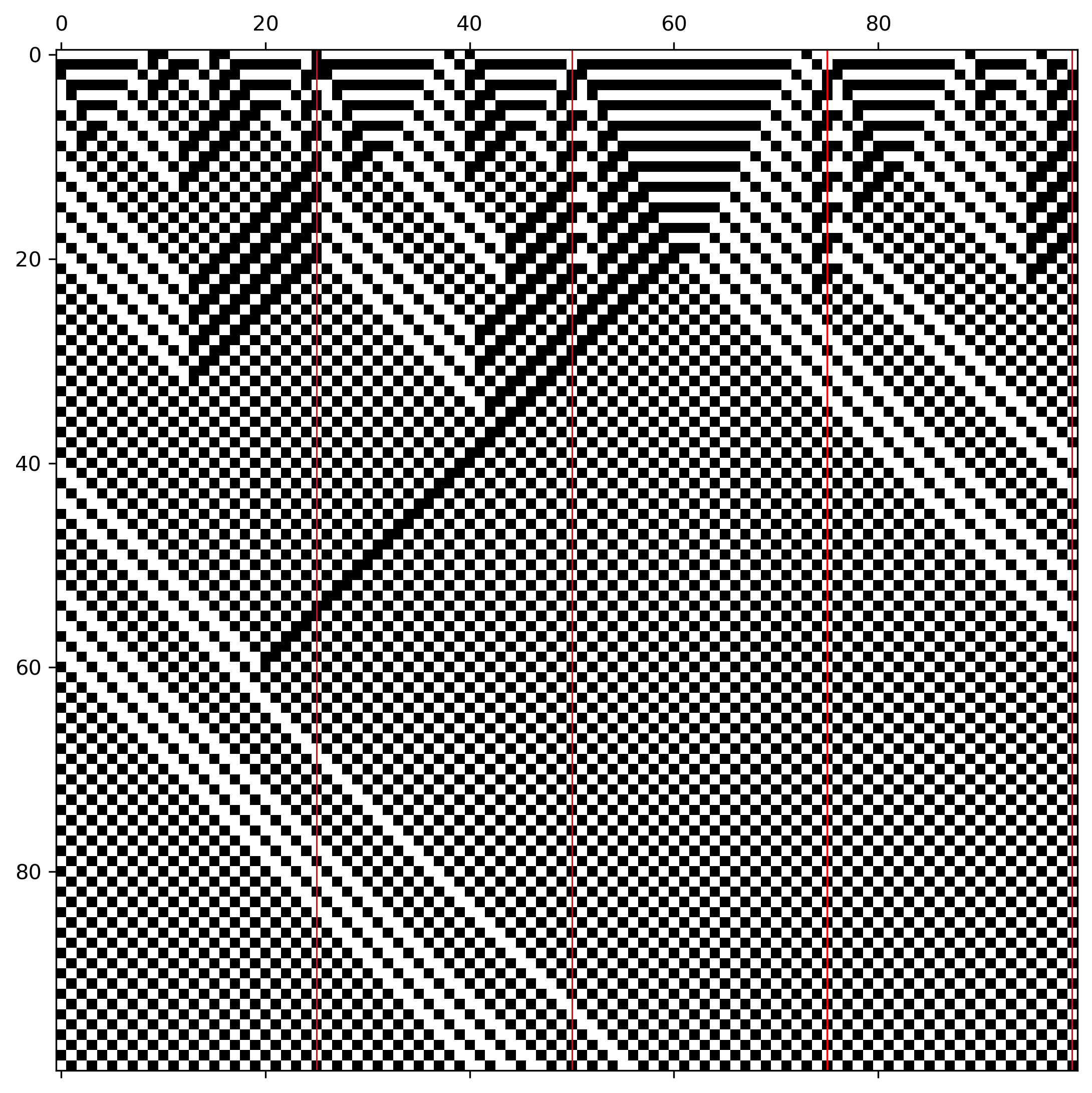}
			\caption{4Wb-c\&r}
			\label{fig:57candr}
		\end{subfigure}
		\hfill
		\begin{subfigure}[b]{.27\textwidth}
			\centering
			\includegraphics[width=\textwidth]{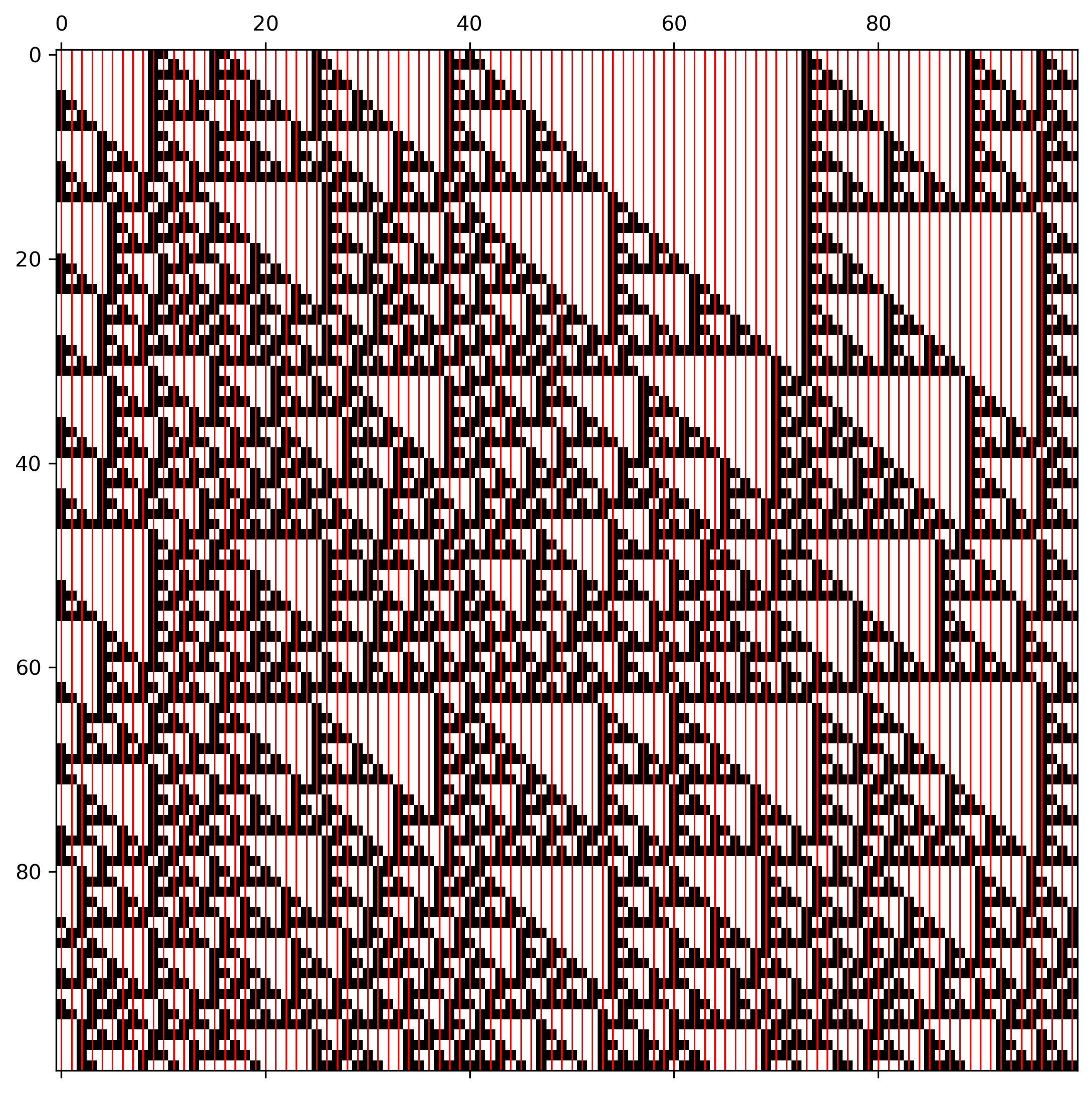}
			\caption{allWb-on}
			\label{fig:57all}
		\end{subfigure}
		\caption{Rule 57 on EFCA affected by different Wb activation. All start from the same initial state (10 ones).}							
		\label{fig:r57parallelallmodes}
	\end{figure}

\begin{figure}[!htb]
		\centering
		\begin{subfigure}[b]{.27\textwidth}
			\centering
			\includegraphics[width= \textwidth]{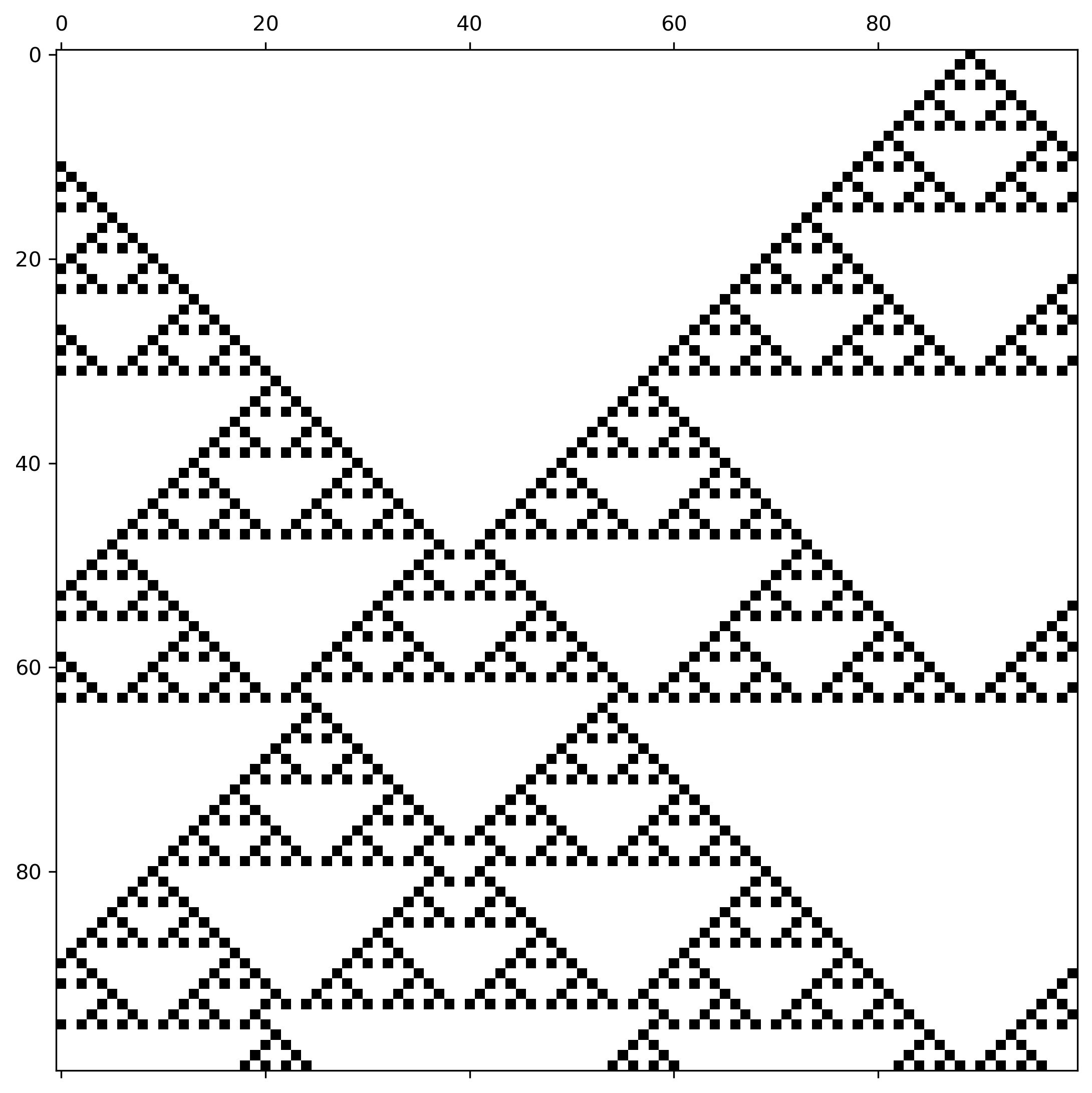}
			\caption{Control rule (none Wb)}
			\label{fig:90ctrl}
		\end{subfigure}
		\hfill
		\begin{subfigure}[b]{.27\textwidth}
			\centering
			\includegraphics[width=\textwidth]{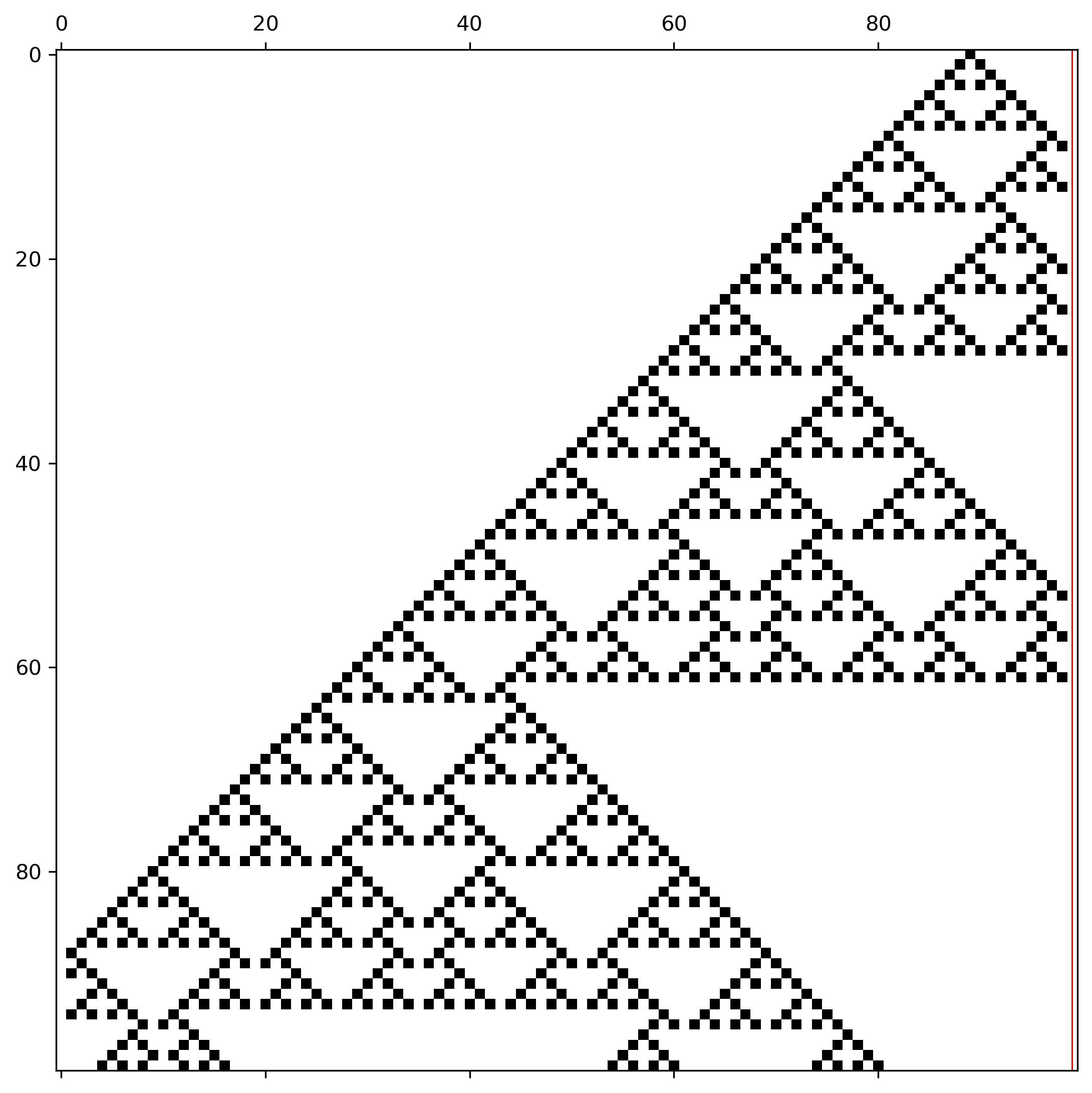}
			\caption{1Wb}
			\label{fig:901wb}
		\end{subfigure}
		\hfill
		\begin{subfigure}[b]{.27\textwidth}
			\centering
			\includegraphics[width=\textwidth]{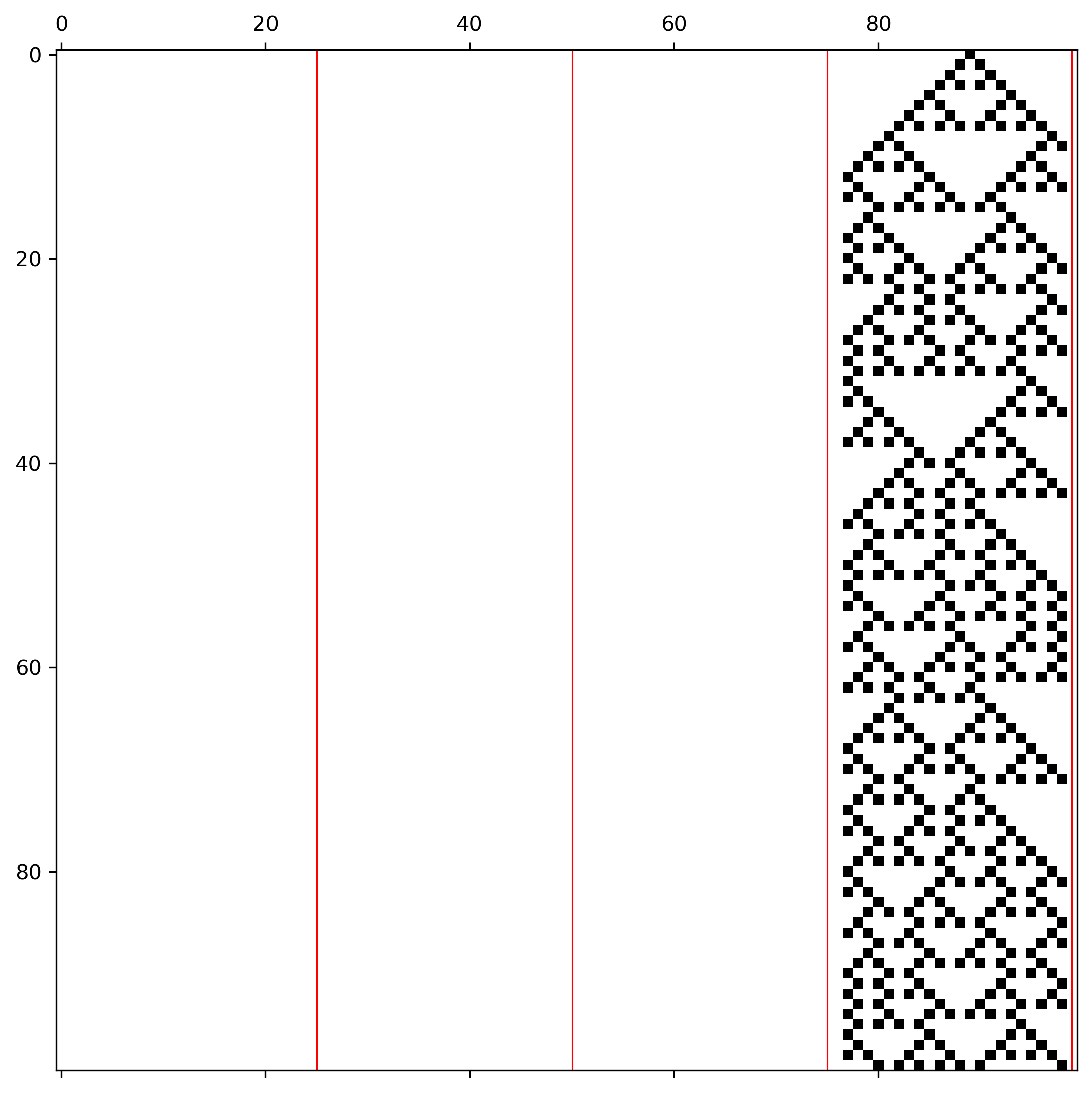}
			\caption{4Wb}
			\label{fig:904wb}
		\end{subfigure}
		\newline
		\begin{subfigure}[b]{.27\textwidth}
			\centering
			\includegraphics[width= \textwidth]{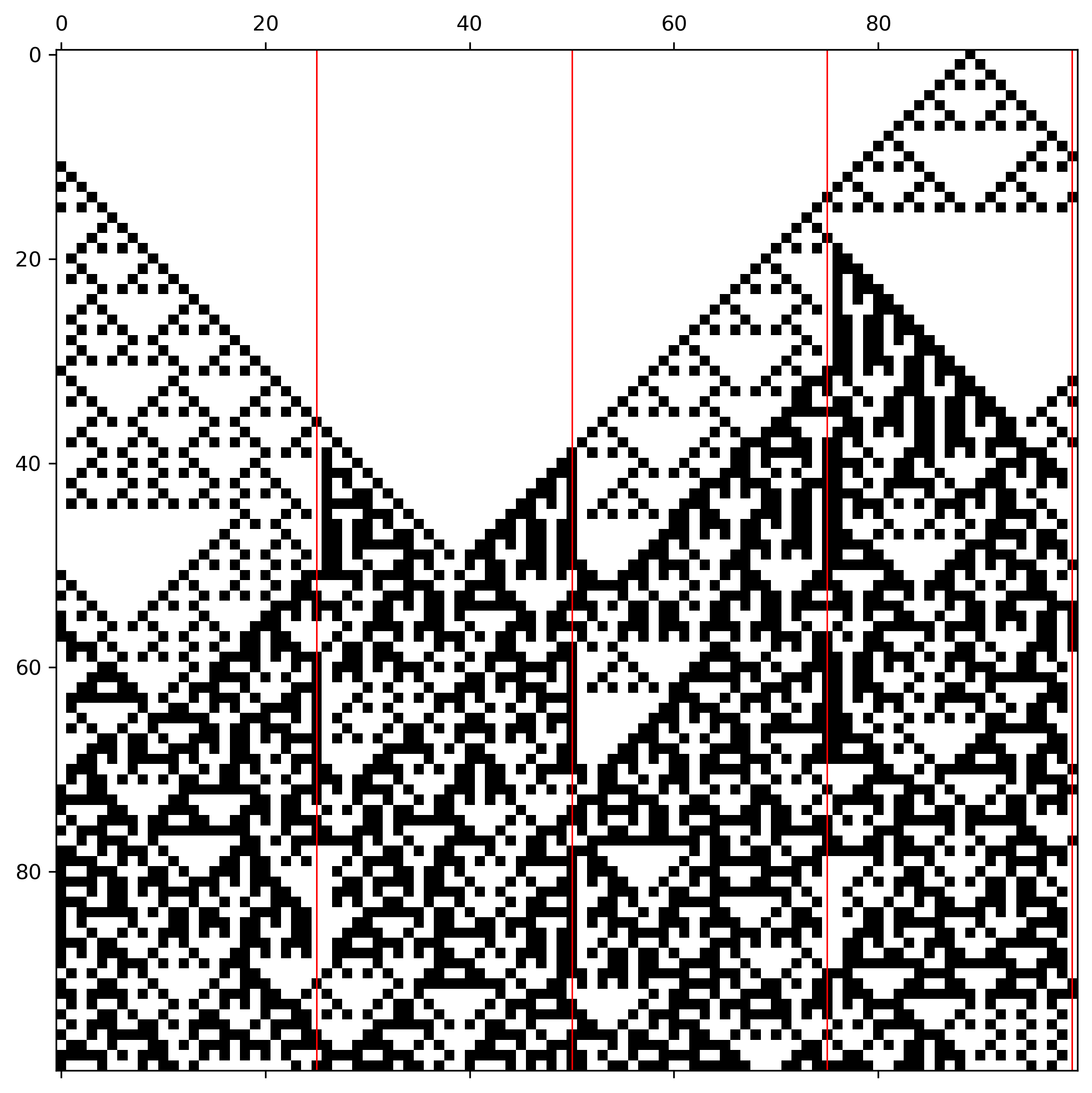}
			\caption{4Wb-mod}
			\label{fig:90mod}
		\end{subfigure}
		\hfill
		\begin{subfigure}[b]{.27\textwidth}
			\centering
			\includegraphics[width=\textwidth]{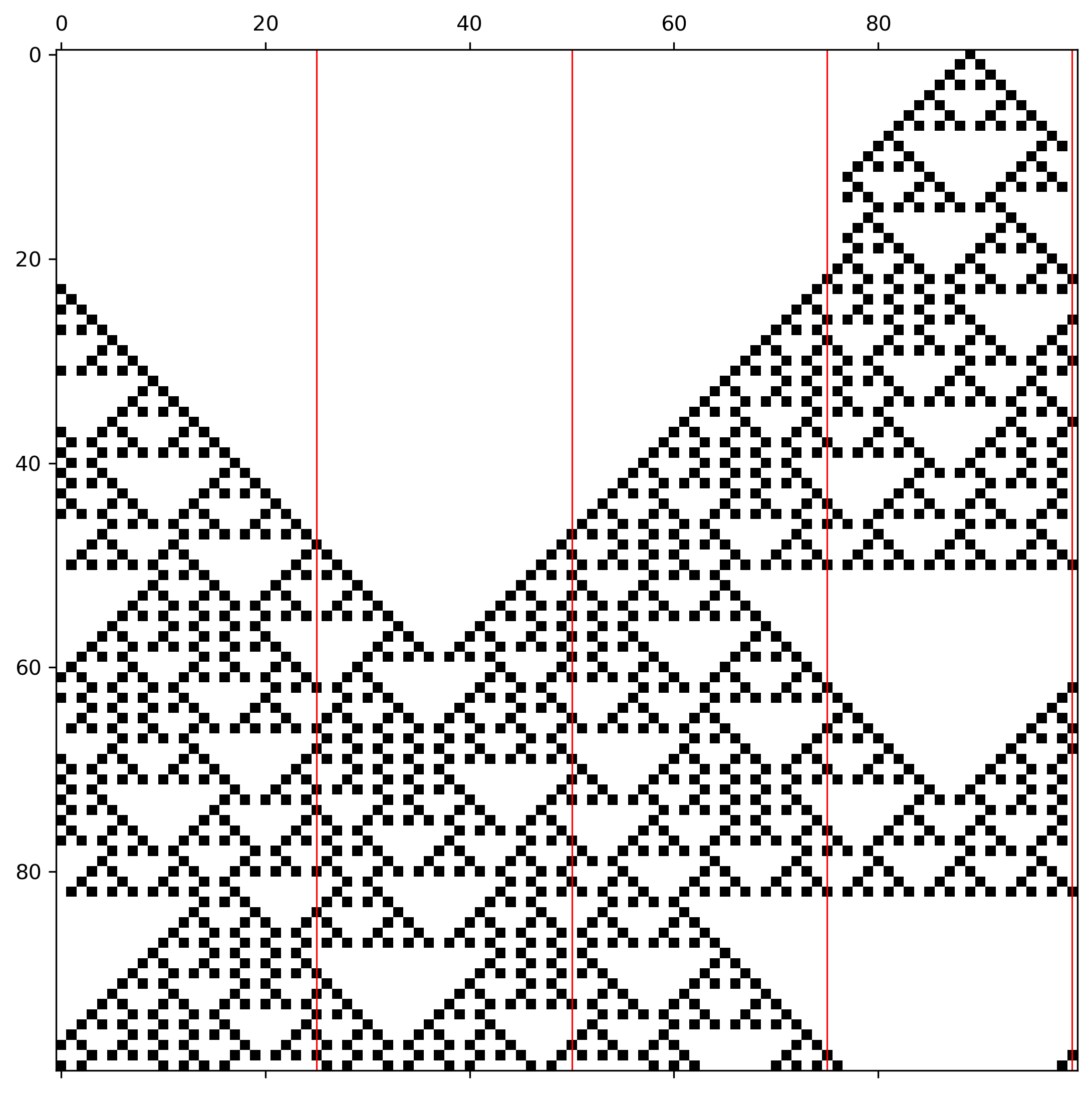}
			\caption{4Wb-c\&r}
			\label{fig:90candr}
		\end{subfigure}
		\hfill
		\begin{subfigure}[b]{.27\textwidth}
			\centering
			\includegraphics[width=\textwidth]{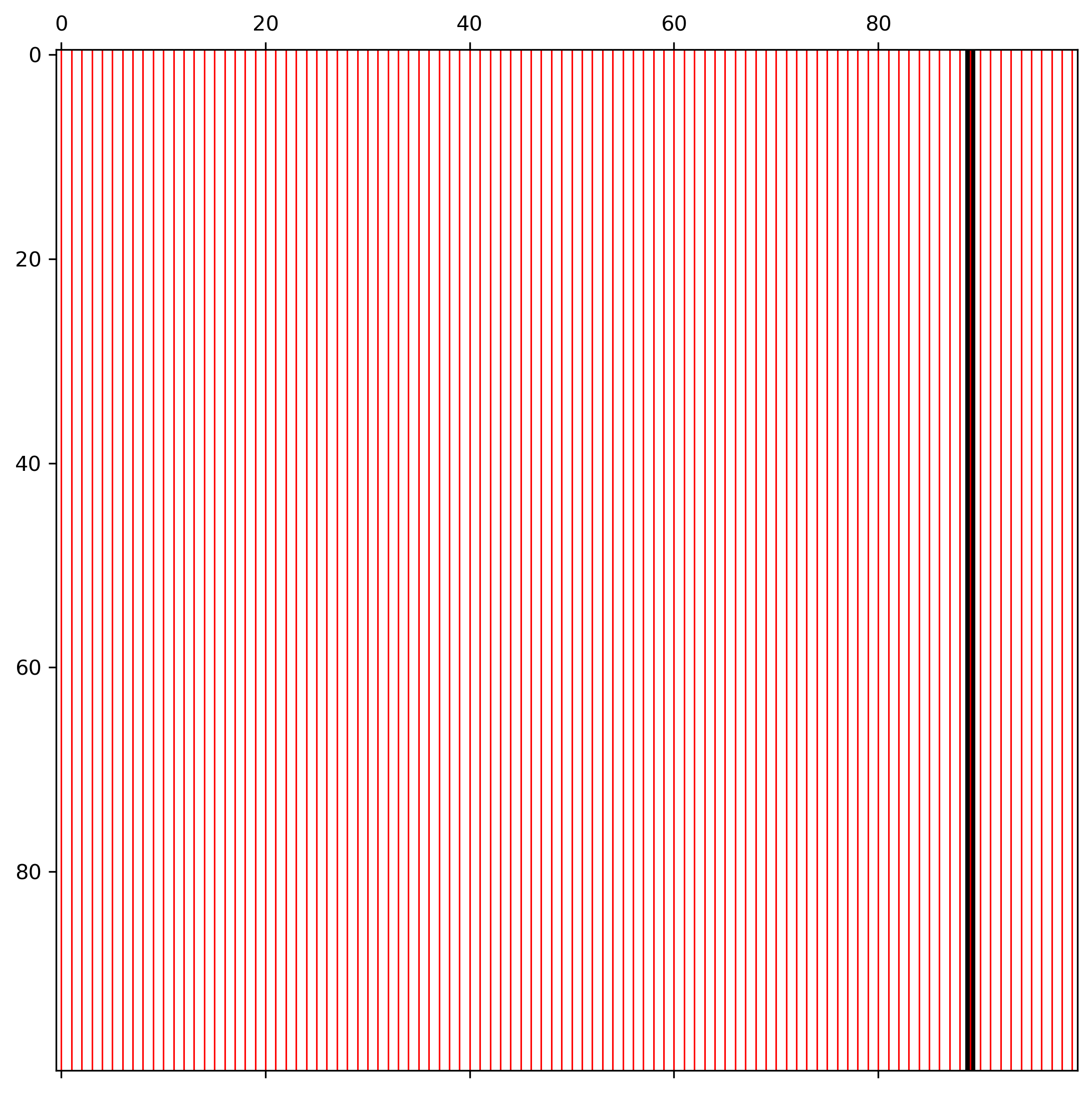}
			\caption{allWb-on}
			\label{fig:90all}
		\end{subfigure}
		\caption{Rule 90 on EFCA affected by different Wb activation. All start from the same initial state.}							
		\label{fig:r90parallelallmodes}
	\end{figure}

\begin{figure}[!htb]
		\centering
		\begin{subfigure}[b]{.27\textwidth}
			\centering
			\includegraphics[width= \textwidth]{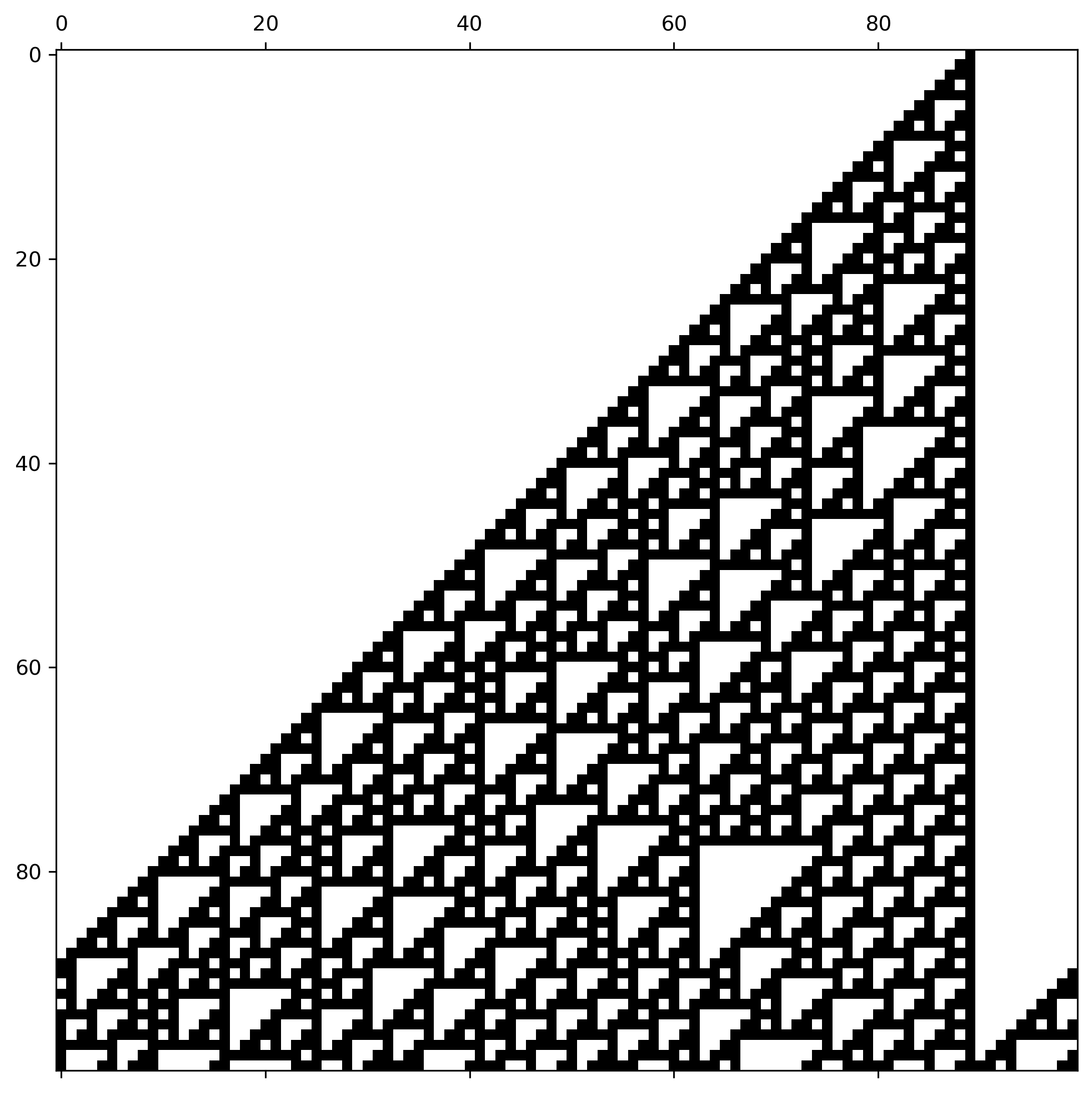}
			\caption{Control rule (none Wb)}
			\label{fig:110ctrl}
		\end{subfigure}
		\hfill
		\begin{subfigure}[b]{.27\textwidth}
			\centering
			\includegraphics[width=\textwidth]{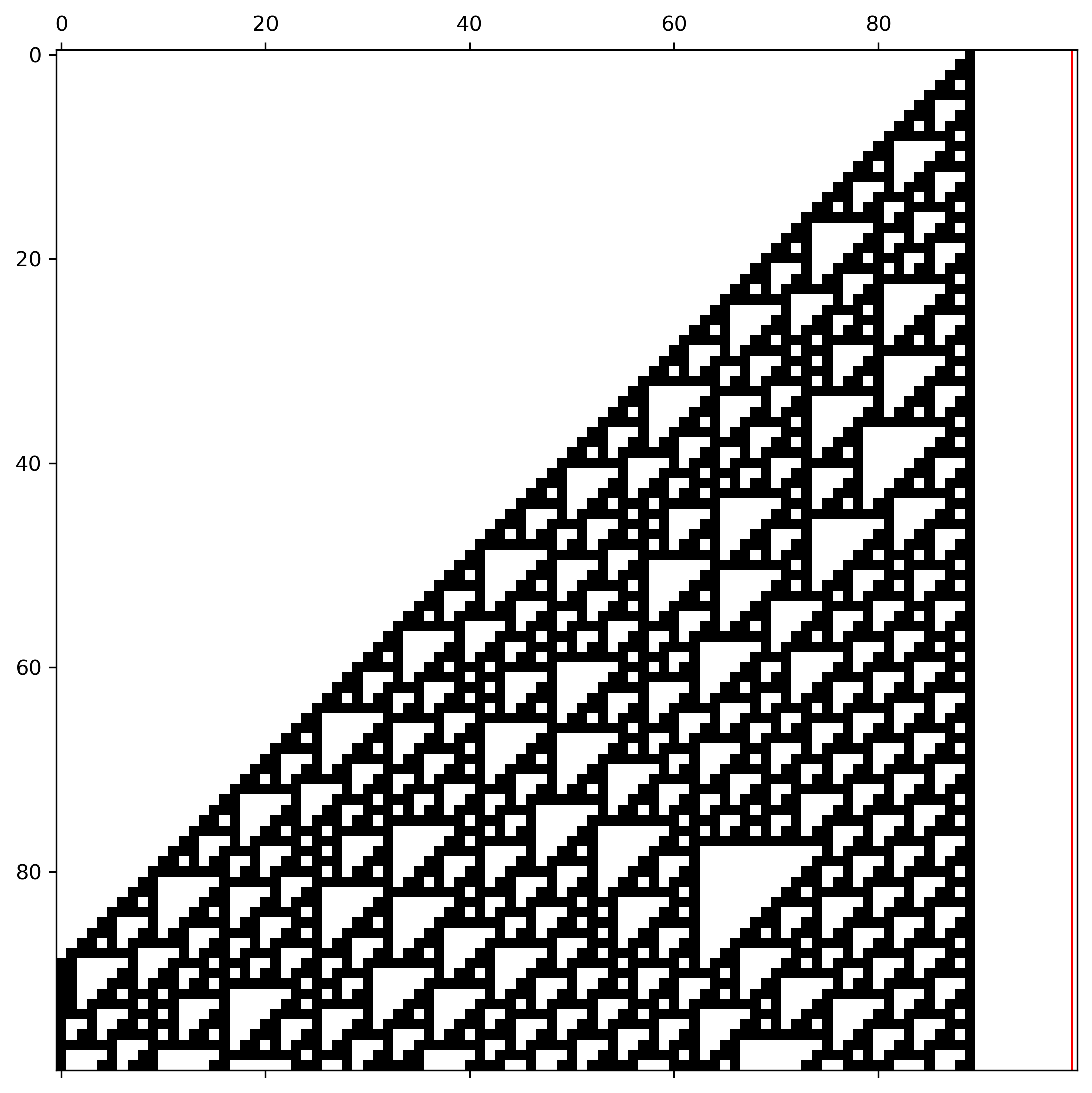}
			\caption{1Wb}
			\label{fig:1101wb}
		\end{subfigure}
		\hfill
		\begin{subfigure}[b]{.27\textwidth}
			\centering
			\includegraphics[width=\textwidth]{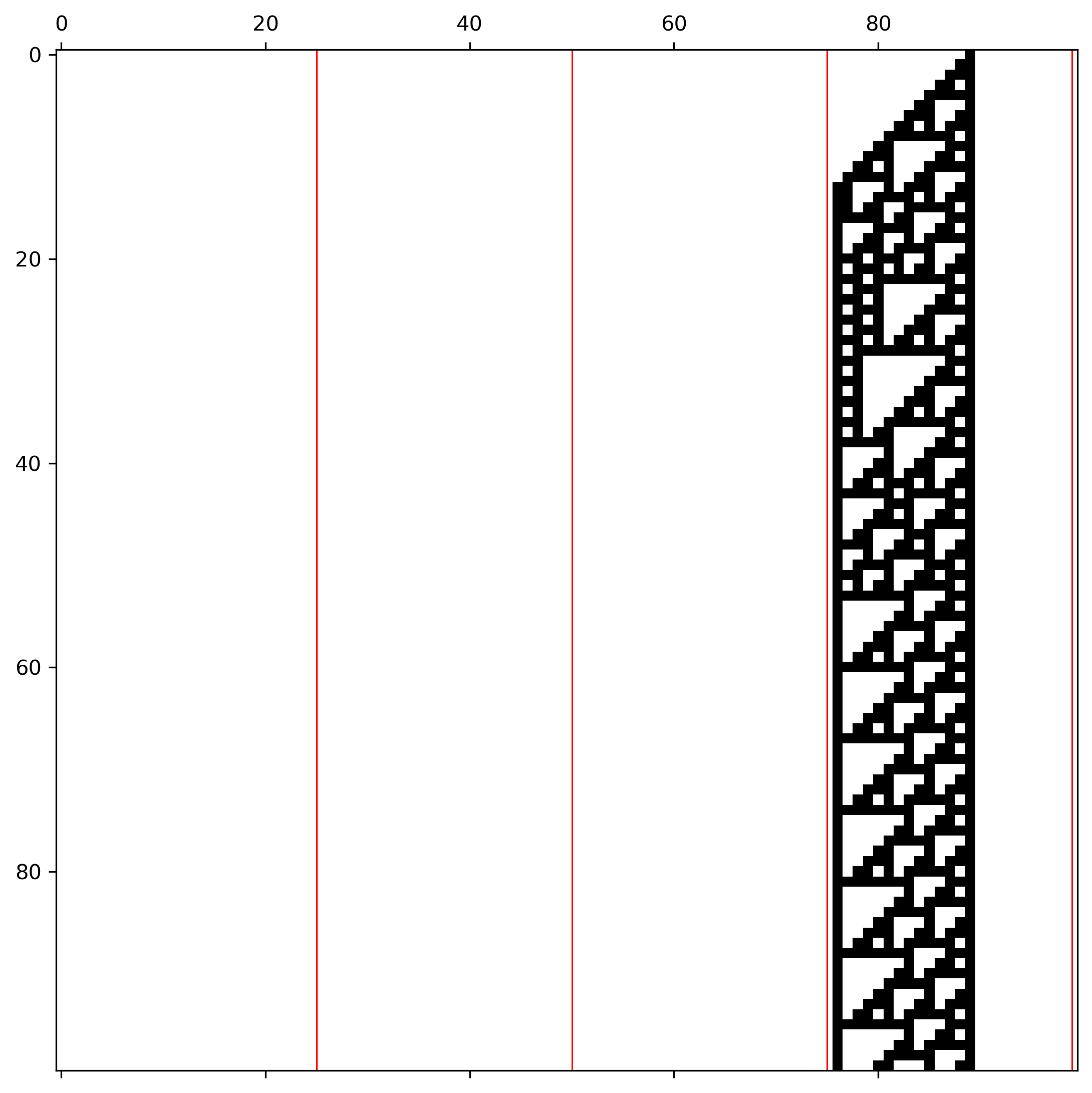}
			\caption{4Wb}
			\label{fig:1104wb}
		\end{subfigure}
		\newline
		\begin{subfigure}[b]{.27\textwidth}
			\centering
			\includegraphics[width= \textwidth]{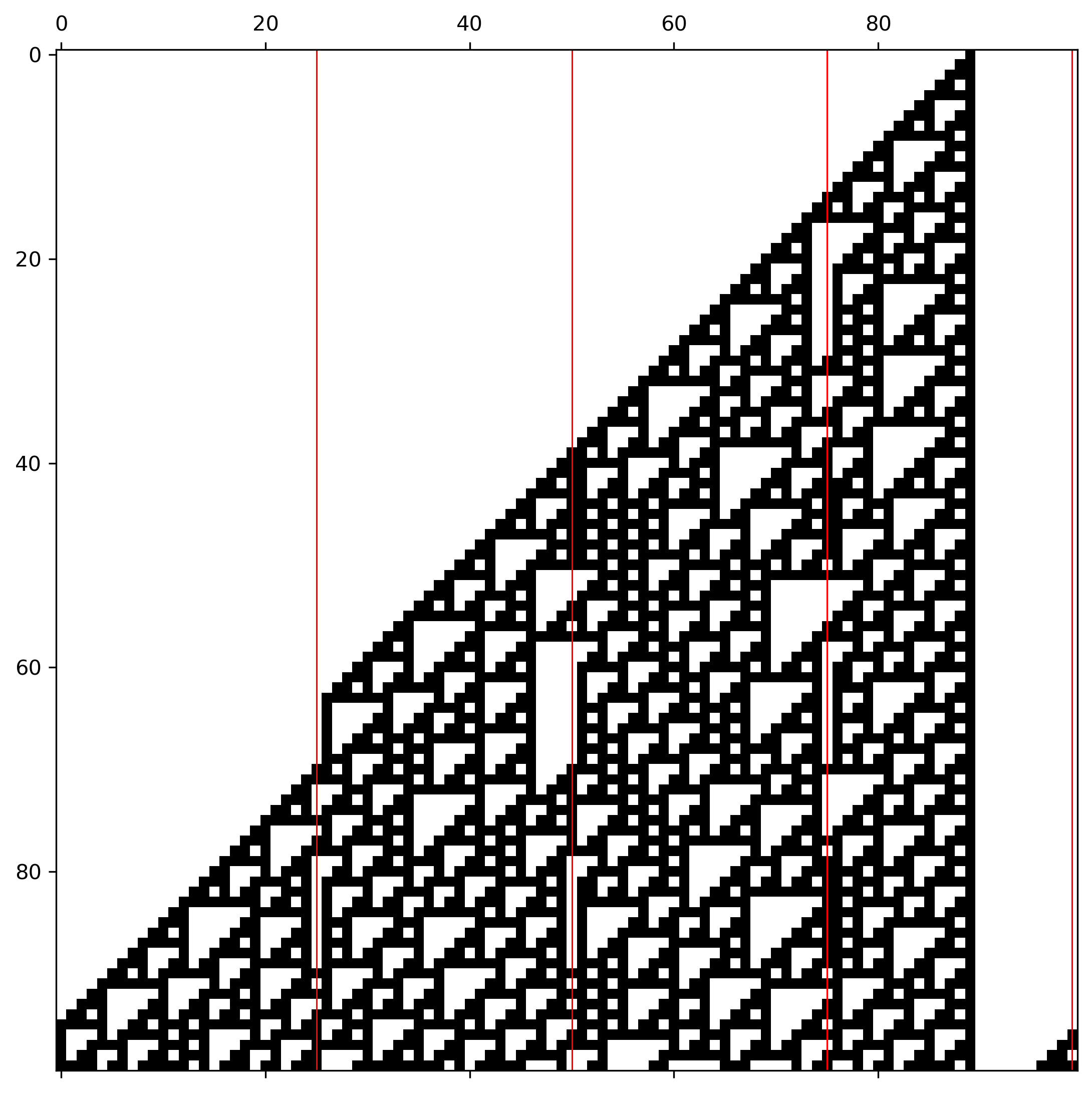}
			\caption{4Wb-mod}
			\label{fig:110mod}
		\end{subfigure}
		\hfill
		\begin{subfigure}[b]{.27\textwidth}
			\centering
			\includegraphics[width=\textwidth]{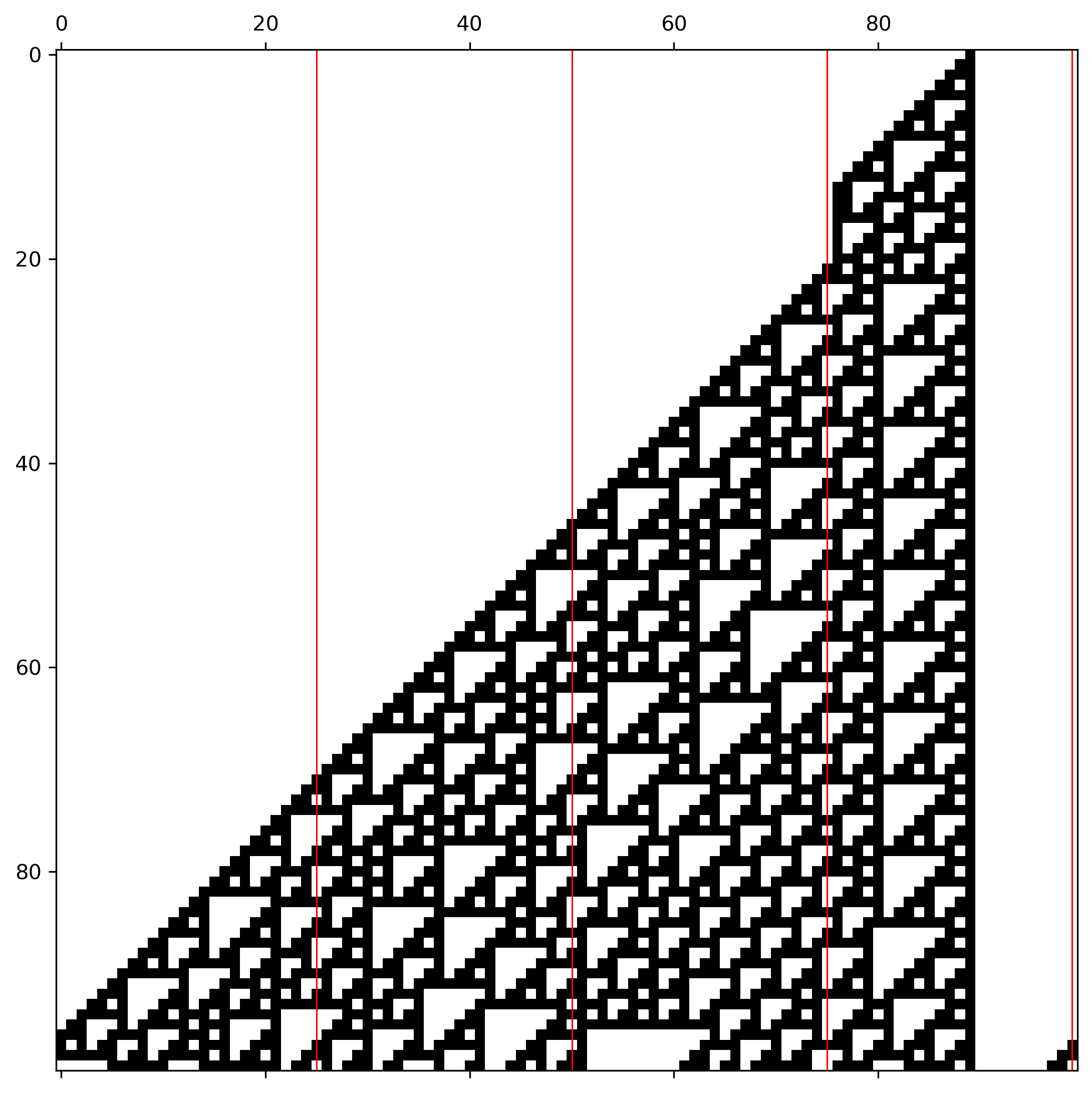}
			\caption{4Wb-c\&r}
			\label{fig:110candr}
		\end{subfigure}
		\hfill
		\begin{subfigure}[b]{.27\textwidth}
			\centering
			\includegraphics[width=\textwidth]{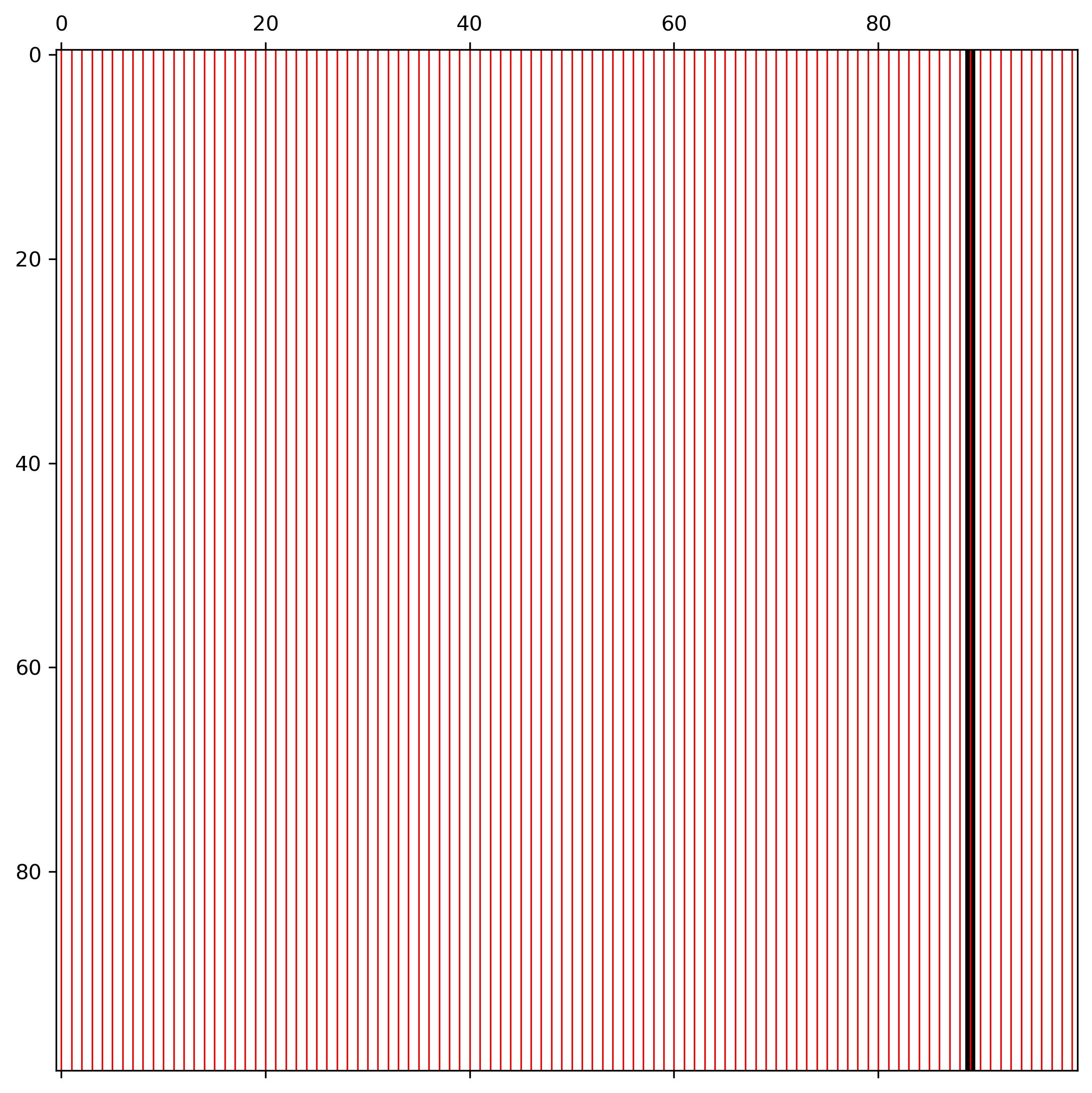}
			\caption{allWb-on}
			\label{fig:110all}
		\end{subfigure}
		\caption{Rule 110 on EFCA affected by different Wb activation. All start from the same initial state.}							
		\label{fig:r110parallelallmodes}
	\end{figure}

\begin{figure}[!htb]
		\centering
		\begin{subfigure}[b]{.27\textwidth}
			\centering
			\includegraphics[width= \textwidth]{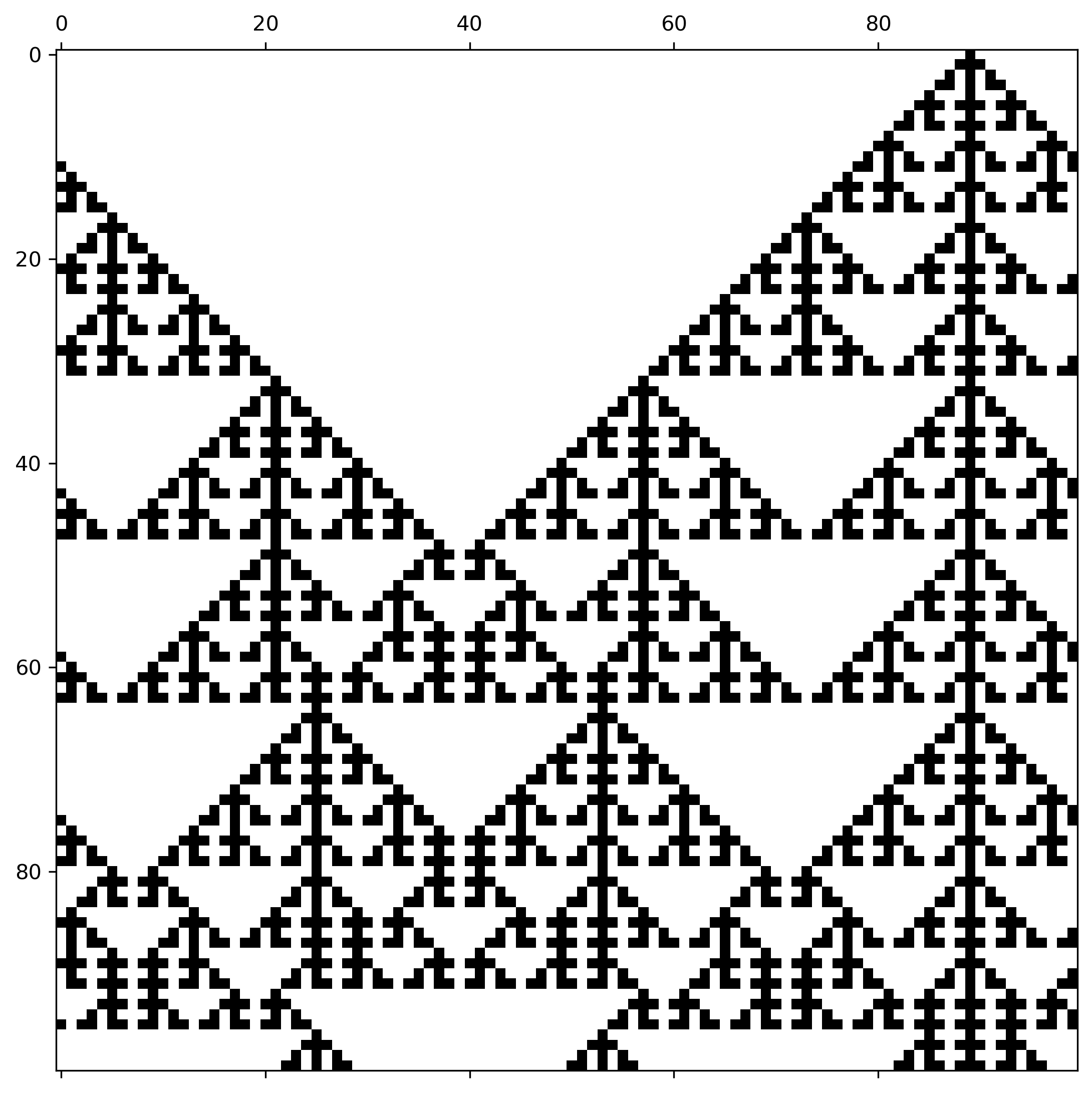}
			\caption{Control rule (none Wb)}
			\label{fig:150ctrl}
		\end{subfigure}
		\hfill
		\begin{subfigure}[b]{.27\textwidth}
			\centering
			\includegraphics[width=\textwidth]{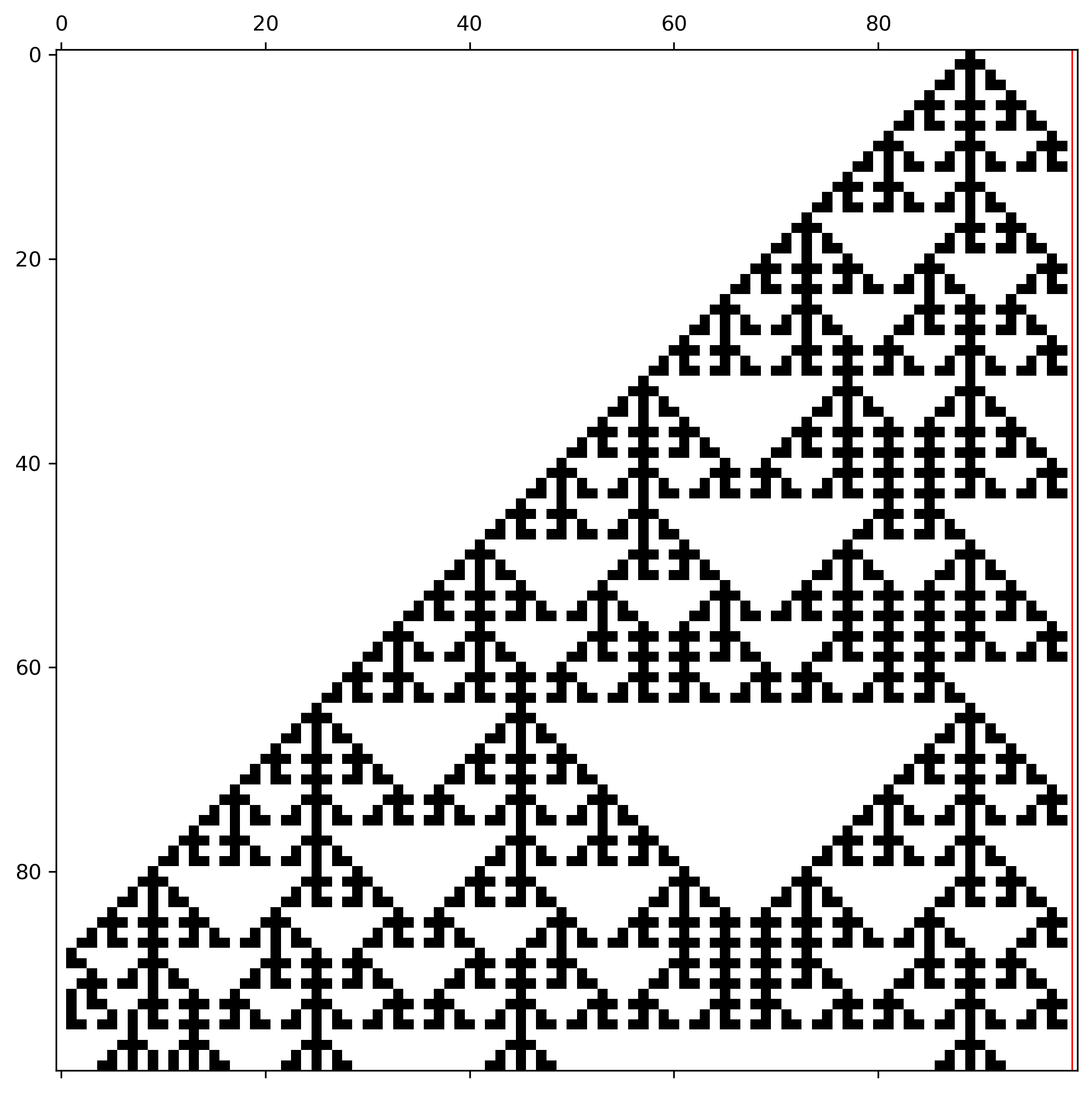}
			\caption{1Wb}
			\label{fig:1501wb}
		\end{subfigure}
		\hfill
		\begin{subfigure}[b]{.27\textwidth}
			\centering
			\includegraphics[width=\textwidth]{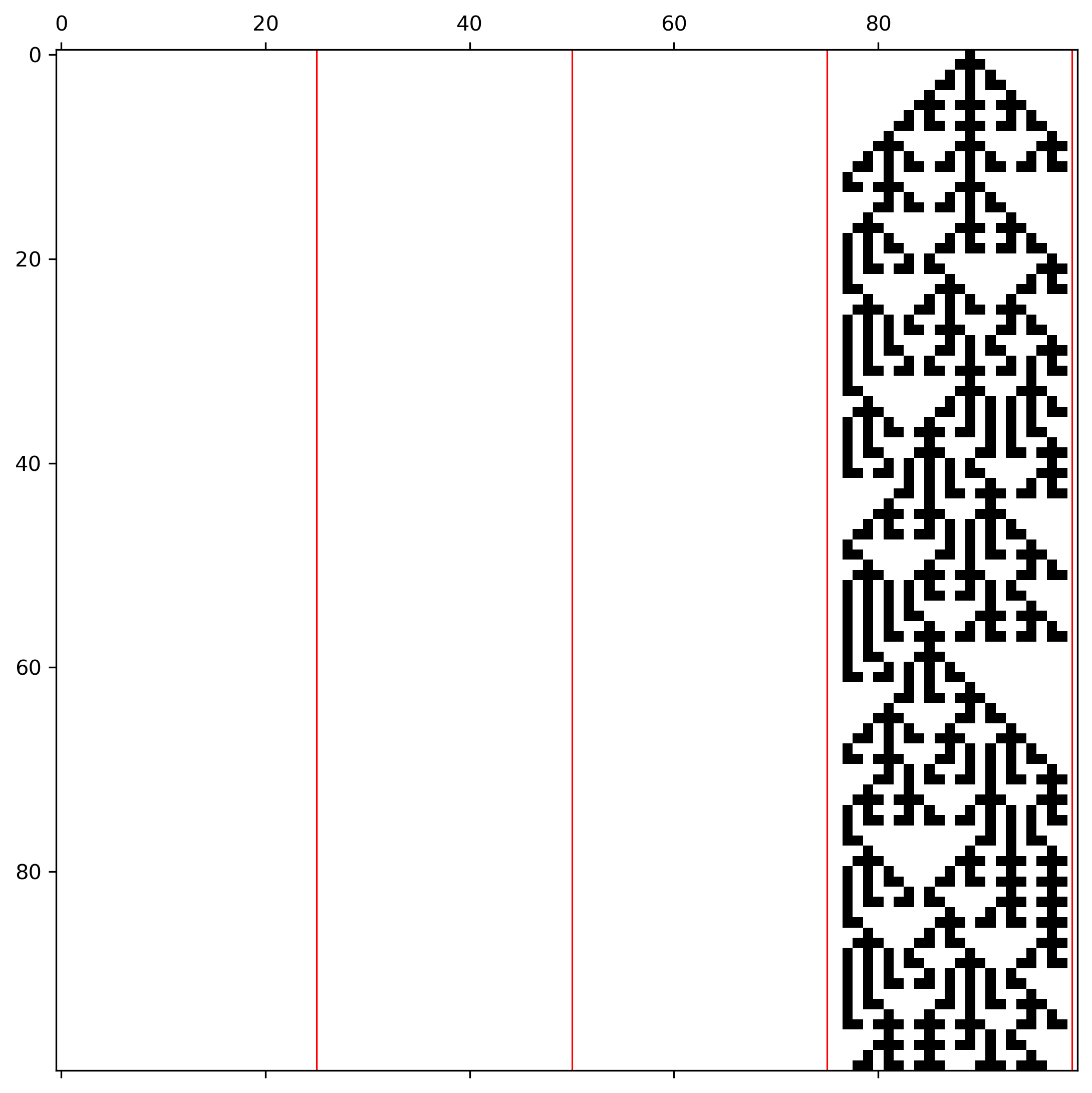}
			\caption{4Wb}
			\label{fig:1504wb}
		\end{subfigure}
		\newline
		\begin{subfigure}[b]{.27\textwidth}
			\centering
			\includegraphics[width= \textwidth]{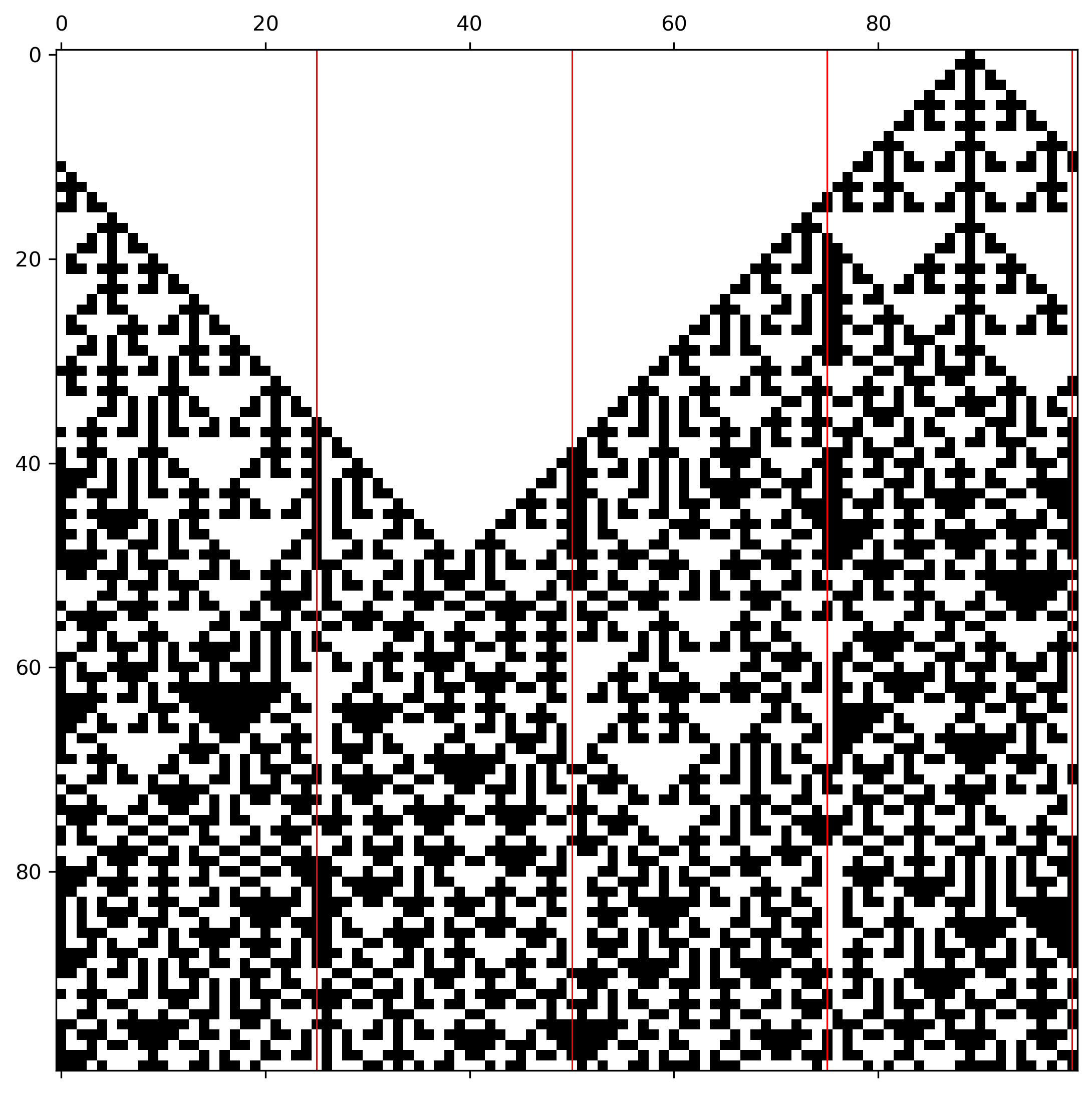}
			\caption{4Wb-mod}
			\label{fig:150mod}
		\end{subfigure}
		\hfill
		\begin{subfigure}[b]{.27\textwidth}
			\centering
			\includegraphics[width=\textwidth]{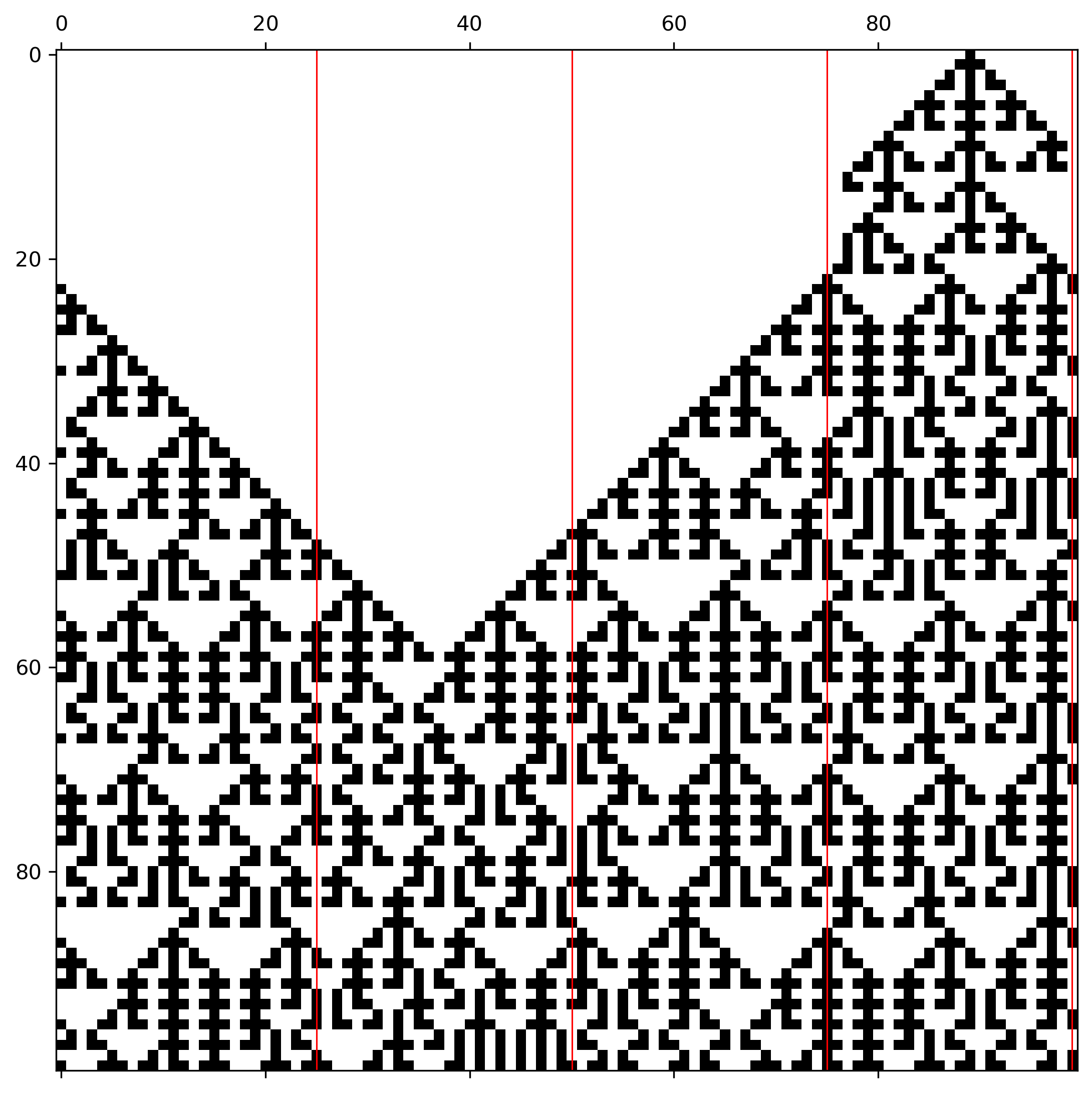}
			\caption{4Wb-c\&r}
			\label{fig:150candr}
		\end{subfigure}
		\hfill
		\begin{subfigure}[b]{.27\textwidth}
			\centering
			\includegraphics[width=\textwidth]{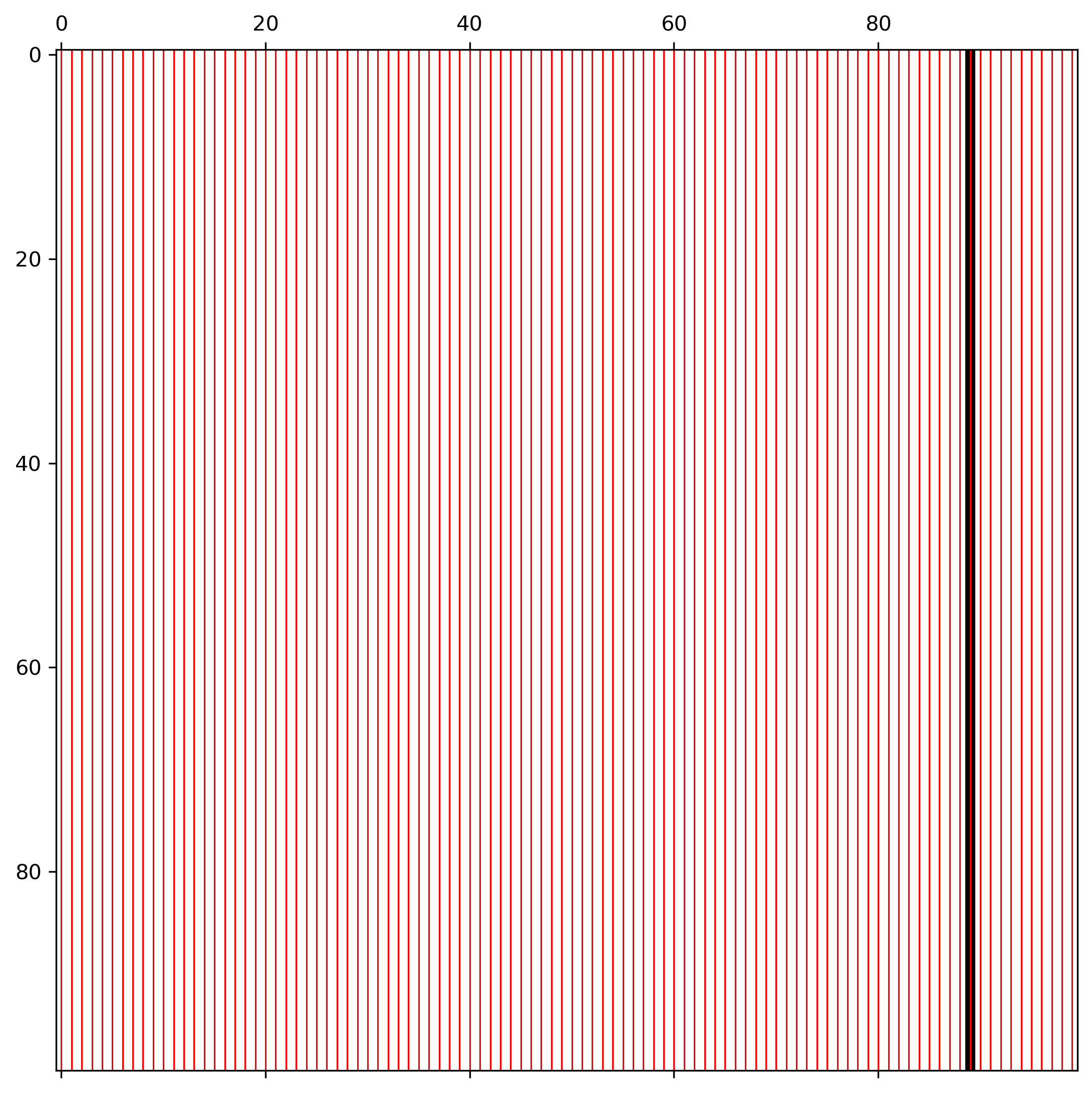}
			\caption{allWb-on}
			\label{fig:150all}
		\end{subfigure}
		\caption{Rule 150 on EFCA affected by different Wb activation. All start from the same initial state.}							
		\label{fig:r150parallelallmodes}
	\end{figure}

	A total of 6336 simulations were run, covering all 255 ECA rules, Figures~\ref{fig:r57parallelallmodes},~\ref{fig:r90parallelallmodes}, ~\ref{fig:r110parallelallmodes} and ~\ref{fig:r150parallelallmodes} are samples of the this numerical experiment. In these figures it is possible to see how the five modes in which we activated Wb affect the normal behavior of ECA rule for the same initial state and we compare against control rule (without Wb activation). 
	
	In  Figure~\ref{fig:57all}, the previous parameters for rule 57 are shown. Observe that this latter rule is one of those rule that increases a complexity with Wb activation. On the other hand, the same can be seen in Figure~\ref{fig:57all}. Figures~\ref{fig:90all}, ~\ref{fig:110all} and ~\ref{fig:150all} exemplify how the activation of all Wb leads the dynamical behavior of rule $90$ to a behavior similar to rule $204$ (the identity rule) which plays the role of a sink on the graph shown in Figure \ref{fig:attractors} .
	
\begin{figure}[!htb]
		\centering
		\includegraphics[width=0.95\linewidth]{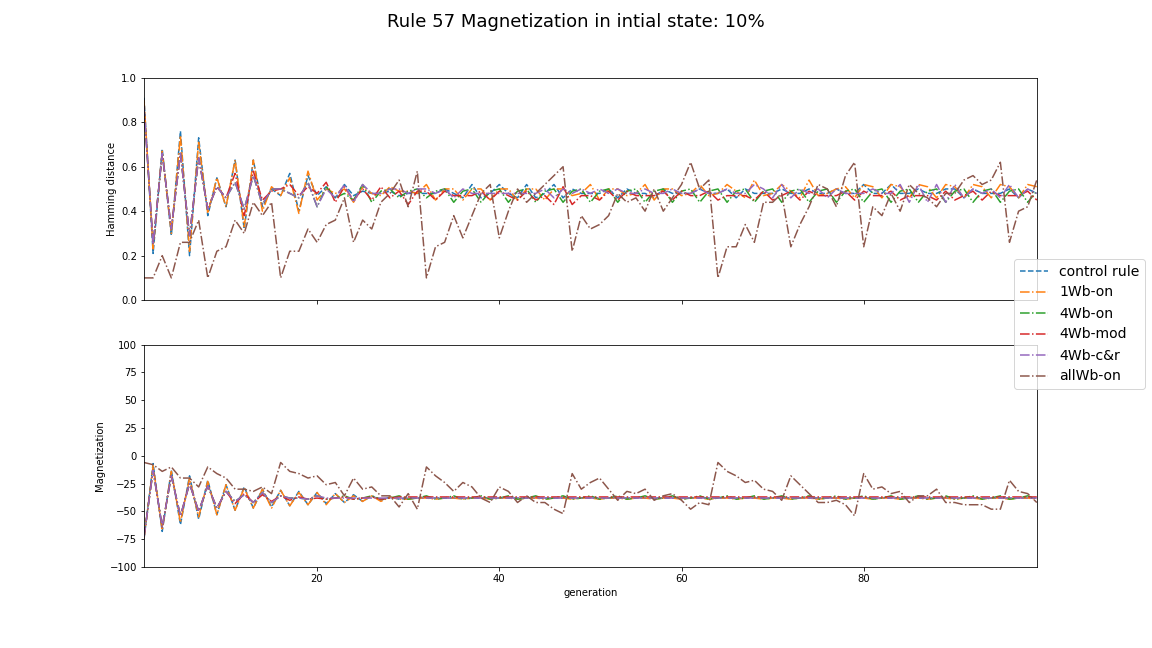}
		\caption{Rule 57. Upper plot shows how relative hamming distance vary according to evolution step for each Woronin bodies activation mode. Lower graph shows how the magnetization index varies accordingly to the same parameters.}
		\label{fig:r57d10}
	\end{figure}

\begin{figure}[!htb]
		\centering
		\includegraphics[width=0.95\linewidth]{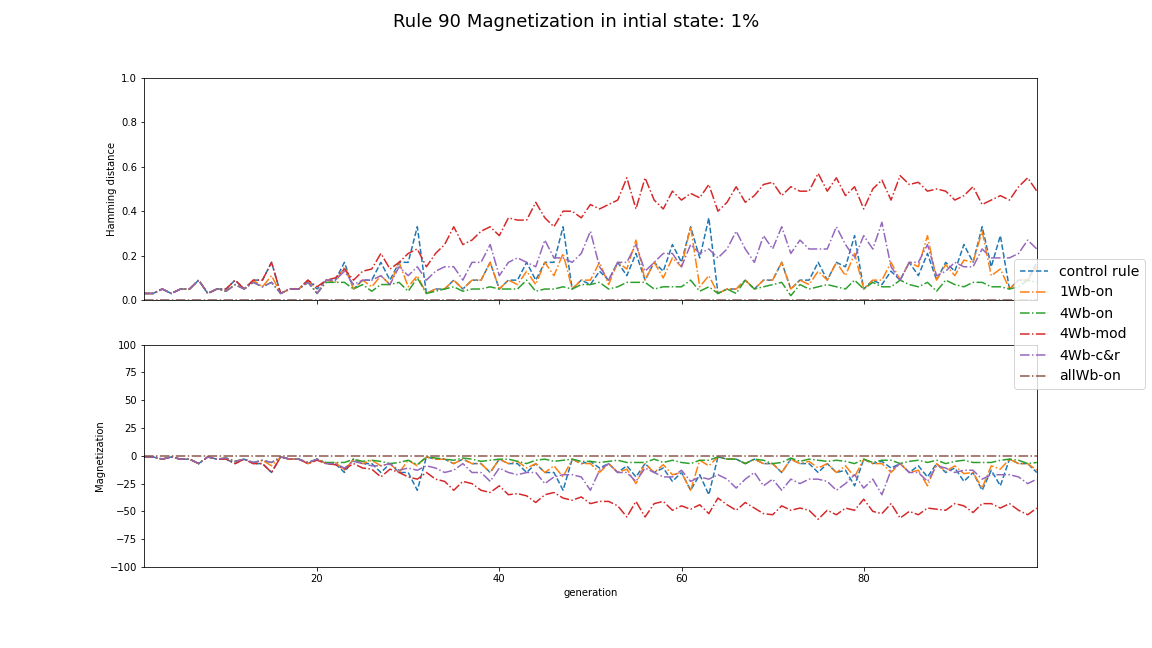}
		\caption{Rule 90. Upper plot shows how relative hamming distance vary according to evolution step for each one of the Woronin bodies activation mode. Lower graph shows how the magnetization index vary according to the same parameters.}
		\label{fig:r90d1}
	\end{figure}

\begin{figure}[!htb]
		\centering
		\includegraphics[width=0.95\linewidth]{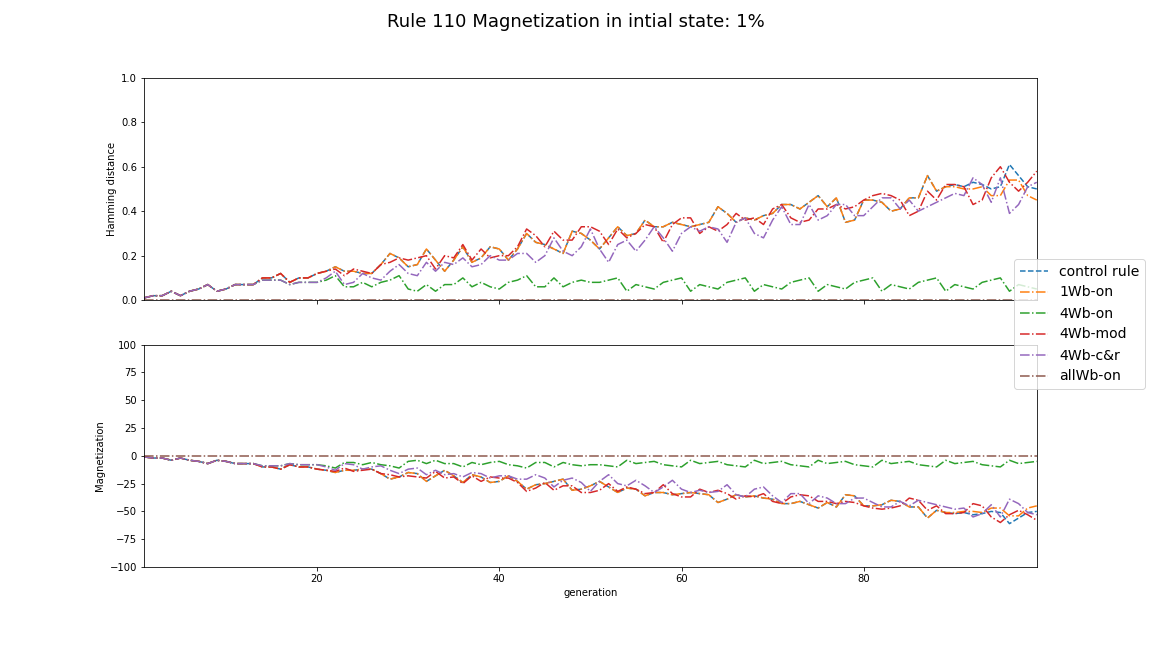}
		\caption{Rule 110. Upper plot shows how relative hamming distance varies accordingly to evolution step for each Wb activation mode. Lower graph shows how the magnetization index vary according to the same parameters.}
		\label{fig:r110d1}
	\end{figure}

\begin{figure}[!htb]
		\centering
		\includegraphics[width=0.95\linewidth]{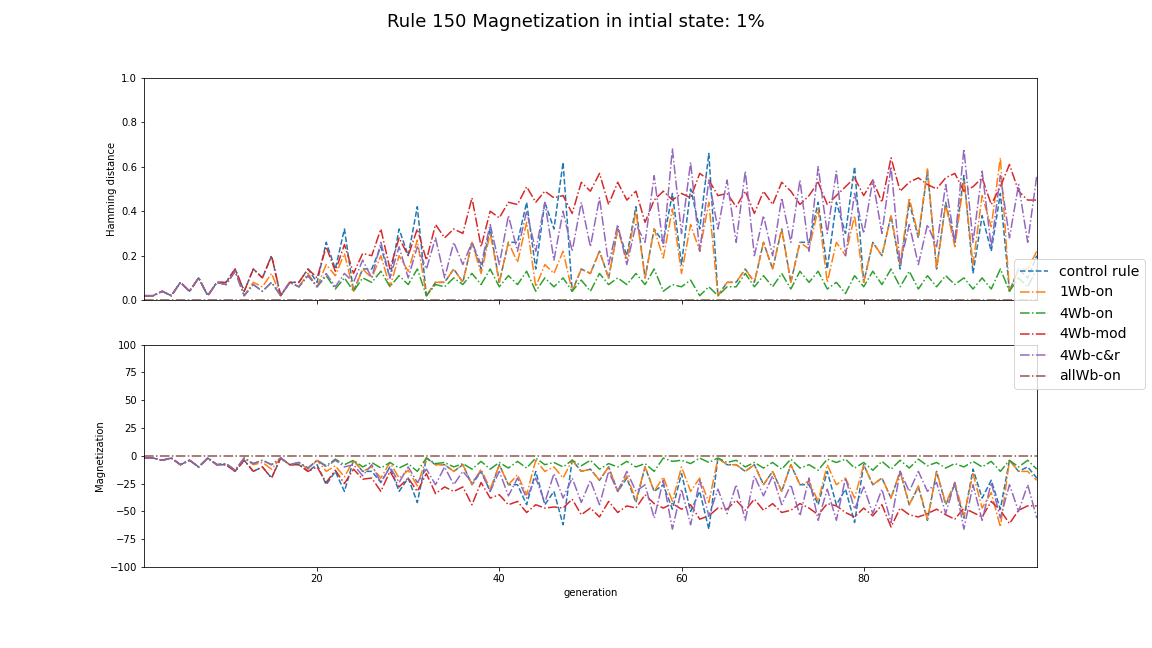}
		\caption{Rule 150. Upper plot shows how relative hamming distance varies accordingly to evolution step for each Wb activation mode. Lower graph shows how the magnetization index vary according to the same parameters.}
		\label{fig:r150d1}
	\end{figure}

	A more quantitative way to see the effects produced by Wb activation on the dynamics of each ECA rule is shown in Figures~\ref{fig:r57d10}, \ref{fig:r90d1}, \ref{fig:r110d1} and ~\ref{fig:r150d1}, where for each rule the Hamming distance and the magnetization index are shown for a given initial state, the same initial state as in  Figures~\ref{fig:r57parallelallmodes} to ~\ref{fig:r150parallelallmodes}. 
	
	\subsubsection{Exhaustive simulation results}
\begin{figure}[!htb]
		\centering
		\begin{subfigure}{.9 \linewidth}
			\centering
			\includegraphics[width=0.95\linewidth]{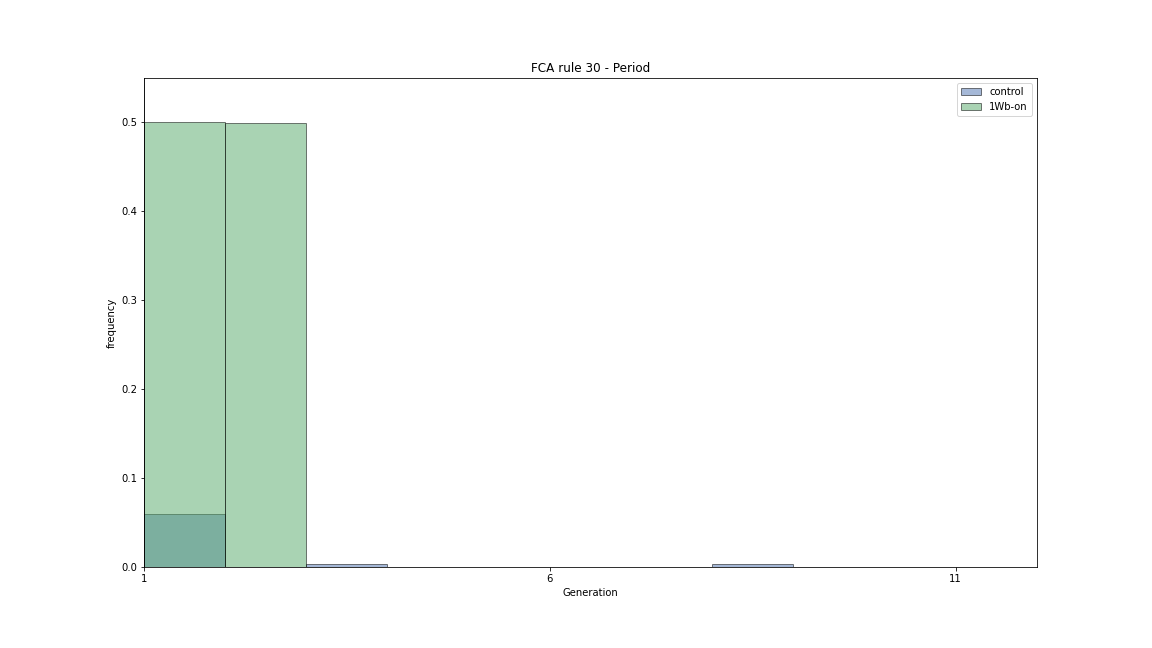}
			\caption{Periods for rule 30 with and without a Wb activated}
			\label{fig:histo-periodo-r-30}
		\end{subfigure}
		\begin{subfigure}{.9 \linewidth}
			\centering
			\includegraphics[width=0.95\linewidth]{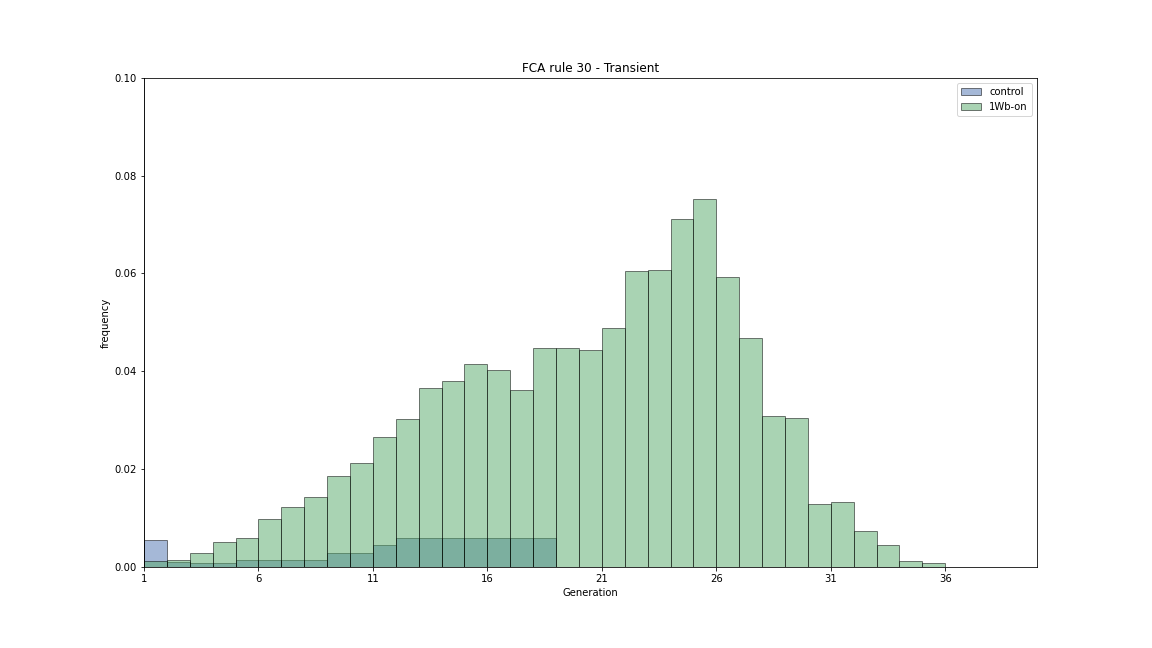}
			\caption{Transient for rule 30 with and without a Wb activated}
			\label{fig:histo-transient-r-30}
		\end{subfigure}
		\caption{Histograms for period and transient for EFCA rule 30 in a ring of size 12}
		\label{fig:histr30}
	\end{figure}

\begin{figure}[!htb]
		\centering
		\begin{subfigure}{.9 \linewidth}
			\centering
			\includegraphics[width=0.95\linewidth]{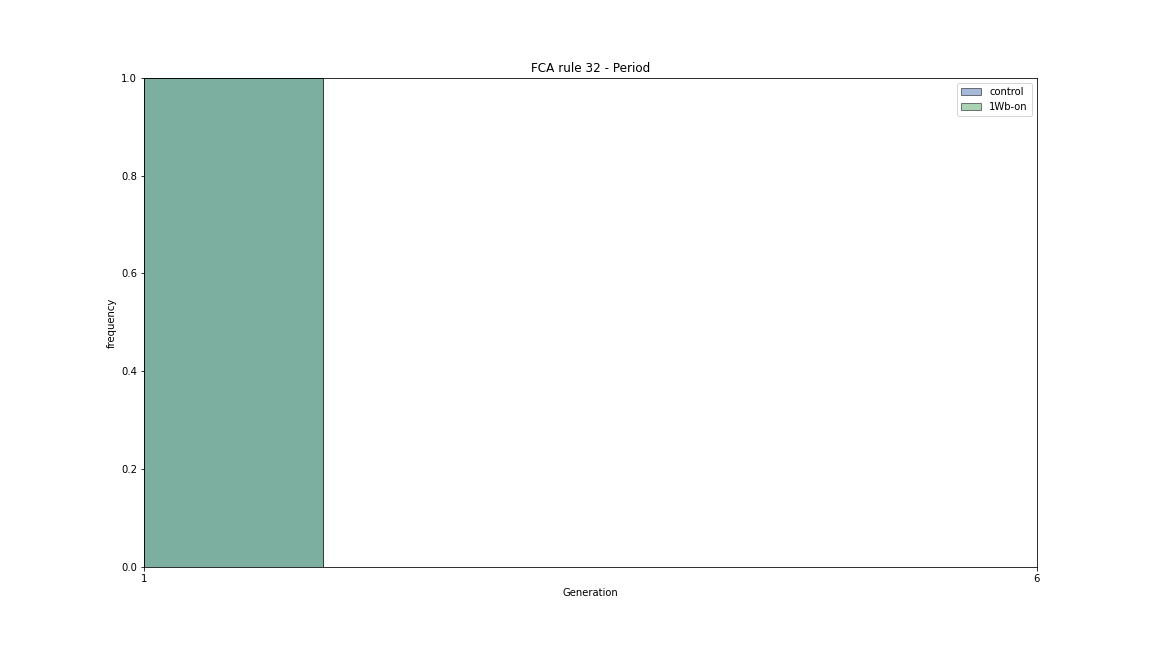}
			\caption{Periods for rule 32 with and without a Wb activated}
			\label{fig:histo-periodo-r-32}
		\end{subfigure}
		\begin{subfigure}{.9 \linewidth}
			\centering
			\includegraphics[width=0.95\linewidth]{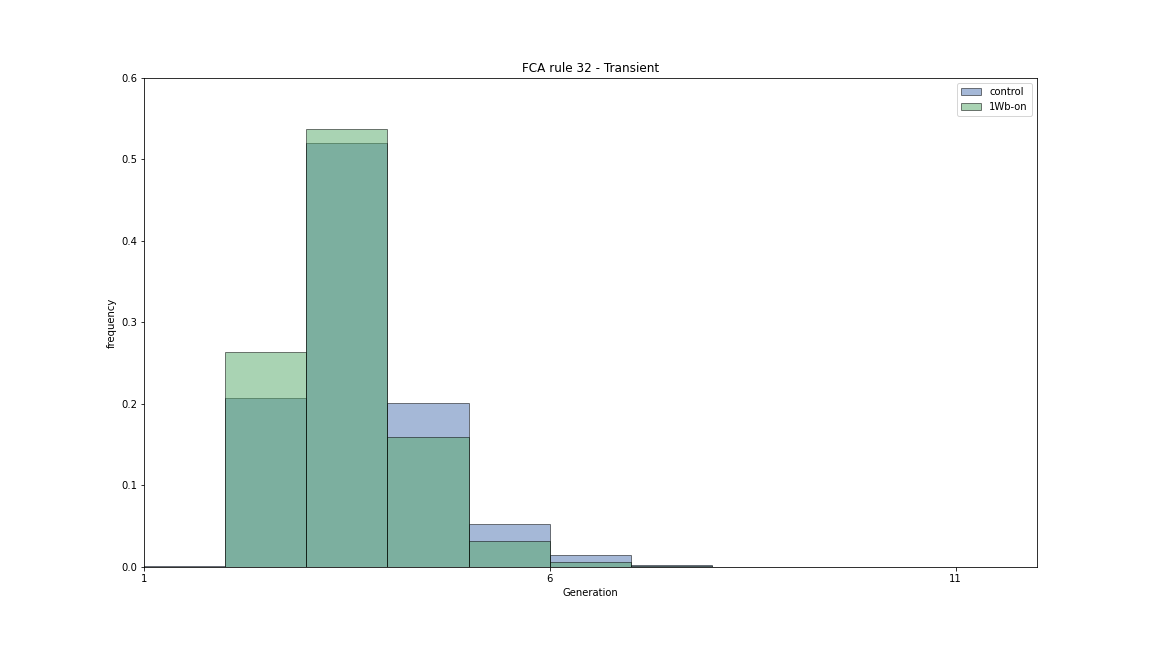}
			\caption{Transient for rule 32 with and without a Woronin body activated}
			\label{fig:histo-transient-r-32}
		\end{subfigure}
		\caption{Histograms for period and transient for EFCA rule 32 in a ring of size 12}
		\label{fig:histr32}
	\end{figure}

\begin{figure}[!htb]
		\centering
		\begin{subfigure}{.9 \linewidth}
			\centering
			\includegraphics[width=0.95\linewidth]{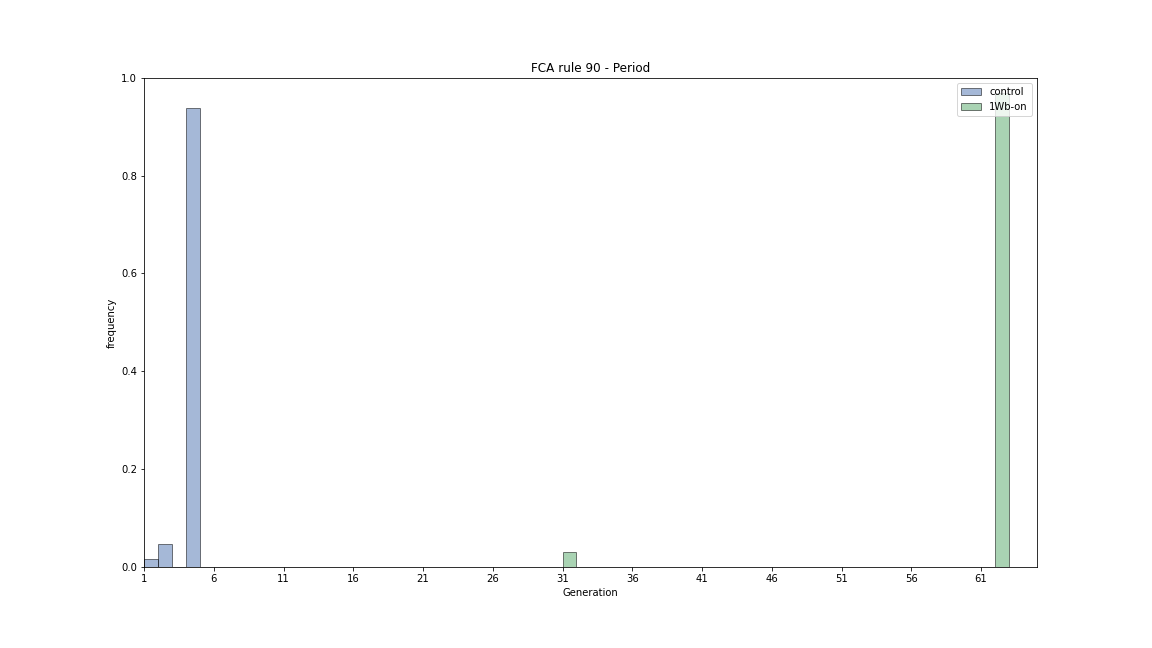}
			\caption{Periods for rule 90 with and without a Wb activated}
			\label{fig:histo-periodo-r-90}
		\end{subfigure}
		\begin{subfigure}{.9 \linewidth}
			\centering
			\includegraphics[width=0.95\linewidth]{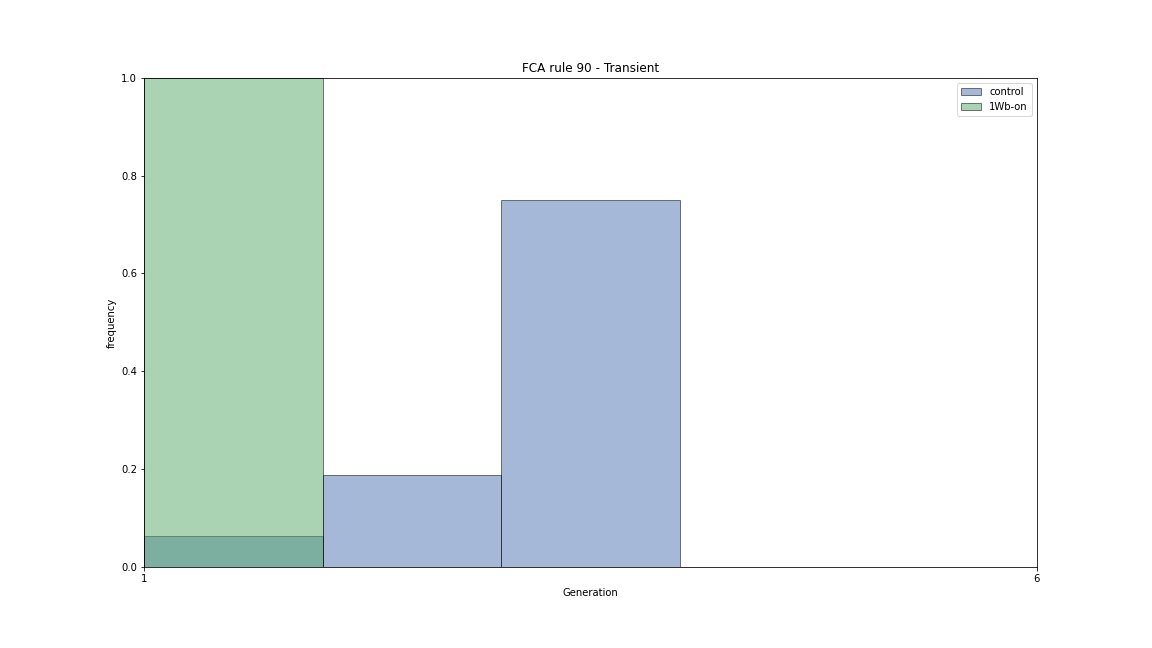}
			\caption{Transient for rule 90 with and without a Wb activated}
			\label{fig:histo-transient-r-90}
		\end{subfigure}
		\caption{Histograms for period and transient for EFCA rule 90 in a ring of size 12}
		\label{fig:histr90}
	\end{figure}

\begin{figure}[!htb]
		\centering
		\begin{subfigure}{.9 \linewidth}
			\centering
			\includegraphics[width=0.95\linewidth]{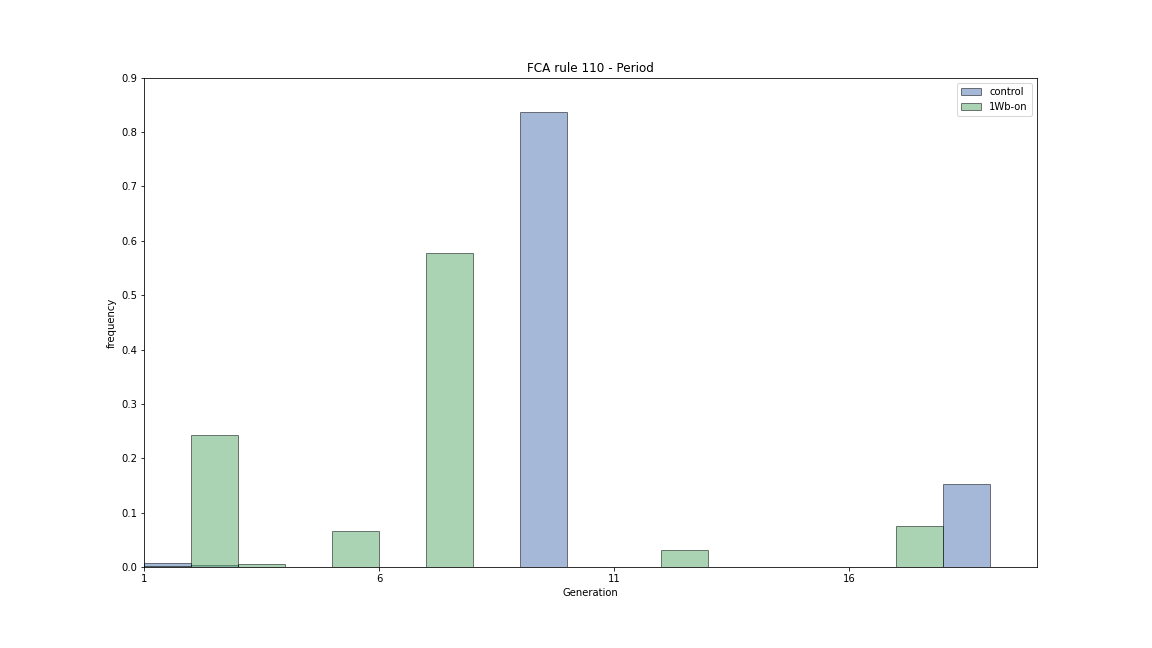}
			\caption{Periods for rule 110 with and without a Wb activated}
			\label{fig:histo-periodo-r-110}
		\end{subfigure}
		\begin{subfigure}{.9 \linewidth}
			\centering
			\includegraphics[width=0.95\linewidth]{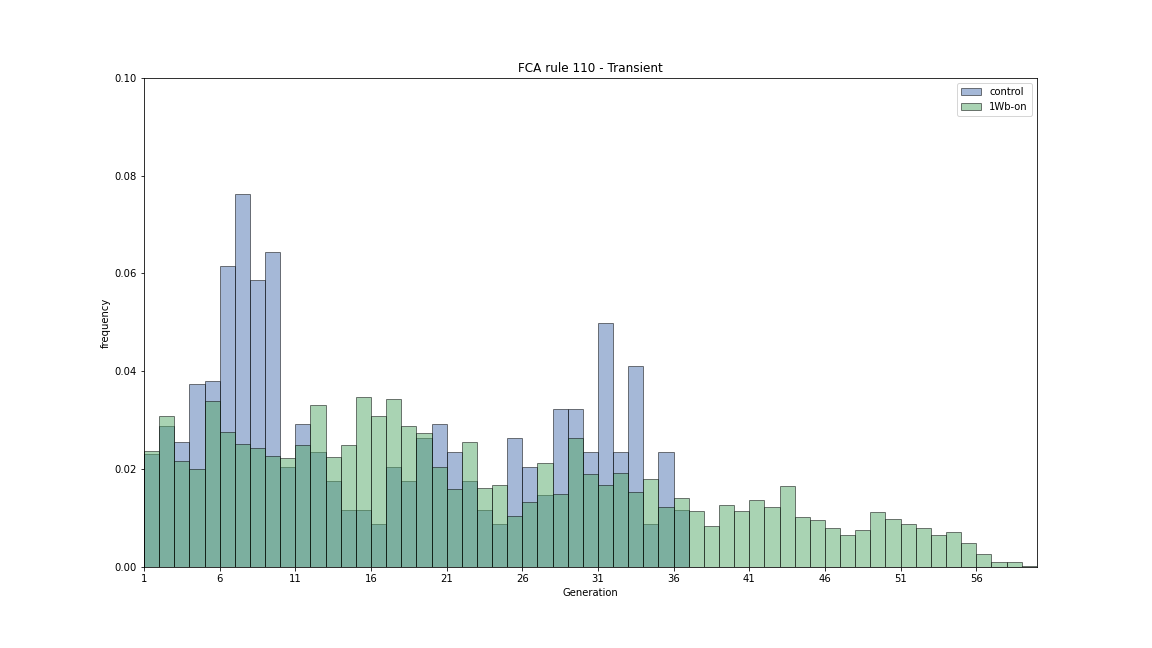}
			\caption{Transient for rule 110 with and without a Wb activated}
			\label{fig:histo-transient-r-110}
		\end{subfigure}
		\caption{Histograms for period and transient for EFCA rule 110 in a ring of size 12}
		\label{fig:histr110}
	\end{figure}

\begin{figure}[!htb]
		\centering
		\begin{subfigure}{.9 \linewidth}
			\centering
			\includegraphics[width=0.95\linewidth]{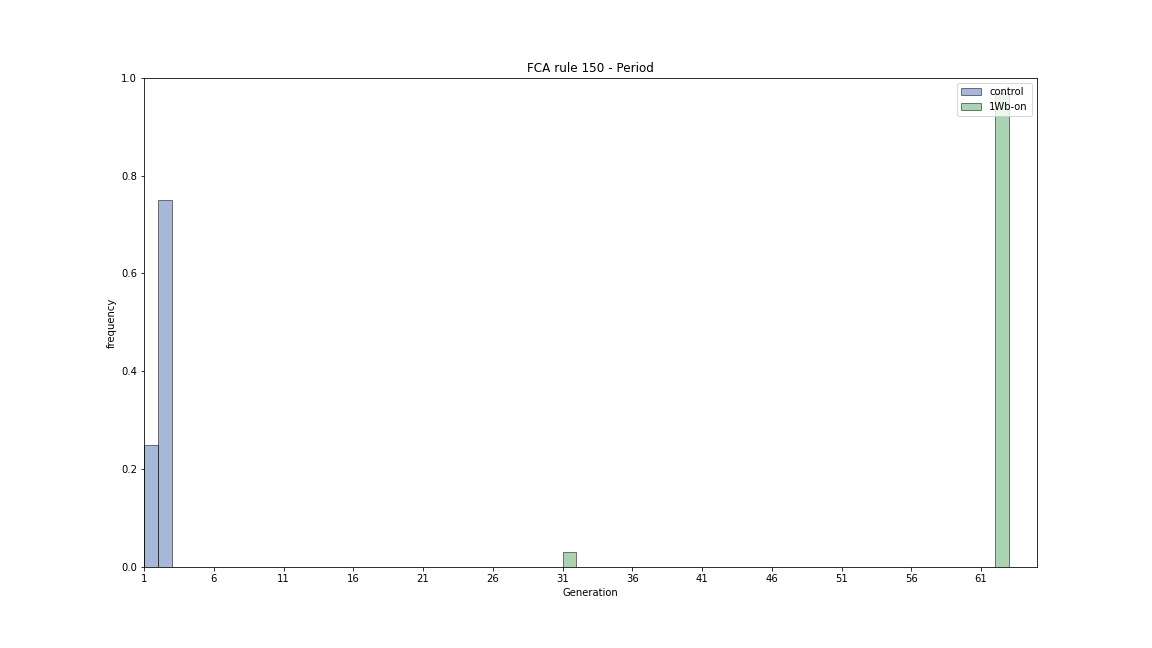}
			\caption{Periods for rule 150 with and without a Wb activated}
			\label{fig:histo-periodo-r-150}
		\end{subfigure}
		\begin{subfigure}{.9 \linewidth}
			\centering
			\includegraphics[width=0.95\linewidth]{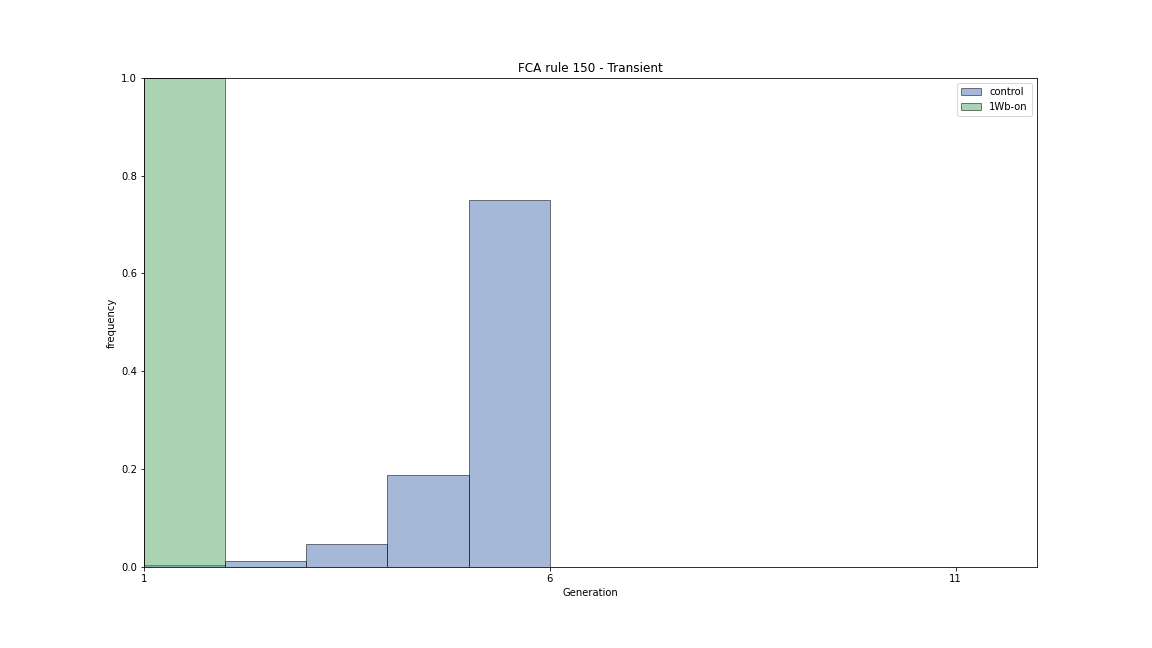}
			\caption{Transient for rule 150 with and without a Wb activated}
			\label{fig:histo-transient-r-150}
		\end{subfigure}
		\caption{Histograms for period and transient for EFCA rule 150 in a ring of size 12}
		\label{fig:histr150}
	\end{figure}

	For all 4096 different initial state and for each rule selected ($\left[30, 32, 90, 110, 150 \right]$) we find cycles or fixed point when they exist and calculate period and transient. In Figures~\ref{fig:histr30} - \ref{fig:histr150} is possible to see the histogram of periods and transient for each rule against the same rule, but with one Wb activated. 
	
\begin{figure}[!htb]
		\centering
		\begin{minipage}{0.4 \textwidth}
			\begin{subfigure}{.45 \linewidth}
				\centering
				\includegraphics[width=0.6 \linewidth]{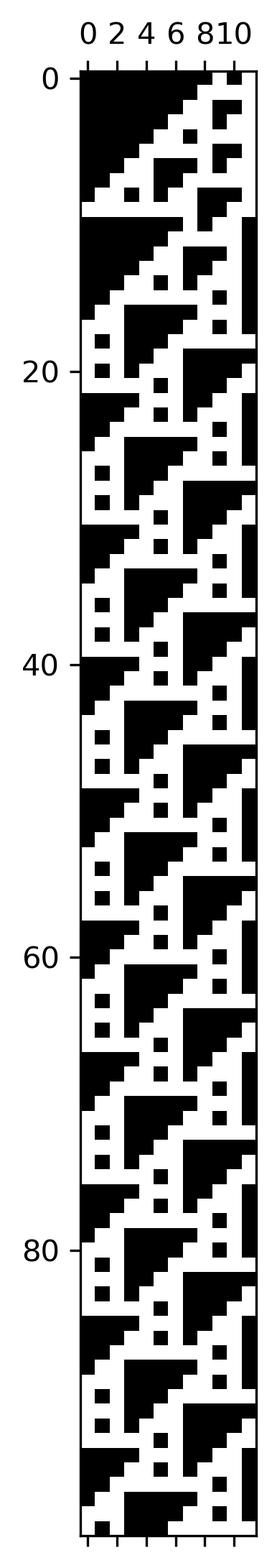}
				\caption{}
				\label{fig:r110n12control}
			\end{subfigure}
			\hfill
			\begin{subfigure}{.45 \linewidth}
				\centering
				\includegraphics[width=0.6 \linewidth]{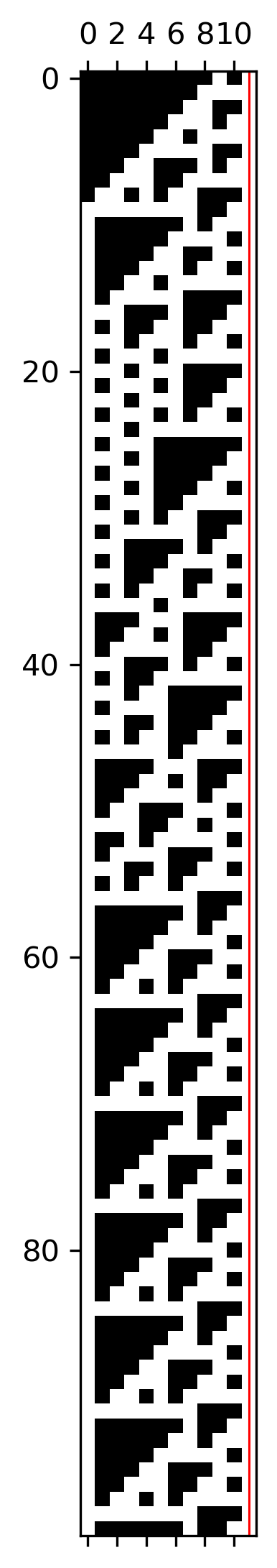}
				\caption{}
				\label{fig:r110n12-1Wb}
			\end{subfigure}	
		\end{minipage}
		\hspace{0.02\linewidth}
		\begin{minipage}{0.5 \textwidth}
			\begin{subfigure}{\linewidth}
				\includegraphics[width=\linewidth]{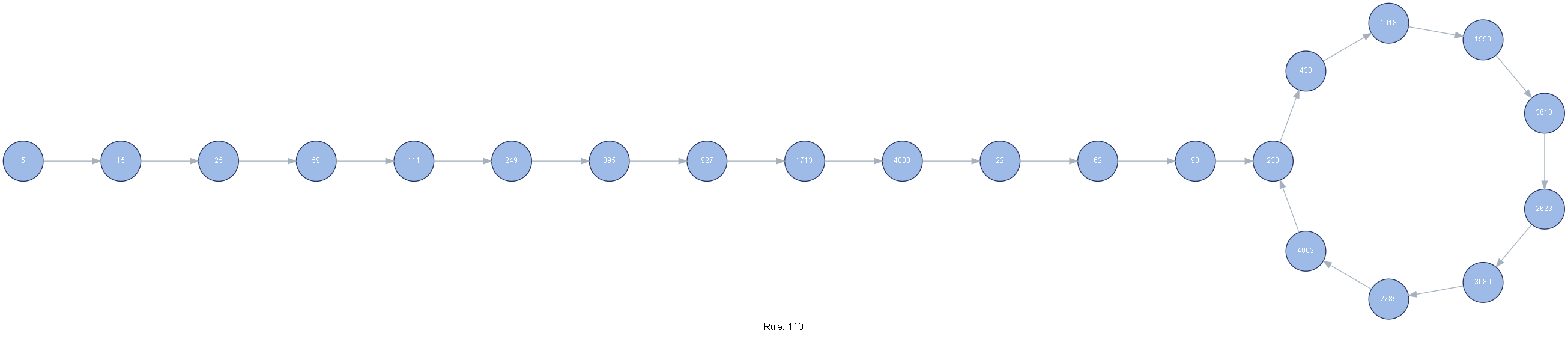}
				\caption{}
				\label{fig:r110_evo}
			\end{subfigure}
			\begin{subfigure}{\linewidth}
				\includegraphics[width=\linewidth]{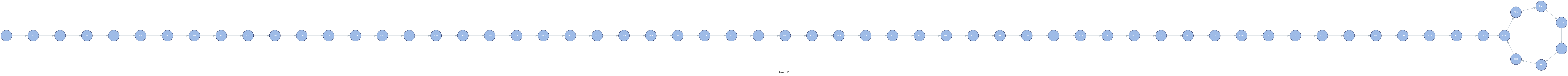}
				\caption{}
				\label{fig:r1101wb_evo}
			\end{subfigure}
			\begin{subfigure}{\linewidth}
				\includegraphics[width=\linewidth]{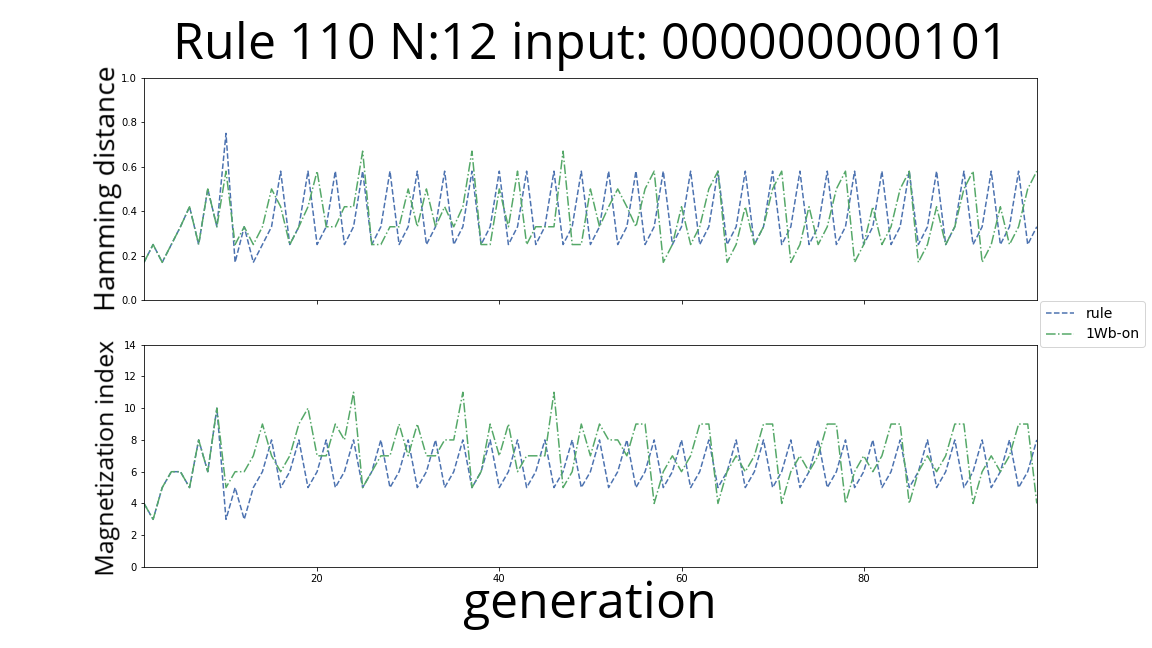}
				\caption{}
				\label{fig:r110_stats}
			\end{subfigure}
			
		\end{minipage}
		\caption{Evolution of  rule 110 from initial state 000000000101. (a) Wbs are passive during first 100 generations, (b) Wbs are active for 100 generations, (c) Evolution graph for rule without Wb, (d) Evolution graph with Wb activated, (e) Variation in relative Hamming distance and magnetization index. }
		\label{fig:110_all}
	\end{figure}

	A detailed example of this process for a given input can be seen in Figure ~\ref{fig:110_all}. The first images show the EFCA evolution for a given initial state with and without Wb activation, next the evolution graph is produced and finally a plot showing the variation in relative Hamming distance and magnetization index. The result shows that the activation of Wb induces changes in the behavior of automata, for example rule 150 generates for almost 95\% of the time cycles with period of 62 in comparison with the scenarios where  no Wbs were activated.
	
	\subsubsection{Simulation with more cells and one Wb activated}
	
\begin{figure}[!htb]
		\centering
		\begin{subfigure}{.9 \linewidth}
			\centering
			\includegraphics[width=0.95\linewidth]{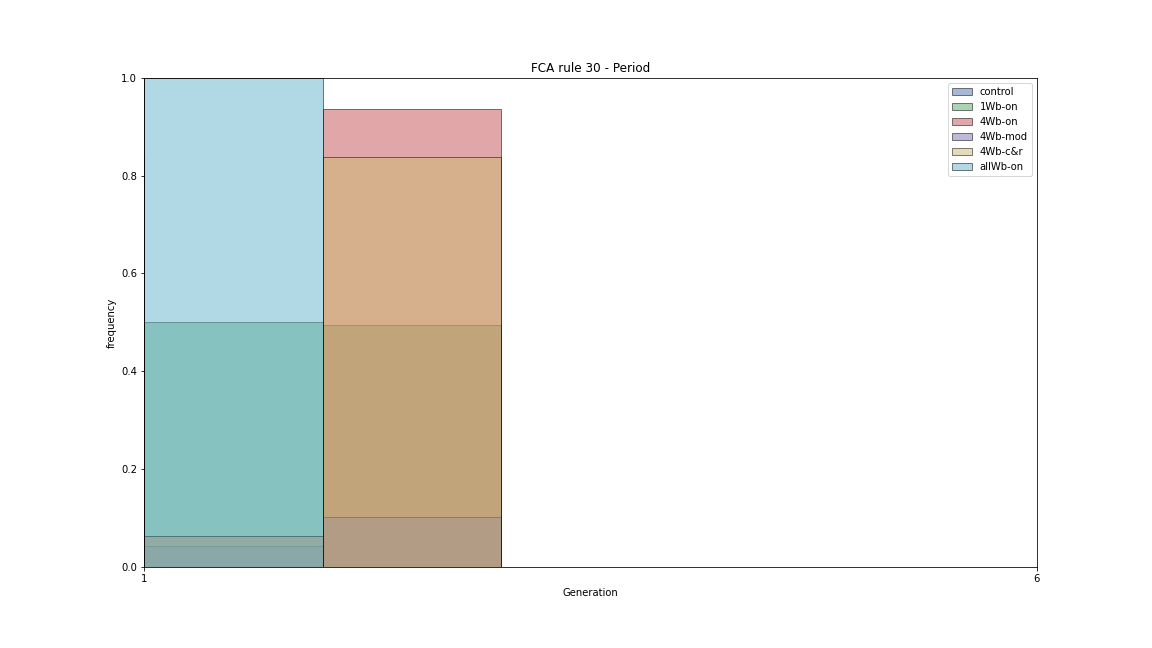}
			\caption{Periods rule 30 for different Wb activation}
			\label{fig:3ehisto-periodo-r-30}
		\end{subfigure}
		\begin{subfigure}{.9 \linewidth}
			\centering
			\includegraphics[width=0.95\linewidth]{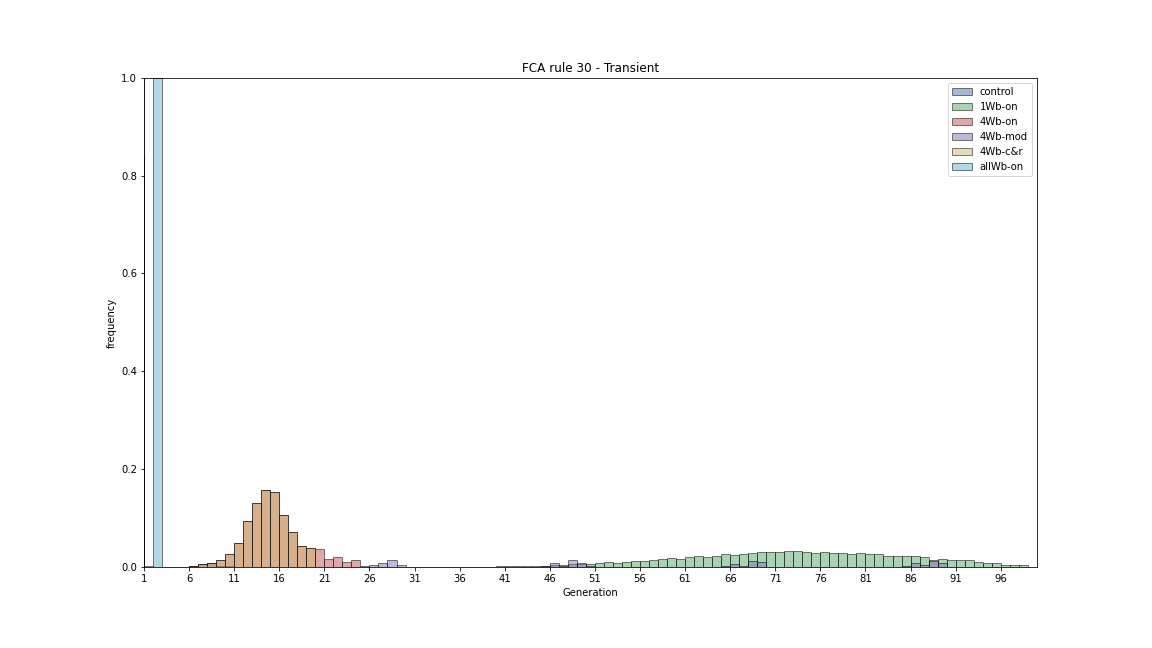}
			\caption{Transient rule 30 for different Wb activation}
			\label{fig:3ehisto-transient-r-30}
		\end{subfigure}
		\caption{Histograms for period and transient for EFCA rule 30 on a ring of size 32. Wb activation 1Wb-on, 4Wb-on, 4Wb-mod, 4Wb-c\&r, allWb-on}
		\label{fig:3eR30}
	\end{figure}

\begin{figure}[!htb]
		\centering
		\begin{subfigure}{.9 \linewidth}
			\centering
			\includegraphics[width=0.95\linewidth]{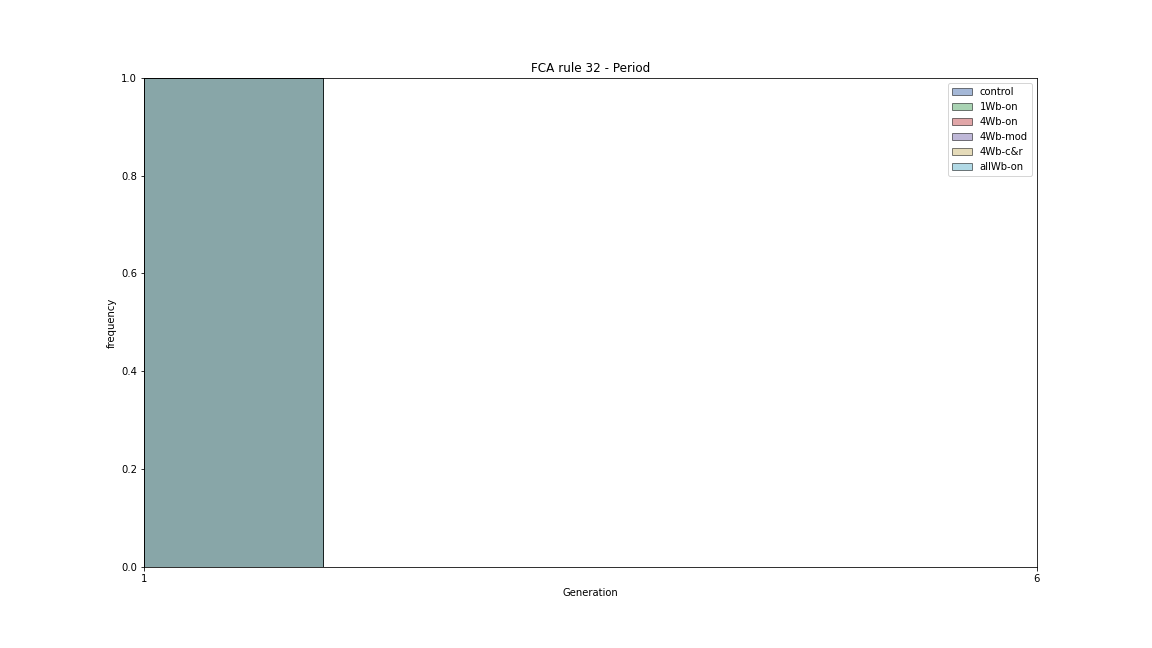}
			\caption{Periods rule 32 for different Wb activation}
			\label{fig:3ehisto-periodo-r-32}
		\end{subfigure}
		\begin{subfigure}{.9 \linewidth}
			\centering
			\includegraphics[width=0.95\linewidth]{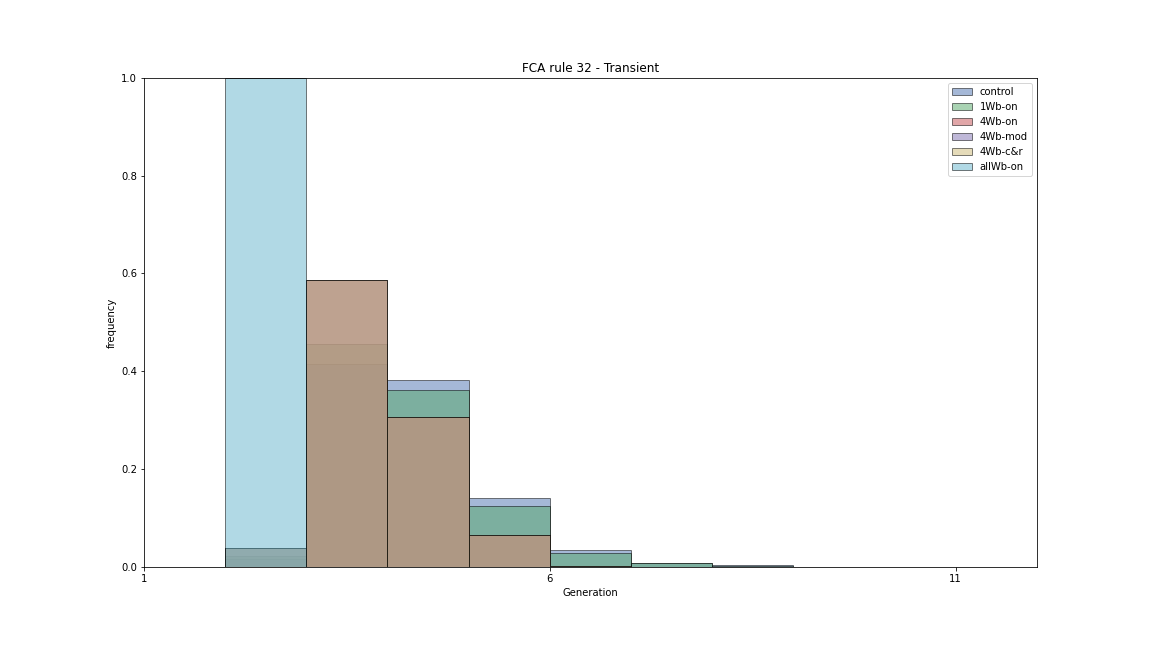}
			\caption{Transient rule 32 for different Wb activation}
			\label{fig:3ehisto-transient-r-32}
		\end{subfigure}
		\caption{Histograms for period and transient for EFCA rule 32 on a ring of size 32. Wb activation 1Wb-on, 4Wb-on, 4Wb-mod, 4Wb-c\&r, allWb-on}
		\label{fig:3eR32}
	\end{figure}

\begin{figure}[!htb]
		\centering
		\begin{subfigure}{.9 \linewidth}
			\centering
			\includegraphics[width=0.95\linewidth]{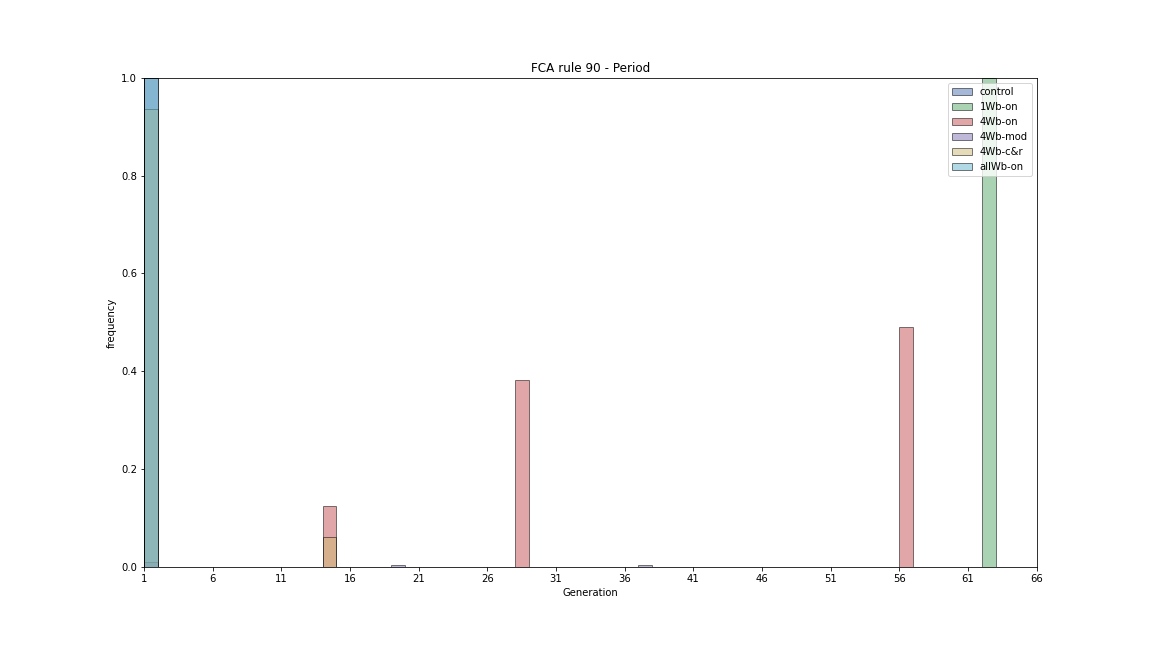}
			\caption{Periods rule 90 for different Woronin body activation}
			\label{fig:3ehisto-periodo-r-90}
		\end{subfigure}
		\begin{subfigure}{.9 \linewidth}
			\centering
			\includegraphics[width=0.95\linewidth]{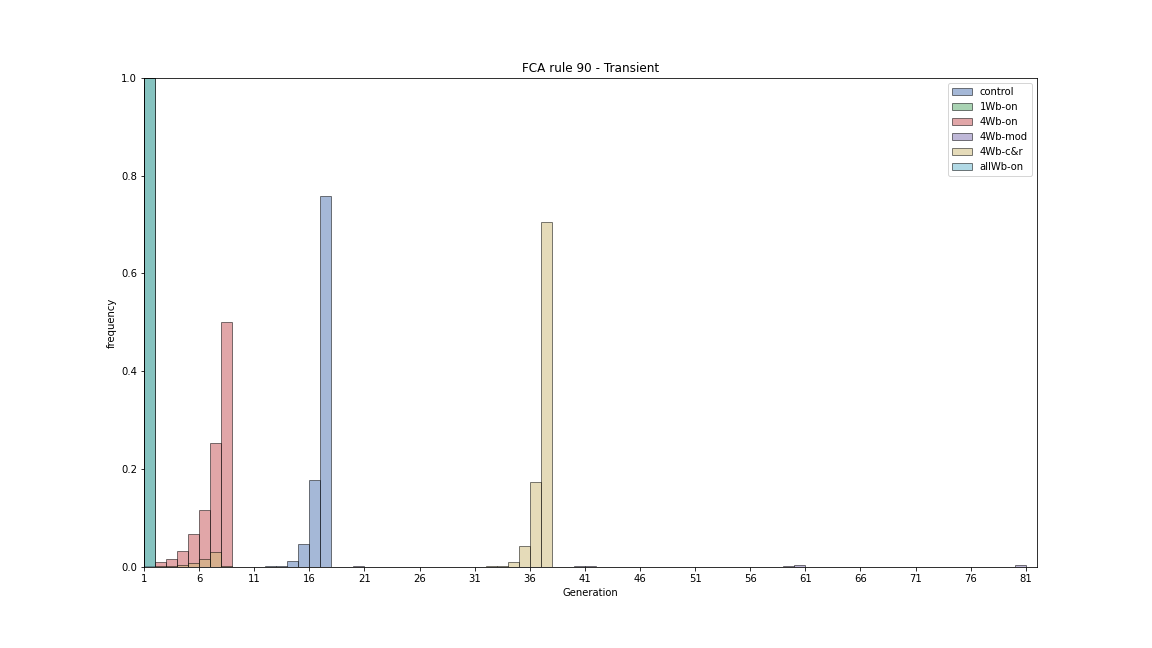}
			\caption{Transient rule 90 for different Wb activation}
			\label{fig:3ehisto-transient-r-90}
		\end{subfigure}
		\caption{Histograms for period and transient for EFCA rule 92 on a ring of size 32. Wb activation 1Wb-on, 4Wb-on, 4Wb-mod, 4Wb-c\&r, allWb-on}
		\label{fig:3eR90}
	\end{figure}

\begin{figure}[!htb]
		\centering
		\begin{subfigure}{.9 \linewidth}
			\centering
			\includegraphics[width=0.95\linewidth]{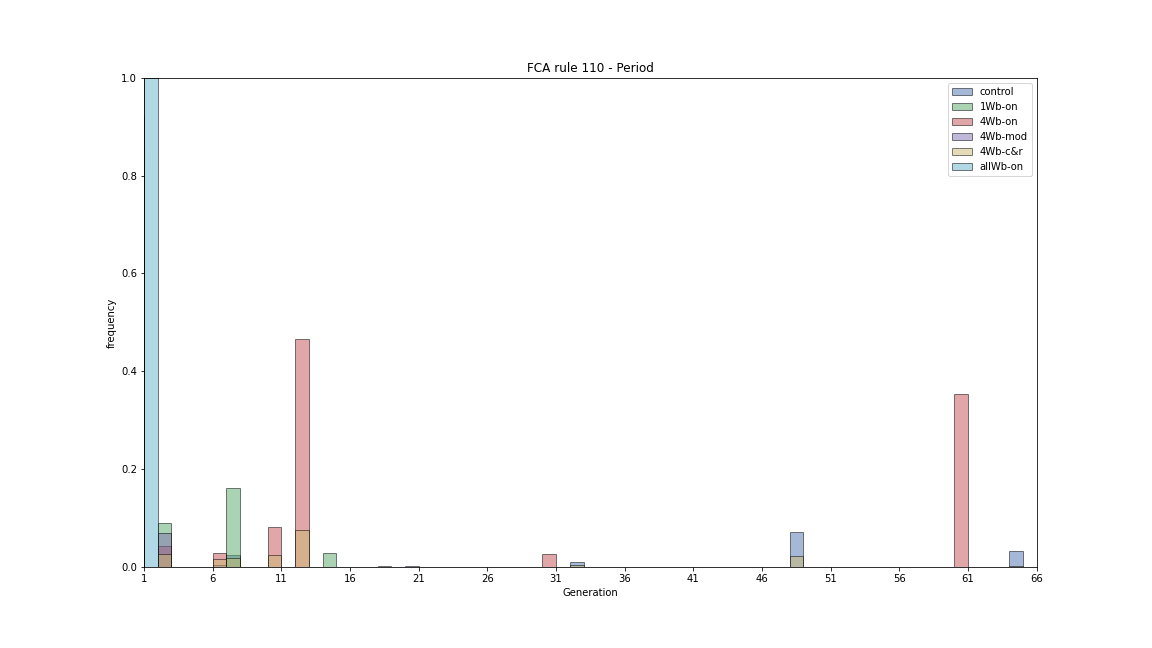}
			\caption{Periods rule 110 for different Wb activation}
			\label{fig:3ehisto-periodo-r-110}
		\end{subfigure}
		\begin{subfigure}{.9 \linewidth}
			\centering
			\includegraphics[width=0.95\linewidth]{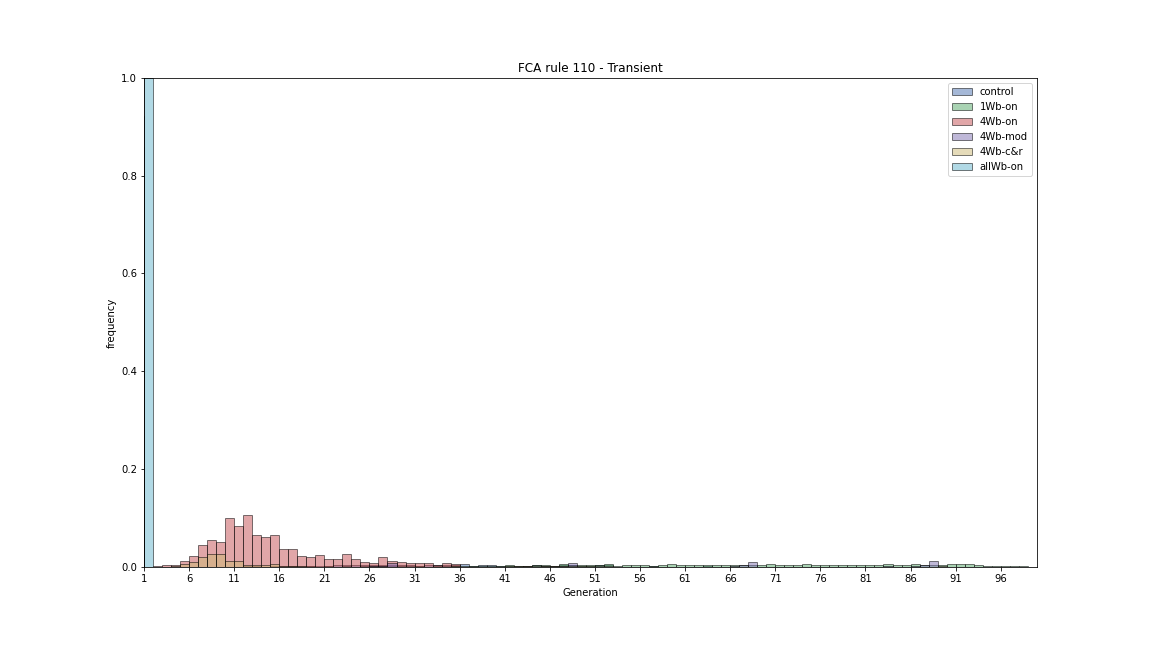}
			\caption{Transient rule 110 for different Wb activation}
			\label{fig:3ehisto-transient-r-110}
		\end{subfigure}
		\caption{Histograms for period and transient for EFCA rule 110 on a ring of size 32. Wb activation 1Wb-on, 4Wb-on, 4Wb-mod, 4Wb-c\&r, allWb-on}
		\label{fig:3eR110}
	\end{figure}

\begin{figure}[!htb]
		\centering
		\begin{subfigure}{.9 \linewidth}
			\centering
			\includegraphics[width=0.95\linewidth]{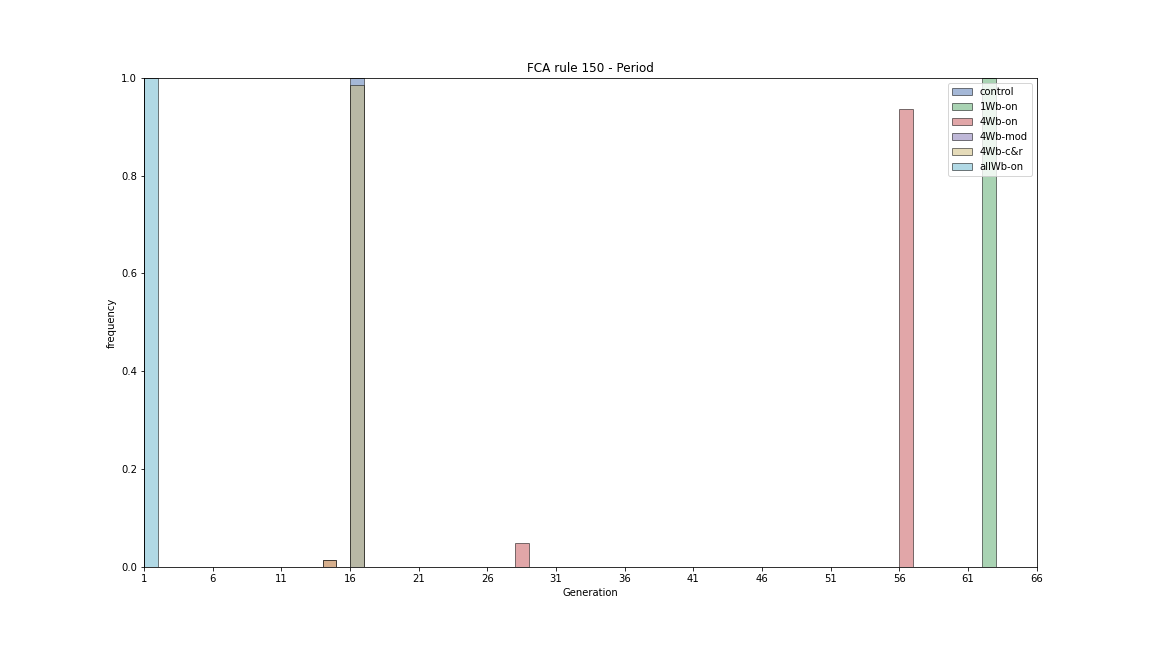}
			\caption{Periods rule 150 for different Wb activation}
			\label{fig:3ehisto-periodo-r-150}
		\end{subfigure}
		\begin{subfigure}{.9 \linewidth}
			\centering
			\includegraphics[width=0.95\linewidth]{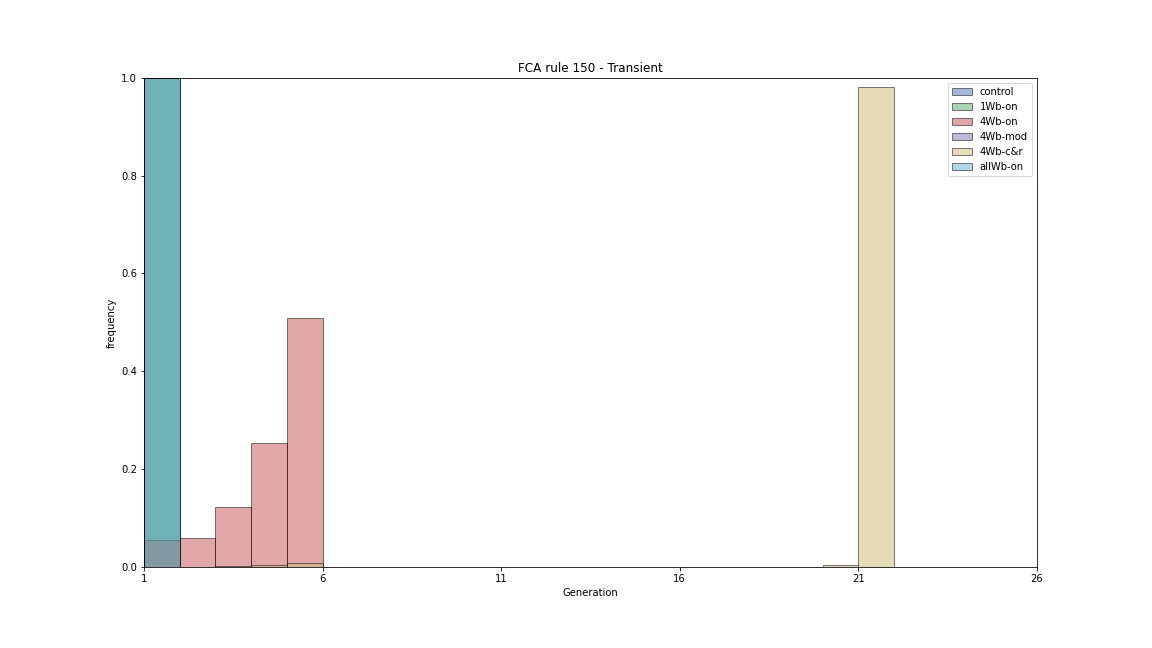}
			\caption{Transient rule 150 for different Wb activation}
			\label{fig:3ehisto-transient-r-150}
		\end{subfigure}
		\caption{Histograms for period and transient for EFCA rule 150 on a ring of size 32. Wb activation 1Wb-on, 4Wb-on, 4Wb-mod, 4Wb-c\&r, allWb-on}
		\label{fig:3eR150}
	\end{figure}

	A total of 195,000 simulations were run. For each of them we find cycles and fixed points. The findings are summarized in histograms shown on Figures~\ref{fig:3eR30} to ~\ref{fig:3eR150}. The aim was to explore if a variation on the ring size could increase or decrease the effects produced by different Wb activation.
	
	\subsection{Majority Fungal Automanta}
\begin{figure}[!htb]
		\centering
		\includegraphics[height=0.40\textheight]{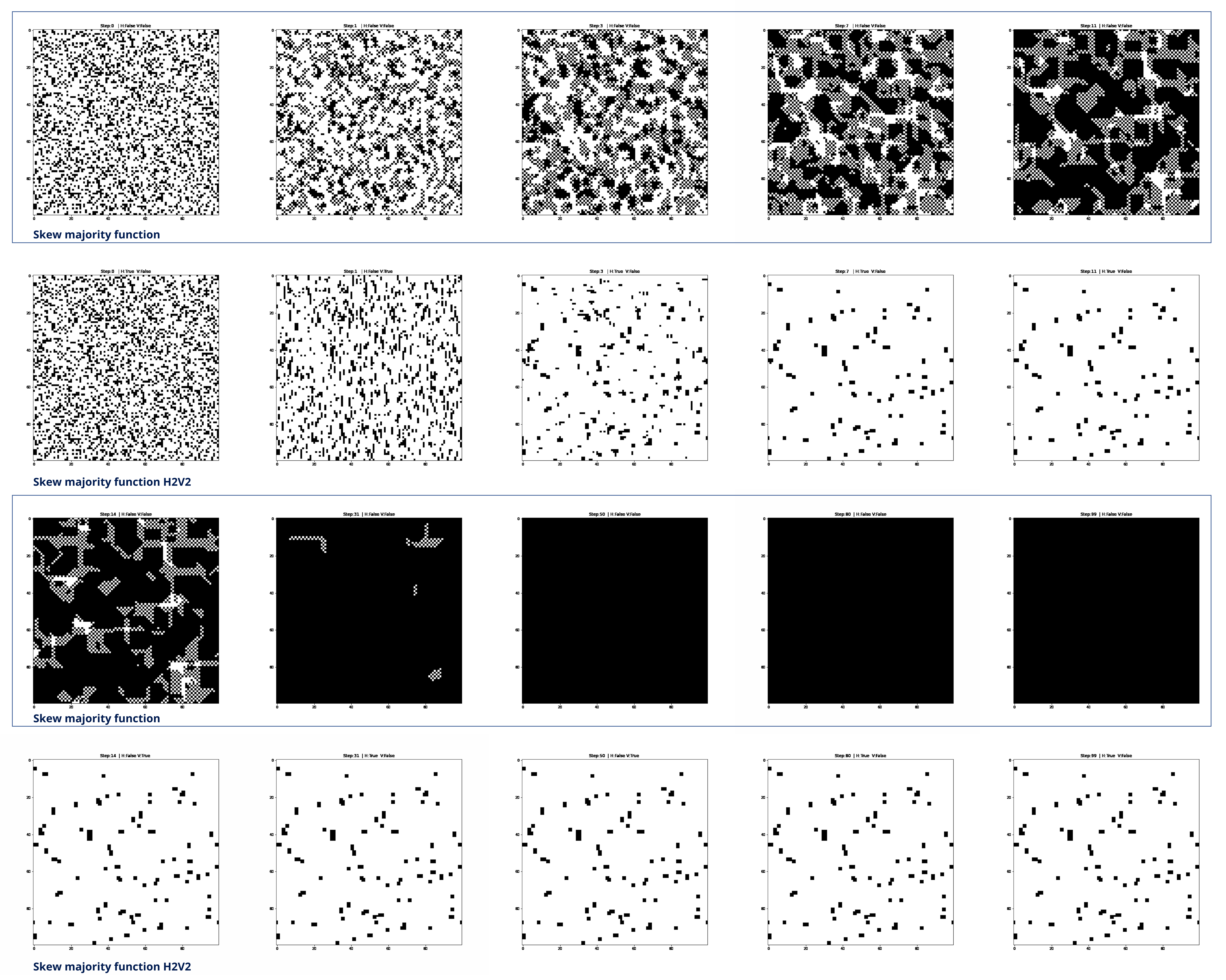}
		\caption{Evolution steps 0,1,3,7,11,14,31,50, 80 y 99 for skew majority global function and same rule with and H2V2 Wbs activation scheme.}
		\label{fig:maj_all}
	\end{figure}

\begin{figure}[!htb]
		\centering
		\includegraphics[height=0.40\textheight]{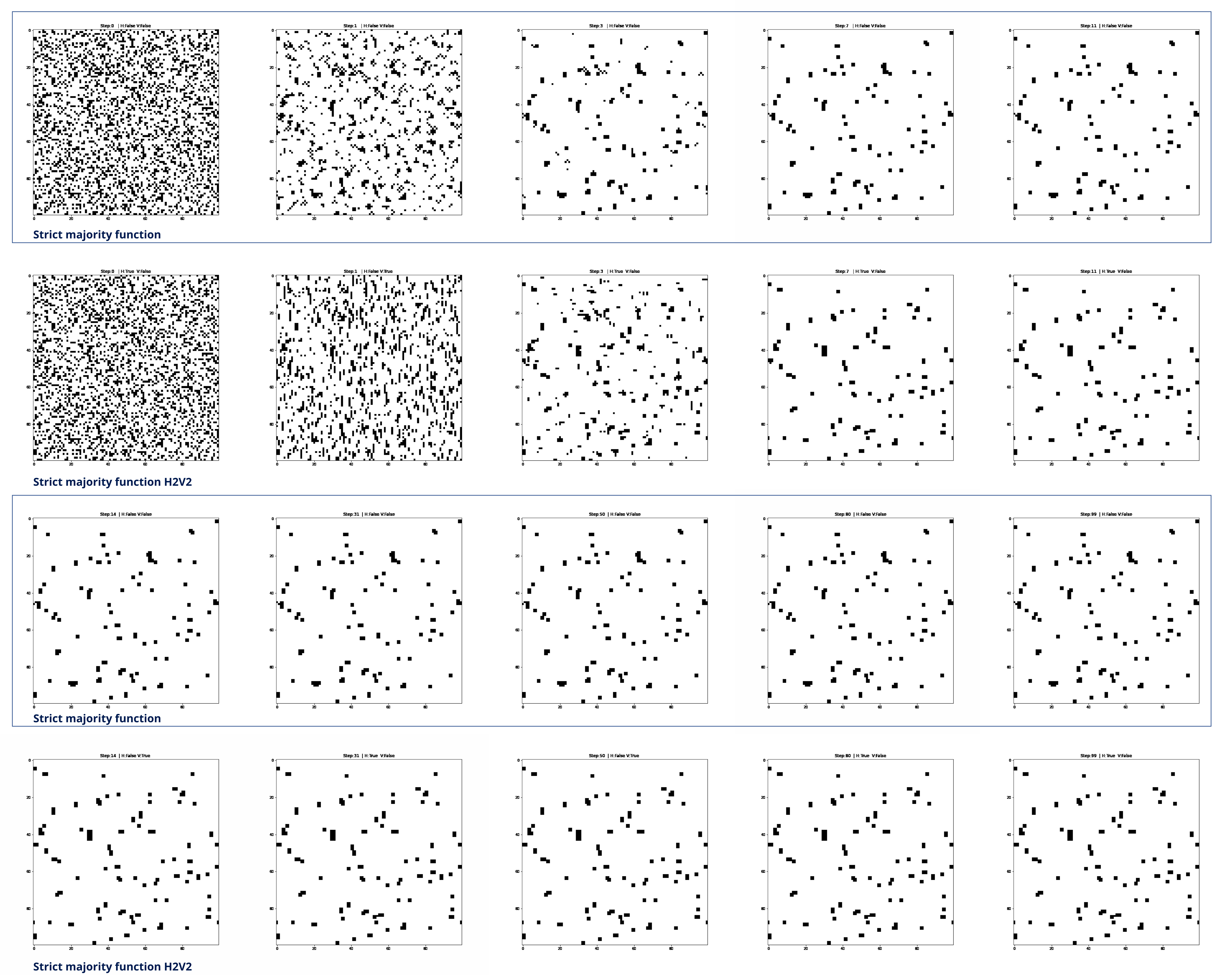}
		\caption{Evolution steps 0,1,3,7,11,14,31,50, 80 y 99 for strict majority global function and same rule with and H2V2 Wbs activation scheme.}
		\label{fig:strict_all}
	\end{figure}

	A total of 12000 simulations with random initial magnetization state were run. Sample of some timestamps for both global rules and H2V2 activation mode are shown in Figures~\ref{fig:maj_all} and ~\ref{fig:strict_all}.

\begin{figure}[!htb]
		\centering
		\includegraphics[height=0.40\textheight]{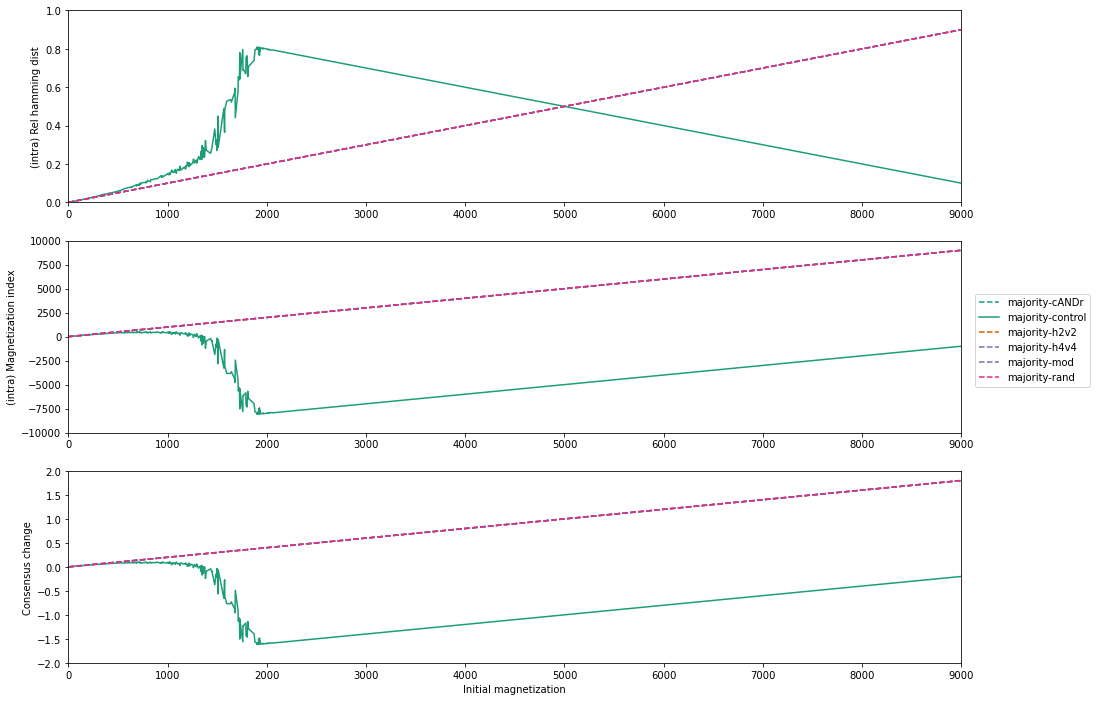}
		\caption{Relative hamming distance, Magnetization index and Consensus Change against initial magnetization for skew-majority.}
		\label{fig:maj_intra}
	\end{figure}

\begin{figure}[!htb]
		\centering
		\includegraphics[height=0.40\textheight]{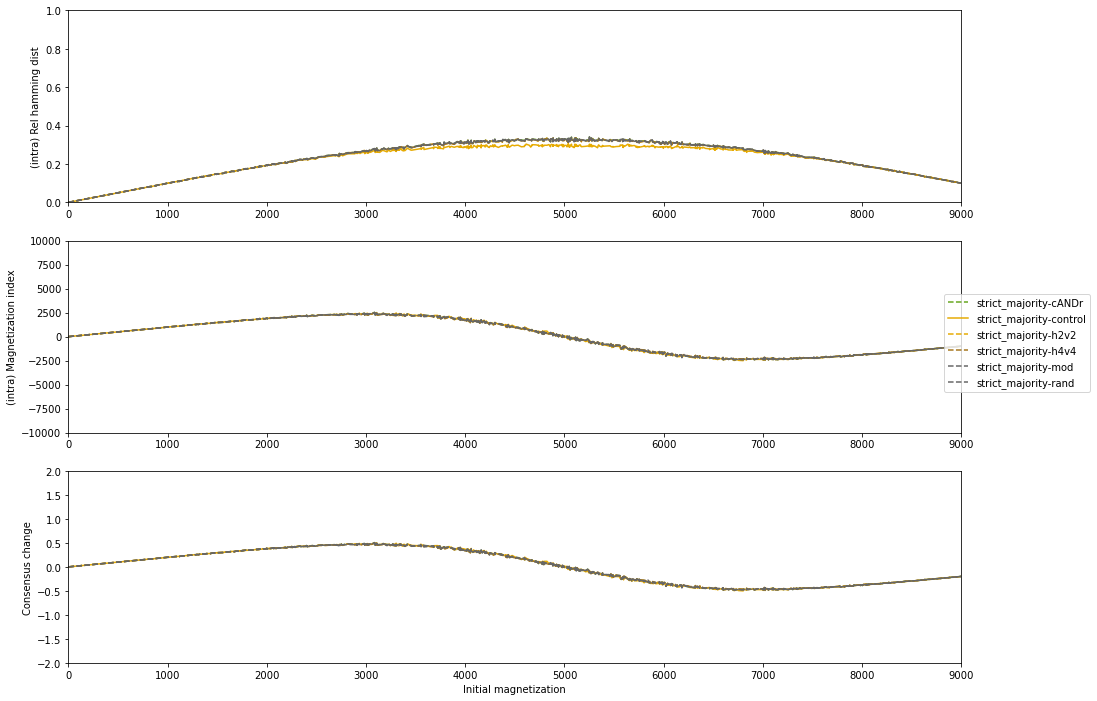}
		\caption{Relative hamming distance, Magnetization index and Consensus Change against initial magnetization for skew-majority.}
		\label{fig:str_majintra}
	\end{figure}

	The general results are summarized in Figures~\ref{fig:maj_intra} - \ref{fig:str_majintra}. This figures show the individual metrics as a function of magnetization in initial state using skew-majority and majority as global function, and different Wb activation sequence. It is possible to observe in Figure~\ref{fig:maj_intra} the effect of Wb activation, transforming the dynamic of skew-majority. In the case of strict majority only irrelevant changes are induced as shown Figure~\ref{fig:str_majintra}.
	
\begin{figure}[!htb]
		\centering
		\includegraphics[width=0.95\linewidth]{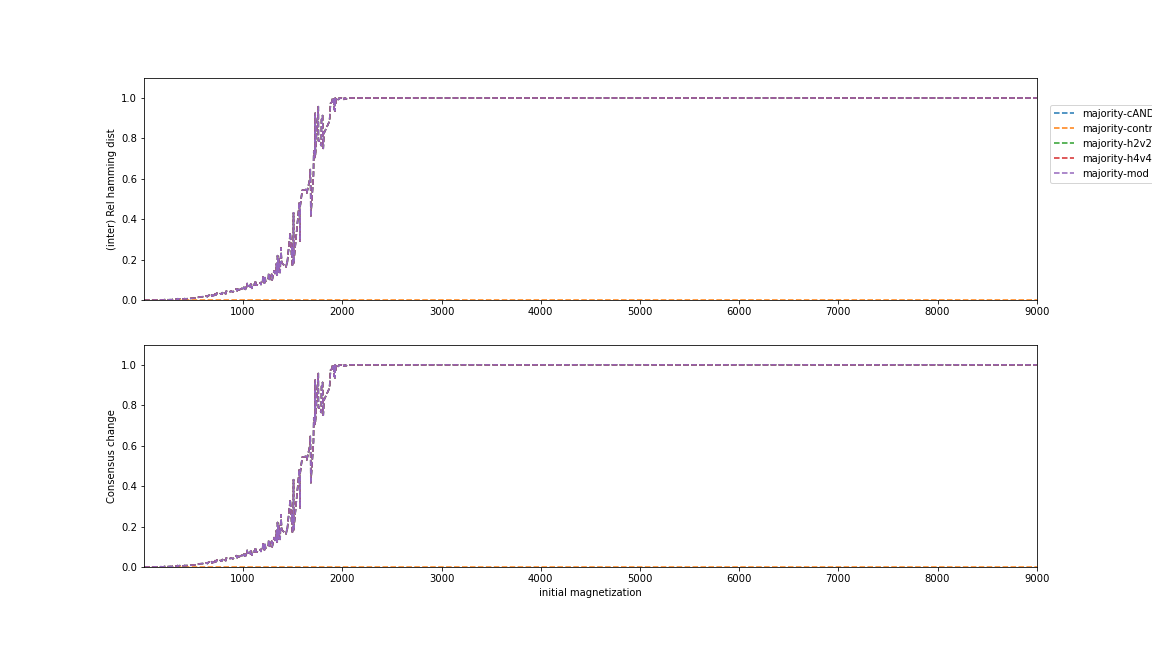}
		\caption{Relative Hamming distance and consensus change between skew-majority global function and same global function with different Woronin bodies activation.}
		\label{fig:maj_inter}
	\end{figure}

\begin{figure}[!htb]
		\centering
		\includegraphics[height=0.40\textheight]{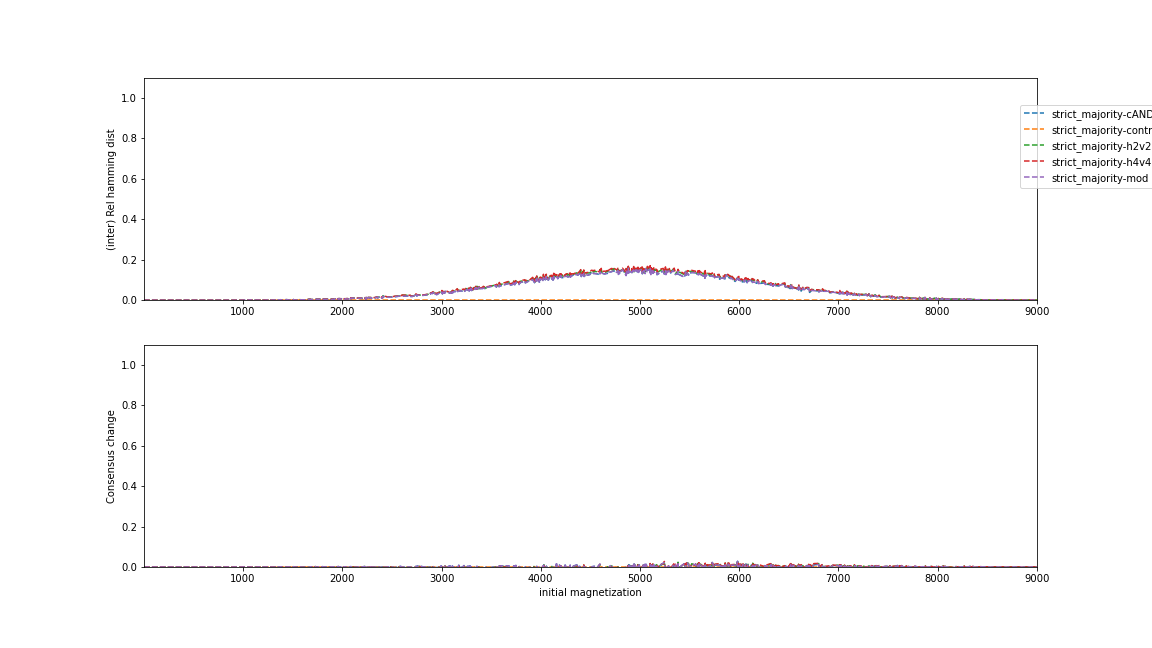}
		\caption{Relative hamming distance and consensus change between strict majority global function and same global function with different Woronin bodies activation.}
		\label{fig:str_maj_inter}
	\end{figure}

	Using MFA final state with global functions without Wb activation as a point of comparison we get metrics to quantify the change between these reference points and the activation of Woronin bodies in sequence previously defined. Figures~\ref{fig:maj_inter} and ~\ref{fig:str_maj_inter} are the plots representing this changes.

\begin{figure}[!htb]
		\centering
		\includegraphics[width=0.95\linewidth]{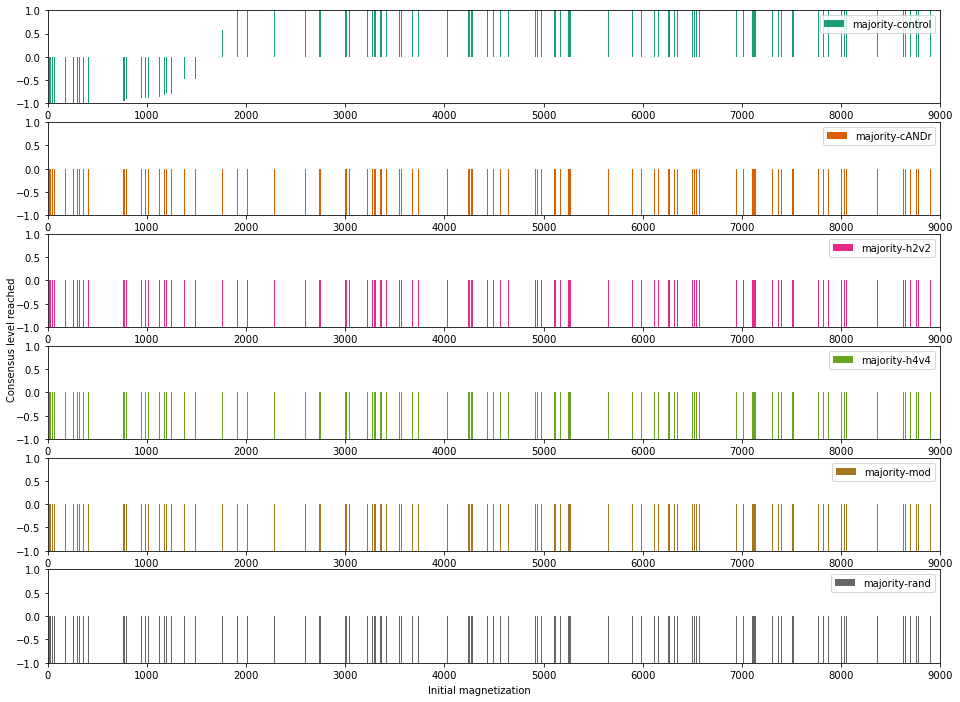}
		\caption{Level of consensus reached in final generation against initial magnetization for skew-majority.}
		\label{fig:maj_con}
	\end{figure}

\begin{figure}[!htb]
		\centering
		\includegraphics[height=0.40\textheight]{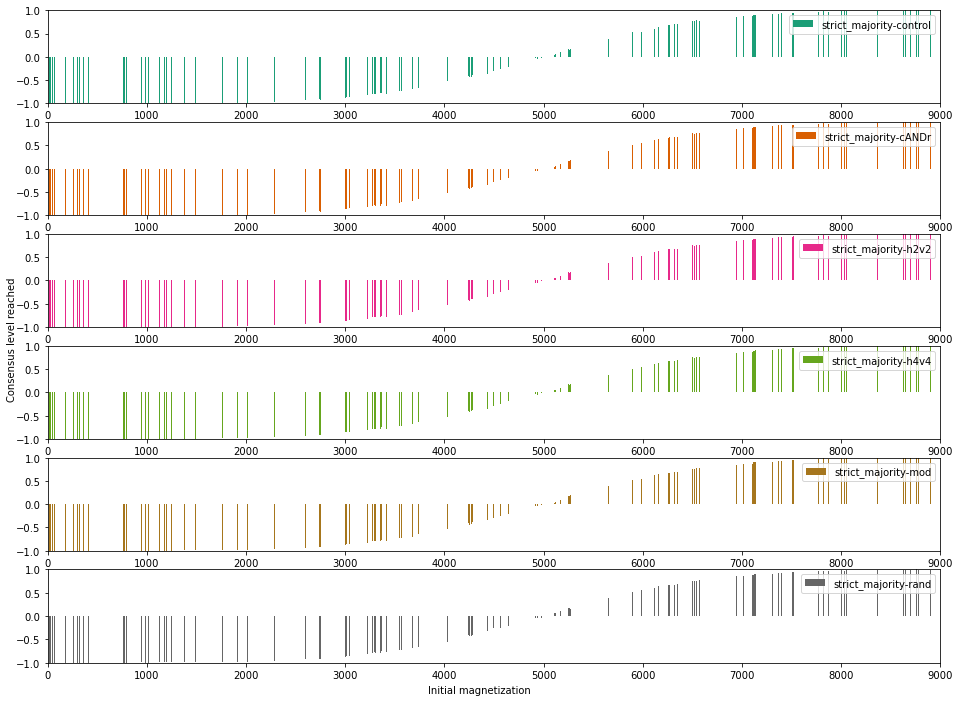}
		\caption{Level of consensus reached in final generation against initial magnetization for strict majority}
		\label{fig:str_maj_con}
	\end{figure}

	To get a better understanding of consensus dynamics due Woronin bodies activation, we plot the consensus achieved in the final step of every simulation getting the Figures ~\ref{fig:maj_con} and ~\ref{fig:str_maj_con}.
	
	\section{Discussion}
	
	The activation of Woronin bodies (wb) in EFCA has direct impact in system dynamics. This is shown for example in rule 32 where the Wbs activation didn't change transient of cycle, but instead changes his period. In rule 90 on a ring of size 32, the Wb activation introduces cycles where previously fixed points exist. Same changes in dynamics happen on size 12 ring and in all rules implemented, and rules 57-99 increase complexity behavior due the activation of Woronin bodies were other group decrease complexity, leading to more trivial behaviors.
	
	In MFA at first glance we have the intuition that Wb could change the consensus dynamics, perhaps transforming a un-skew function as strict majority into skew one or kind of behaviors. The data shows strict majority is not affected in a significant way by the Wbs activation, achieving same dynamics of the global function without Wb activated. In the case of skew-majority the skew is augmented making that all initial configuration tried, quickly converge to zero (-1 when in consensus), which means that no mater what we do, in this network the negative alternative/opinion will always win.
	
	\section{Aknowledgements}
	This work was supported by Centro de Modelamiento Matem\'atico (CMM), FB210005 BASAL funds for centers of excellence from ANID-Chile, FONDECYT 1200006 (E.G.) and ANID FONDECYT Postdoctorado  3220205 (M.R-W).

	\appendix
	
	\section{APPENDIX}\label{doc:app1}

	\begin{longtable}{ r r r | r r r | r r r | r r r }
		\caption{Rules with his right and left rule equivalent.} \\
		\hline 
		rule & Rrule & Lrule &rule & Rrule & Lrule &rule & Rrule & Lrule &rule & Rrule & Lrule \\ 
		\hline
		\endfirsthead
		\multicolumn{12}{c}
		{{\bfseries \tablename\ \thetable{} -- continued from previous page}} \\ 
		rule & Rrule & Lrule &rule & Rrule & Lrule &rule & Rrule & Lrule &rule & Rrule & Lrule \\
		\endhead
		\hline \multicolumn{6}{r}{{Continued on next page $\ldots$}} \\ \hline
		\endfoot
		
		\endlastfoot

		0 & 0 & 0 & 64 & 192 & 68 & 128 & 192 & 136 & 192 & 192 & 204 \\  
		\hline 
		1 & 0 & 0 & 65 & 192 & 68 & 129 & 192 & 136 & 193 & 192 & 204 \\  
		\hline 
		2 & 0 & 0 & 66 & 192 & 68 & 130 & 192 & 136 & 194 & 192 & 204 \\  
		\hline 
		3 & 3 & 0 & 67 & 195 & 68 & 131 & 195 & 136 & 195 & 195 & 204 \\  
		\hline 
		4 & 12 & 68 & 68 & 204 & 68 & 132 & 204 & 204 & 196 & 204 & 204 \\  
		\hline 
		5 & 12 & 68 & 69 & 204 & 68 & 133 & 204 & 204 & 197 & 204 & 204 \\  
		\hline 
		6 & 12 & 68 & 70 & 204 & 68 & 134 & 204 & 204 & 198 & 204 & 204 \\  
		\hline 
		7 & 15 & 68 & 71 & 207 & 68 & 135 & 207 & 204 & 199 & 207 & 204 \\  
		\hline 
		8 & 12 & 136 & 72 & 204 & 204 & 136 & 204 & 136 & 200 & 204 & 204 \\  
		\hline 
		9 & 12 & 136 & 73 & 204 & 204 & 137 & 204 & 136 & 201 & 204 & 204 \\  
		\hline 
		10 & 12 & 136 & 74 & 204 & 204 & 138 & 204 & 136 & 202 & 204 & 204 \\  
		\hline 
		11 & 15 & 136 & 75 & 207 & 204 & 139 & 207 & 136 & 203 & 207 & 204 \\  
		\hline 
		12 & 12 & 204 & 76 & 204 & 204 & 140 & 204 & 204 & 204 & 204 & 204 \\  
		\hline 
		13 & 12 & 204 & 77 & 204 & 204 & 141 & 204 & 204 & 205 & 204 & 204 \\  
		\hline 
		14 & 12 & 204 & 78 & 204 & 204 & 142 & 204 & 204 & 206 & 204 & 204 \\  
		\hline 
		15 & 15 & 204 & 79 & 207 & 204 & 143 & 207 & 204 & 207 & 207 & 204 \\  
		\hline 
		16 & 0 & 0 & 80 & 192 & 68 & 144 & 192 & 136 & 208 & 192 & 204 \\  
		\hline 
		17 & 0 & 17 & 81 & 192 & 85 & 145 & 192 & 153 & 209 & 192 & 221 \\  
		\hline 
		18 & 0 & 0 & 82 & 192 & 68 & 146 & 192 & 136 & 210 & 192 & 204 \\  
		\hline 
		19 & 3 & 17 & 83 & 195 & 85 & 147 & 195 & 153 & 211 & 195 & 221 \\  
		\hline 
		20 & 12 & 68 & 84 & 204 & 68 & 148 & 204 & 204 & 212 & 204 & 204 \\  
		\hline 
		21 & 12 & 85 & 85 & 204 & 85 & 149 & 204 & 221 & 213 & 204 & 221 \\  
		\hline 
		22 & 12 & 68 & 86 & 204 & 68 & 150 & 204 & 204 & 214 & 204 & 204 \\  
		\hline 
		23 & 15 & 85 & 87 & 207 & 85 & 151 & 207 & 221 & 215 & 207 & 221 \\  
		\hline 
		24 & 12 & 136 & 88 & 204 & 204 & 152 & 204 & 136 & 216 & 204 & 204 \\  
		\hline 
		25 & 12 & 153 & 89 & 204 & 221 & 153 & 204 & 153 & 217 & 204 & 221 \\  
		\hline 
		26 & 12 & 136 & 90 & 204 & 204 & 154 & 204 & 136 & 218 & 204 & 204 \\  
		\hline 
		27 & 15 & 153 & 91 & 207 & 221 & 155 & 207 & 153 & 219 & 207 & 221 \\  
		\hline 
		28 & 12 & 204 & 92 & 204 & 204 & 156 & 204 & 204 & 220 & 204 & 204 \\  
		\hline 
		29 & 12 & 221 & 93 & 204 & 221 & 157 & 204 & 221 & 221 & 204 & 221 \\  
		\hline 
		30 & 12 & 204 & 94 & 204 & 204 & 158 & 204 & 204 & 222 & 204 & 204 \\  
		\hline 
		31 & 15 & 221 & 95 & 207 & 221 & 159 & 207 & 221 & 223 & 207 & 221 \\  
		\hline 
		32 & 0 & 0 & 96 & 192 & 68 & 160 & 192 & 136 & 224 & 192 & 204 \\  
		\hline 
		33 & 0 & 0 & 97 & 192 & 68 & 161 & 192 & 136 & 225 & 192 & 204 \\  
		\hline 
		34 & 0 & 34 & 98 & 192 & 102 & 162 & 192 & 170 & 226 & 192 & 238 \\  
		\hline 
		35 & 3 & 34 & 99 & 195 & 102 & 163 & 195 & 170 & 227 & 195 & 238 \\  
		\hline 
		36 & 12 & 68 & 100 & 204 & 68 & 164 & 204 & 204 & 228 & 204 & 204 \\  
		\hline 
		37 & 12 & 68 & 101 & 204 & 68 & 165 & 204 & 204 & 229 & 204 & 204 \\  
		\hline 
		38 & 12 & 102 & 102 & 204 & 102 & 166 & 204 & 238 & 230 & 204 & 238 \\  
		\hline 
		39 & 15 & 102 & 103 & 207 & 102 & 167 & 207 & 238 & 231 & 207 & 238 \\  
		\hline 
		40 & 12 & 136 & 104 & 204 & 204 & 168 & 204 & 136 & 232 & 204 & 204 \\  
		\hline 
		41 & 12 & 136 & 105 & 204 & 204 & 169 & 204 & 136 & 233 & 204 & 204 \\  
		\hline 
		42 & 12 & 170 & 106 & 204 & 238 & 170 & 204 & 170 & 234 & 204 & 238 \\  
		\hline 
		43 & 15 & 170 & 107 & 207 & 238 & 171 & 207 & 170 & 235 & 207 & 238 \\  
		\hline 
		44 & 12 & 204 & 108 & 204 & 204 & 172 & 204 & 204 & 236 & 204 & 204 \\  
		\hline 
		45 & 12 & 204 & 109 & 204 & 204 & 173 & 204 & 204 & 237 & 204 & 204 \\  
		\hline 
		46 & 12 & 238 & 110 & 204 & 238 & 174 & 204 & 238 & 238 & 204 & 238 \\  
		\hline 
		47 & 15 & 238 & 111 & 207 & 238 & 175 & 207 & 238 & 239 & 207 & 238 \\  
		\hline 
		48 & 48 & 0 & 112 & 240 & 68 & 176 & 240 & 136 & 240 & 240 & 204 \\  
		\hline 
		49 & 48 & 17 & 113 & 240 & 85 & 177 & 240 & 153 & 241 & 240 & 221 \\  
		\hline 
		50 & 48 & 34 & 114 & 240 & 102 & 178 & 240 & 170 & 242 & 240 & 238 \\  
		\hline 
		51 & 51 & 51 & 115 & 243 & 119 & 179 & 243 & 187 & 243 & 243 & 255 \\  
		\hline 
		52 & 60 & 68 & 116 & 252 & 68 & 180 & 252 & 204 & 244 & 252 & 204 \\  
		\hline 
		53 & 60 & 85 & 117 & 252 & 85 & 181 & 252 & 221 & 245 & 252 & 221 \\  
		\hline 
		54 & 60 & 102 & 118 & 252 & 102 & 182 & 252 & 238 & 246 & 252 & 238 \\  
		\hline 
		55 & 63 & 119 & 119 & 255 & 119 & 183 & 255 & 255 & 247 & 255 & 255 \\  
		\hline 
		56 & 60 & 136 & 120 & 252 & 204 & 184 & 252 & 136 & 248 & 252 & 204 \\  
		\hline 
		57 & 60 & 153 & 121 & 252 & 221 & 185 & 252 & 153 & 249 & 252 & 221 \\  
		\hline 
		58 & 60 & 170 & 122 & 252 & 238 & 186 & 252 & 170 & 250 & 252 & 238 \\  
		\hline 
		59 & 63 & 187 & 123 & 255 & 255 & 187 & 255 & 187 & 251 & 255 & 255 \\  
		\hline 
		60 & 60 & 204 & 124 & 252 & 204 & 188 & 252 & 204 & 252 & 252 & 204 \\  
		\hline 
		61 & 60 & 221 & 125 & 252 & 221 & 189 & 252 & 221 & 253 & 252 & 221 \\  
		\hline 
		62 & 60 & 238 & 126 & 252 & 238 & 190 & 252 & 238 & 254 & 252 & 238 \\  
		\hline 
		63 & 63 & 255 & 127 & 255 & 255 & 191 & 255 & 255 & 255 & 255 & 255 \\  
		\hline
		\label{tab:eqRule} 
	\end{longtable}

\bibliographystyle{unsrtnat}
\bibliography{Exploring_the_dynamics_of_fungal_cellular_automata_arxiv}

\end{document}